\documentclass{article}

\RequirePackage{amsthm,amsmath,amsfonts,amssymb}
\RequirePackage[authoryear]{natbib}
\RequirePackage[dvipsnames]{xcolor}
\RequirePackage{graphicx}%% uncomment this for including figures
\RequirePackage{tikz}
\usetikzlibrary{arrows.meta, positioning, shapes, fit, backgrounds}
\RequirePackage{booktabs}
\RequirePackage{makecell}
\RequirePackage{algorithm,algorithmic} 
\RequirePackage{multirow} 
\RequirePackage{makecell} 
\RequirePackage{pifont} 
\RequirePackage{pdflscape}
\RequirePackage{url}
\RequirePackage{tabularx}
\RequirePackage{float}
\RequirePackage{placeins}
\RequirePackage{array}
\RequirePackage{subcaption}
\RequirePackage{geometry}

\usetikzlibrary{positioning, arrows.meta}

\newcolumntype{L}[1]{>{\raggedright\arraybackslash}m{#1}}
\newcolumntype{C}[1]{>{\centering\arraybackslash}p{#1}}
\DeclareMathOperator*{\argmax}{arg\,max}

\makeatletter
\newcommand{\equalcontrib}{\csxappto{author@\the\c@author @ead@marks}{,t1}}
\makeatother

\newcommand{\indep}{\perp \!\!\! \perp}
\newcommand{\Prob}{\mathbb{P}}
\newcommand{\E}{\mathbb{E}}

\definecolor{Goldenrod}{RGB}{218,165,32}
\definecolor{orange}{RGB}{204,85,0}
\definecolor{DarkEmerald}{RGB}{0,100,70}

\begin{document}

\title{Toward Joint Prediction of a Longitudinal Marker and a Terminal Event: A bivariate discrete-time framework}

\author{
Stephanie Armbruster\thanks{Department of Biostatistics, Harvard T.H. Chan School of Public Health, 677 Huntington Avenue, Boston, MA 02115. Email: sarmbruster@g.harvard.edu} 
\and 
Daniel Kramer\thanks{Richard A. and Susan F. Smith Center for Outcomes Research, Beth Israel Deaconess Medical Center, 375 Longwood Road, Suite 440, Boston, MA 02215. Email: dkramer@bidmc.harvard.edu} 
\and 
Rui Duan\thanks{Department of Biostatistics, Harvard T.H. Chan School of Public Health, 677 Huntington Avenue, Boston, MA 02115. Email: rduan@hsph.harvard.edu} 
\and 
Rajarshi Mukherjee\thanks{Department of Biostatistics, Harvard T.H. Chan School of Public Health, 677 Huntington Avenue, Boston, MA 02115. Email: ram521@mail.harvard.edu} 
\and 
Sebastien Haneuse\thanks{Department of Biostatistics, Harvard T.H. Chan School of Public Health, 677 Huntington Avenue, Boston, MA 02115. Email: shaneuse@hsph.harvard.edu. Contributed equally as senior author.} 
\and 
Harrison Reeder\thanks{Biostatistics, Massachusetts General Hospital, 399 Revolution Drive, Somerville, MA 02145; and Department of Medicine, Harvard Medical School, 25 Shattuck Street, Boston, MA 02115. Email: hreeder@mgh.harvard.edu. Contributed equally as senior author.}
}

\date{}
\maketitle

\begin{abstract}
Sudden cardiac death (SCD) is a leading cause of death in the U.S. 
Patients at elevated risk of SCD are primarily treated with an implantable cardioverter-defibrillator (ICD) which may prevent death from cardiovascular causes but may cause severe side effects, such as reduced quality of life from shock-induced pain. 
Decisions about ICD treatment, therefore, involve complex personal trade-offs across multiple health events, including mortality and quality of life.
While prediction tools could help patients and providers weigh these trade-offs, they commonly focus on univariate outcomes.
At best, prediction tools treat other clinical endpoints as inputs or features, so that trade-offs cannot be directly informed.

To address this, we propose a novel general Bayesian framework that jointly models a terminal event and a longitudinal marker as a bivariate process over discrete time, for settings where prediction is the primary goal. 
Discretization of study time lets the bivariate discrete-time framework capture the dynamic interplay between outcomes over time while avoiding implicit extrapolation beyond truncation by a terminal event, such as death.
The framework flexibly accommodates a range of phenomena that may arise in applied contexts; this includes global time-invariant and local time-dependent dependence structures between the terminal event and the longitudinal marker and latent association via a shared frailty term. 
Estimation proceeds via the Bayesian paradigm, which yields patient-specific joint posterior predictions for the time to terminal event and the future longitudinal marker trajectory. 
We introduce the bivariate discrete-time framework with a focus on its modeling flexibility, provide guidance on choosing the discretization of study time as well as on Bayesian model construction and selection, and discuss insights gained from the generated joint posterior predictions. 
Furthermore, we demonstrate the framework's clinical relevance and practicality in the context of SCD and ICD therapy using data from the Sudden Cardiac Death in Heart Failure Trial (SCD-HeFT), an important ICD-related benchmark trial. 

\end{abstract}


 \clearpage

\section{Introduction}
\label{sec: introduction}
Sudden cardiac death (SCD) is caused by an unexpected fatal arrhythmia which occurs within an hour of symptom onset \citep{zipes1998sudden}. 
In 2023, approximately 700,000 SCD were reported, making it the primary cause of death in the U.S.\citep{ahmad2024mortality}. 
Patients at an increased risk of SCD from either inherited or acquired cardiac conduction may receive implantable cardioverter-defibrillators (ICDs), an electronic device that continuously monitors cardiac rhythm and, upon registering an irregular heartbeat, delivers electric shocks to the heart to restore a normal heart rhythm \citep{dimarco2003implantable}.
Multiple randomized primary-prevention trials have demonstrated that ICD therapy reduces mortality in well-selected recipients \citep{bardy2005amiodarone}.
Side effects to ICD implantation, however, may include anxiety, depression, and diminished quality of life from shock-induced pain and trauma, especially among elderly patients nearing the end of their life \citep{kramer2017d}. 
% goldstein2004management, magyar2011prevalence, 
As a consequence, current guidelines recommend patients, their families, and their physicians to engage in a comprehensive discussion when deciding for or against an ICD implantation and when managing ICD care post implantation \citep{alkhatib_2018_aha_guidelines, CMS_ICD_decision_memo_2018, icd_patient_decision_aid}.
% Such conversation might include questions like: What is my predicted quality-of-life trajectory and mortality risk, with an ICD and without? What consequences does my predicted health outlook have for managing my ICD implant; should it be turned off or extracted? 

Medical decision making commonly involves striking a balance across multiple health-related outcomes \citep{haneuse2020invited}. 
Yet, most clinical prediction tools focus on one single `primary' clinical outcome, such as mortality or quality of life, univariately. 
At best, mortality is treated as a nuisance for the prediction of longitudinal markers, and longitudinal markers are leveraged only as inputs or features when predicting mortality \citep{van2007dynamic}. 
In each case, the prediction tools are misaligned with a patient reality in which both endpoints are joint outcomes of interest, and, hence, insufficiently inform patients and physicians as they strike trade-offs across multiple dimensions of health and competing health risks. 

For joint outcomes comprising a non-terminal and terminal time-to-event, also known as semi-competing risks, a well-developed body of statistical methods exists to jointly predict joint incidence curves \citep{fine2001semi,haneuse2016semi}. 
More generally, multi-state models seek to capture disease progression by modeling a patient's health journey through multiple health states \citep{andersen2002multi, beyersmann2011competing}. 
From these, one can jointly predict patient-specific cumulative incidence curves for relevant health states, conditional on prior disease status and patient-specific latent frailty values \citep{klein1993plotting, reeder2019joint}.

To the best of our knowledge, however, a similarly comprehensive and interpretable framework for joint prediction is not available in settings with a general longitudinal trajectory and terminal event, when both outcomes are of intrinsic clinical interest.
Existing statistical methods, which jointly analyze longitudinal measures and time-to-event outcomes, focus on either estimating the impact of the longitudinal marker on the terminal-event outcome, prioritize prediction of a terminal event using longitudinal measures as dynamic features, or result in longitudinal predictions with limitations to clinical interpretability. 
In recent years, joint longitudinal-survival models \citep{rizopoulos2012joint} and joint latent-class models \citep{proust2014joint} have comprised an active area of research; they construct submodels for a longitudinal marker and terminal event and link them statistically by some dependence structure such as shared random effects or latent classes.
% The standard models have been expanded to capture complex and diverse clinical contexts; e.g. issues of latent disease onset in ALS diagnosis \citep{ortholand2026joint} by modeling latent disease age, impacts of biomarker variability on Alzheimer risk \citep{courcoul2026joint} or on cardio- and cerebrovascular disease risk \citep{courcoul2025location} and adjustments for informative measurement processes \citep{gupta2021flexible} or informative drop-out \citep{doms2025joint}. 
However, any joint longitudinal-survival or latent-class model adopts a mixed model for the longitudinal trajectory, which results in longitudinal predictions that implicitly extrapolate beyond any observed terminal event; in a sense, such models predict the trajectory for an `immortal cohort' in which the terminal event cannot occur \citep{chaix2012commentary, rizopoulos2012joint, wen2018methods}. 
A more interpretable longitudinal trajectory would reflect the marker value across time among those who are actually still alive, or more generally, a dynamic cohort of patients who have not yet experienced the terminal event \citep{dufouil2004analysis}. 
This quantity is referred to as a `partly conditional' mean and can be modeled using generalizing estimating equations (GEE) \citep{kurland2005directly}. 
Yet, this approach is oriented towards studies of association rather than joint prediction, as it considers truncation by the terminal event as a observable nuisance rather than an outcome of interest. 
To-date, there have not been joint modeling approaches that facilitate joint prediction of time to terminal event and a longitudinal trajectory while retaining its natural `partly conditional' interpretation.

With this background, this paper proposes a novel general Bayesian framework to jointly predict a terminal event process and a longitudinal marker trajectory over time. 
The proposed bivariate discrete-time framework uses time discretization to eliminate implicit extrapolation of the longitudinal trajectory beyond truncation by the terminal event.
It admits flexible regression structures to capture patterns in the longitudinal trajectory, including random effects and autoregressive dependence.
It accommodates flexible specification of the terminal event hazard, including global time-invariant and local time-dependent dependence on the longitudinal trajectory \citep{nevo2022modeling}, as well as additional latent dependence between the outcomes via a shared frailty term. 
The clinical relevance of the proposed method is motivated by the influential cardiovascular trial, SCD-HeFT, in Section \ref{sec: data}. 
This trial, though published in 2005 \citep{bardy2005amiodarone}, continues to influence professional practice guidelines and thus remains an important benchmark for considering patient trajectories after ICD implantation.
The framework is formally introduced in Section \ref{sec: method}. 
Section \ref{sec: bayes} elaborates on estimation via the Bayesian paradigm, including how to set prior distributions and how, by leveraging the probabilistic programming language Stan, the framework provides joint predictions of the terminal event time and the future longitudinal trajectory. 
Section \ref{sec: model selection} addresses how to construct a set of candidate models and select a model with best goodness-of-fit. 
Section \ref{sec: practical considerations} discusses the choice of discrete-time partition, and practical recommendations for implementation. 
In Section \ref{sec: simulation}, we show unbiased estimation of regression parameters and investigate model selection performance based on simulation studies.  
In Section \ref{sec: application}, we apply the bivariate discrete-time framework to the SCD-HeFT data, and jointly predict quality of life and time to death.
% We find that quality-of-life measurements inform time to death beyond indirect associations based on a patient's age and biological sex.
Section \ref{sec: discussion} elaborate on patient-relevant insight that is gained from the joint predictions and discusses avenues for future work.

\section{SCD-HeFT data: treatment to reduce risk of sudden death}
\label{sec: data}

\subsection{Informed decision making based on joint prediction}
When an individual is identified to be at high risk of SCD, they may consider a range of interventions, including non-invasive interventions, e.g. lifestyle changes and medication, or invasive interventions, in particular an ICD implantation \citep{icd_patient_decision_aid}. 
A decision around ICD therapy requires balancing the potential to avert SCD against the risk of device-induced pain,  diminished quality of life and disease progression; yet, frequently, such discussions are framed as requiring patients to prioritize longevity versus quality of life. 
Hence, we aim to develop a prediction tool which better informs a patient how their mortality risk and quality of life will jointly develop over time if they received an ICD implant. 

\subsection{The Sudden Cardiac Death in Heart Failure Trial}
The SCD-HeFT (Sudden Cardiac Death in Heart Failure Trial) was a prospective randomized controlled trial whcih investigated the treatment effect of an ICD, compared to a pharmacological intervention, on preventing heart failure events, including cardiovascular death \citep{bardy2005amiodarone}. 
The trial was described in detail elsewhere \citep{bardy2005amiodarone}. 
Here, we focus on the subpopulation of $N = 817$ patients who received an ICD device and survived beyond their first follow-up appointment after one week. 
The baseline survey recorded biological sex, age at randomization, blood markers and comorbidities at time of randomization (Table A1 in Appendix A). 
Following randomization, patients were assessed at regular follow-up appointment up to 72 months \citep{bardy2005amiodarone}. 
The median follow up was 44.5 months among ICD patients, and $N = 171$ ($20.8\%$) patients died (all-cause death) during follow-up.
The SCD-HeFT patient survey queried quality of life at week 1 and every 6 months after ICD implantation, using the Minnesota Living with Heart Failure Questionnaire (MLHFQ) \citep{mark2008quality}.
The MLHFQ questionnaire produces a validated clinical score with range $[0, 105]$; a higher score value indicates diminished quality of life.

\subsection{ICD treatment, mortality risk and quality-of-life trajectoryies in the SCD-HeFT data}
\begin{figure}[h!]
    \centering
    \begin{subfigure}[c]{0.52\linewidth}
        \centering
        \includegraphics[width=\linewidth]{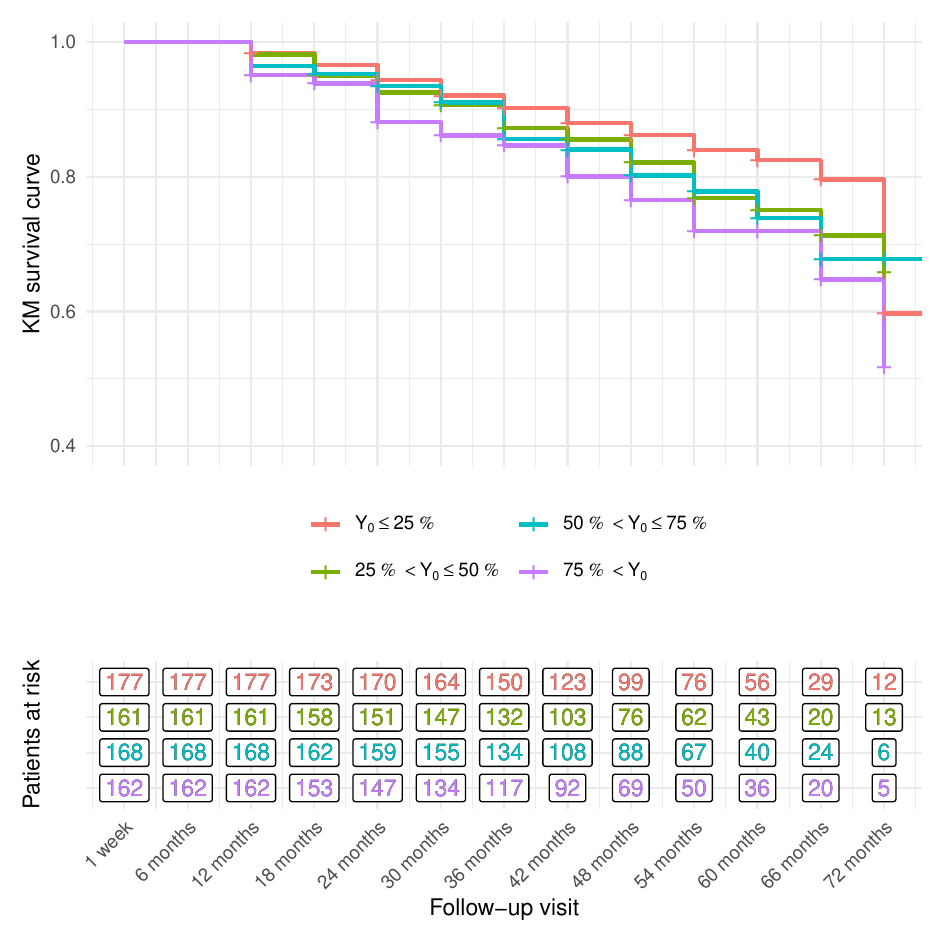}
        \caption{Stratified Kaplan-Meier estimate.}
        \label{fig:strat_KM_estimate}
    \end{subfigure}
    \hfill
    \begin{subfigure}[c]{0.46\linewidth}
        \centering
        \includegraphics[width=\linewidth]{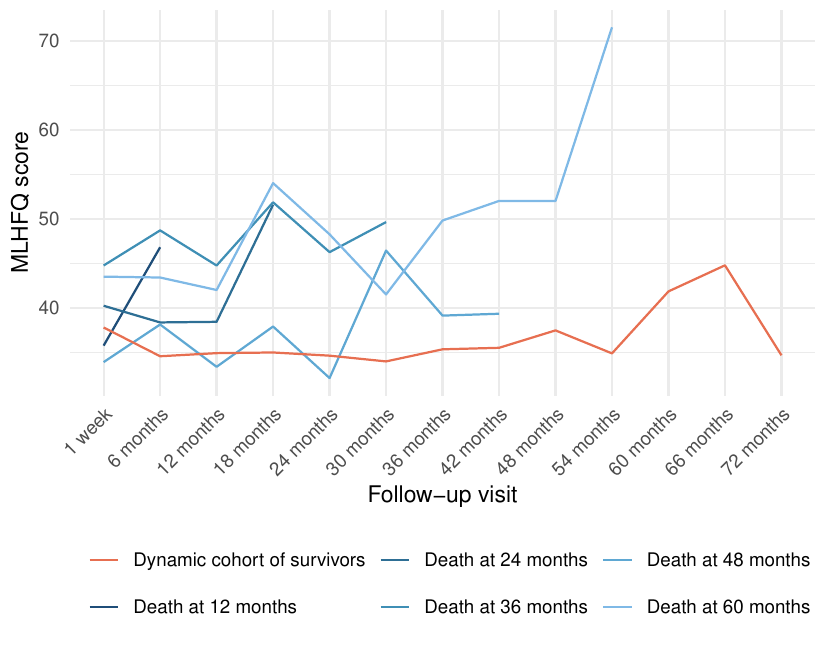}
        \caption{Partly-conditional mean MLHFQ score trajectory.}
        \label{fig:long_mean_traj}
        % \vspace{1em}
        \includegraphics[width=\linewidth]{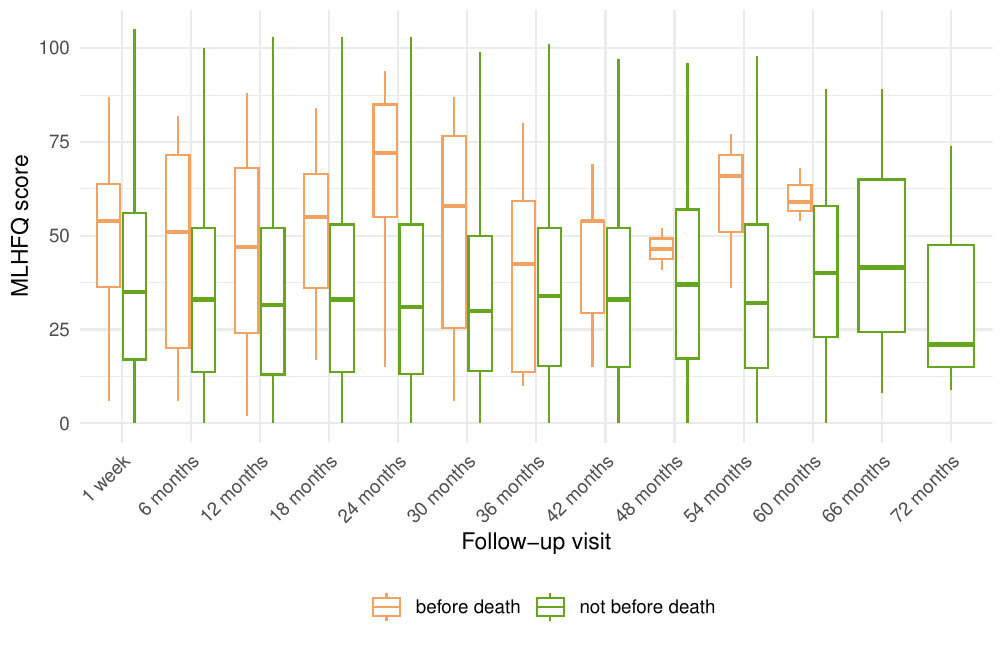}
        \caption{Boxplot for MLHFQ score values prior and not prior to death.}
        \label{fig:density_before_death}
    \end{subfigure}
    \caption{Left: stratified Kaplan-Meier estimate for survival probability for all-cause mortality and the number of patients at risk over discrete time, stratified by quartiles of MLHFQ score measured at baseline. Right: comparison of mean MLHFQ score trajectories for the dynamic cohort of survivors (in red), mean MLHFQ score trajectory for the cohort of patients who do not die (in orange), fully-conditional mean MLHFQ score trajectories according to time to death (in blue); and boxplots for MLHFQ score values recorded at last follow-up visit before death vs. other follow-up visits not prior to a patient's death.}
    \label{fig:combined}
\end{figure}

Prior analysis of the SCD-HeFT data evaluated quality of life and mortality univariately, suggesting that an ICD implant improved quality of life, and also prolonged all-cause survival \citep{mark2008quality}. 
Here, we consider a joint exploratory data analysis for time to death and quality of life to explicitly investigate their dependence.  
Figure \ref{fig:strat_KM_estimate} visualizes Kaplan-Meier survival curves, stratified by MLHFQ score quartiles at baseline. 
The mortality risk over time was higher among patients who had poorer quality of life at baseline, suggesting that quality of life contains prognostic information for survival.

Figure \ref{fig:long_mean_traj} depicts the mean MLHFQ score trajectory for several cohorts defined in various ways based on mortality.  
The red curve describes the `partly-conditional' MLHFQ score trajectory among the the dynamic cohort of survivors; that is, the mean score at each particular time point among individuals who are still alive at said time point \citep{kurland2009longitudinal}. 
Each blue curve shows the `fully-conditional' MLHFQ score trajectory for a set of death-times; that is, the mean MLHFQ score trajectory that was computed across time for a cohort of individual who all died at a particular point in time. 
We consider death time at month 12, month 24, month 36, month 48 or month 60. 
Each fully-conditional MLHFQ score trajectory increases until time to death, and lies well above the partly-conditional trajectory; this shows that the quality of life degrades more strongly before a patient dies than it would if the patient remained alive.
Notably, fully-conditional MLHFQ score mean at the time point preceding death varies depending on the time of death: among those who die at 12 months, the 6-month mean is 46, while among those who die at 60 months, the 54-month mean is 72. 
This heterogeneity in MLHFQ score trajectories by time to death  suggests that the dependence between MLHFQ score and mortality risk may change over time. 
Similarly, Figure \ref{fig:density_before_death} reports for each time point the distribution of MLHFQ scores among those who die within 6 months relative to those who do not. 
At each visit, MLHFQ marker values are consistently higher among those who die at the subsequent visit, reflecting a clear proximal relationship between declining quality of life and mortality.
The distribution among those who do not die in the following interval has a similar shape and location across time, indicating relative consistency in quality of life in periods more distant from mortality. 
By contrast, the distribution of MLHFQ scores for patients who die at the subsequent visit shifts over time, indicates that the dependence between mortality risk and the MLHFQ score may itself vary over time. 

Overall, these observations highlight that the mortality risk and the MLHFQ score trajectory are associated via a complex, time-varying dependence structure. 
Motivated by these preliminary analyses, we propose the novel bivariate discrete-time framework that generates patient-specific joint predictions of a partly-conditional outcome trajectory and time to death, while acknowledging the complex dependence structure. 
Based in the SCD-HeFT data, we further illustrate how such joint predictions can support shared healthcare decision making for ICD therapy.

\section{The bivariate discrete-time framework} 
\label{sec: method}
\subsection{Outcome notation for the longitudinal marker trajectory and the time to terminal event}
\label{subsec: joint outcome}
The terminal event and the longitudinal marker arise in continuous time. 
Without loss of generality, we set observation time to start at some natural `origin'; in SCD-HeFT, the time origin would be the end of week 1. 
Furthermore, without loss of generality, we take the goal to be joint prediction until some time horizon, $\tau$.   
The time horizon, taken to be pre-specified and context-specific, is chosen to be 6 years for SCD-HeFT. 
With that, let $T_i \in (0, \tau]$ denote the time-to-terminal-event outcome and let $C_i \in (0, \tau]$ denote the time-to-right-censoring outcome for patient $i = 1, \dots, n$. 
In continuous time, we observe the event time $\tilde{T}_i = \min(T_i, C_i)$ and an event indicator $\Delta_i  = \mathbb{I}(T_i < C_i)$. 
The terminal event counting process $N_i(t) = \mathbb{I}(T_i \leq t)$ indicates if a patient has experienced the terminal event before or at time $t \in (0, \tau]$; in SCD-HeFT, $N_i(t) = 1$ encodes that a patient died at or before time $t$. 
The univariate longitudinal marker trajectory, measured over time, is denoted by $Y_i(t) \in \mathbb{R}$; in the SCD-HeFT context, we consider the MLHFQ score. 

Practically, capturing the dynamic between the longitudinal marker and terminal event over time is complex. 
To resolve this challenge, one way forward is to transform continuous time into discrete time \citep{suresh2022survival}. 
Time discretization is ubiquitous in causal inference when estimating treatment effects for time-to-event outcomes in the presence of time-varying confounding; a popular choice is marginal structural modeling to estimate discrete-time hazard function \citep{dagostino1990relation,hernan2000marginal,hernan2016using}. 
Following this tradition, the proposed framework also partitions continuous time into discrete intervals which retains the partly-conditional interpretability of the joint predictions. 

\subsection{A discrete time representation of observed data}
\begin{figure}[h!]
\centering
\tiny
\begin{tikzpicture}

% Timeline i = 1
\node at (-0.5,10) {$i = 1$};

\draw[very thick, color = gray] (0,10) -- (13,10);
\foreach \x/\label in {0/$0$, 2/$\tau_1$, 4/$\tau_2$, 6/$\tau_3$, 8/$\tau_4$, 10/$\tau_5$, 12/$\tau_6$} {
    \draw[thick] (\x,9.85) -- (\x,10.15);
    \node at (\x,9.6) {\label};
}

\draw[very thick, color = black] (0,10) -- (12,10); 

% \node[black] at (0, 10.5) {$L_i$};
\node[gray] at (0, 9.2) {$(N_{0, i}, Y_{0, i})$}; 
\node[RoyalBlue] at (2, 9.2) {$(N_{1, i}, Y_{1, i})$}; 
\node[RoyalBlue] at (4, 9.2) {$(N_{2, i}, Y_{2, i})$}; 
\node[RoyalBlue] at (6, 9.2) {$(N_{3, i}, Y_{3, i})$}; 
\node[RoyalBlue] at (8, 9.2) {$(N_{4, i}, Y_{4, i})$};
\node[RoyalBlue] at (10, 9.2) {$(N_{5, i}, Y_{5, i})$};
\node[Goldenrod, font=\scriptsize] at (12, 10.3) {$(k_i^r = 6)$};
% \node[black] at (12, 10.5) {$T_i$};
\node[BrickRed] at (12, 9.2) {$N_{6, i} = 1$};

% Timeline i = 2
\node at (-0.5,8.5) {$i = 2$};

\draw[very thick, color = gray] (0,8.5) -- (13,8.5);
\foreach \x/\label in {0/$0$, 2/$\tau_1$, 4/$\tau_2$, 6/$\tau_3$, 8/$\tau_4$, 10/$\tau_5$, 12/$\tau_6$} {
    \draw[thick] (\x,8.35) -- (\x,8.65);
    \node at (\x,8.1) {\label};
}

\draw[very thick, color = black] (0,8.5) -- (8,8.5); 

% \node[black] at (0, 8.0) {$L_i$};
\node[gray] at (0, 7.7) {$(N_{0, i}, Y_{0, i})$}; 
\node[RoyalBlue] at (2, 7.7) {$(N_{1, i}, Y_{1, i})$}; 
\node[RoyalBlue] at (4, 7.7) {$(N_{2, i}, Y_{2, i})$}; 
\node[RoyalBlue] at (6, 7.7) {$(N_{3, i}, Y_{3, i})$}; 
\node[RoyalBlue] at (8, 7.7) {$(N_{4, i}, Y_{4, i})$};
\node[Goldenrod, font=\scriptsize] at (10, 8.8) {$(k_i^r = 5)$};
\node[DarkEmerald] at (10, 7.7) {$C_i$};
\end{tikzpicture}
\caption{Bivariate discrete-time process for terminal event indicator $N_{k, i}$ and longitudinal marker $Y_{k, i}$ (in blue), truncated by terminal event occurrence (in red), or right censored (in green); baseline marker value at time origin (in gray). The upper boundary index $k_i^r$ is indicated in yellow.}
\label{fig:example_patients}
\end{figure}
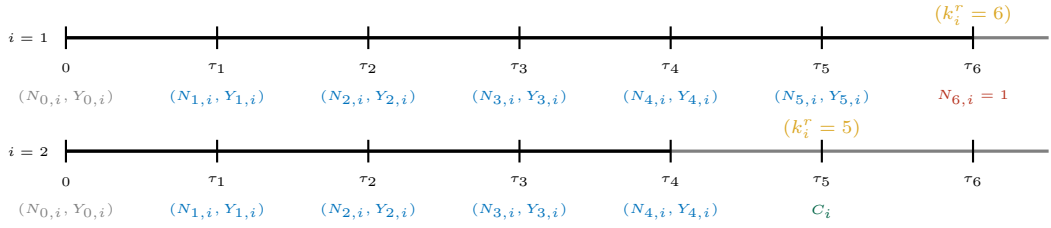

The discrete-time partition is determined by a sequence of boundary points, $\{0, \dots, \tau_{K}=\tau\}$, where $0 < \tau_{k-1} < \tau_k$ ($k = 2, \dots, K$); its intervals are right closed, $\forall k > 0: \; (\tau_{k-1}, \tau_k]$, and are referred to by their index $k$. 
In some settings, a discrete-time partition arises naturally; e.g. the 6-month increments of the SCD-HeFT study visit schedule.
In other settings, the discrete-time partition is defined by the analyst, with considerations detailed in Section \ref{subsec: discrete-time partition}.
For a given partition, patient $i$ contributes information from the time origin to an upper boundary point $k_i^r$ corresponding to the first boundary point $\tau_k$ that arises after the observed event time, 
\begin{align*}
    k_i^r = \text{argmin}_k(\tilde{T}_i \leq \tau_k).
\end{align*}
We transform the counting process into a discrete-time terminal indicator by setting $N_{k, i} = N_i(\tau_k)$ for $k \in \{0, \dots, k_i^r\}$. 
Similarly, the discrete-time longitudinal trajectory is denoted by $Y_{k, i} = Y_i(\tau_k)$.
Conceptually, we take the longitudinal marker to be measured at the end of each partition interval, and only if patient $i$ is neither truncated by the terminal event, nor censored within the $k^{th}$ partition interval; that is $\tilde{T}_i > \tau_k$. 
Formally, $Y_{k, i}$ is defined as a measurement `at' $\tau_k$, as long as $N_{k, i} = 0$ and $C_i > \tau_k$. 
To exemplify, consider measuring a biomarker for two patients in regular increments according to the study design until the patients are dead or censored (Figure \ref{fig:example_patients}).
Patient $i = 1$ experienced the terminal event between $\tau_5$ and $\tau_6$. 
Patient $i = 2$ is censored at time point $\tau_4$ ($k_i^r = 5$) without having experienced the terminal event. 
% Both patients contribute information to the partition intervals $k \in \{ 1, 2, 3, 4 \}$ and patient $i = 1$ additionally contributes information to partition intervals $k \in \{5, 6 \}$. 

Finally, let $\mathbf{X}_{i}\in \mathbb{R}^p$ denote covariates which were measured at baseline and represent the features used for prediction. 
Collectively, denote the $i$th individual's history of covariates and longitudinal trajectory before time $k$ as $\mathcal{H}_{k, i} = \left\{\mathbf{X}_{i} \, , (Y_{l, i})_{l = 0}^{k-1} \right\}$.

\subsection{The observed data likelihood}
\label{subsec: observed data likelihood}
Under the discrete time representation of the joint outcomes, consider the following factorization of the joint density for the discrete-time terminal event counting process value $n_{k, i} \in \{0, 1\}$ and longitudinal marker value $y_{k, i} \in \mathbb{R}$ at discrete time interval $k$,
\begin{equation*} \label{eqn:general_likelihood}
    p\left( N_{k, i} = n_{k, i}, Y_{k, i} = y_{k, i} \; | \; N_{k-1, i} = 0, \mathcal{H}_{k,i} \right) = \begin{cases}
    \begin{aligned}
        & f_Y\left(y_{k, i} \; | \;  N_{k, i} = 0, \mathcal{H}_{k,i} \right) \\
        & \times (1 - \pi_{k, i})
    \end{aligned}
          & \text{for } n_{k, i} = 0 \\
        \pi_{k, i} & \text{for } n_{k, i} = 1
    \end{cases}
    \end{equation*}
where $\pi_{k, i} := \Prob \left(N_{k, i} = 1 \; | \; N_{k-1, i} = 0, \mathcal{H}_{k, i} \right)$, and the longitudinal marker $Y_{k, i}$ follows a parametric distribution, $[Y_{k, i} \, | \, N_{k, i} = 0, \mathcal{H}_{k, i}] \sim f_Y\left(.\right)$, which is parameterized by the mean trajectory $\mu_{k, i} \, := \E\left(Y_{k, i} \; | \; N_{k, i} = 0, \mathcal{H}_{k, i}\right)$ and additional parameters. 
Appendix B describes some examples of possible outcome types and corresponding distributions. 

In each interval $k$, $\pi_{k, i}$ denotes the discrete-time hazard function representing the probability of experiencing terminal event in interval, conditional on not having experienced it in any earlier interval
The estimand $\mu_{k, i}$ is the expected longitudinal marker value at the end of interval $k$, conditional on not having experienced the terminal event in interval $k$ or earlier and given history $\mathcal{H}_{k, i}$. 
This quantity is directly interpretable as the partly-conditional longitudinal mean trajectory \citep{kurland2005directly}.

\subsection{Regression structures for the terminal event submodel and longitudinal submodel}
\label{subsec: regression structures}
\begin{figure}[ht]
\centering
\begin{tikzpicture}[
scale=0.65, transform shape,
    obs/.style={
    rectangle, draw, 
    minimum width=10mm, 
    minimum height=8mm,
    align=center},
    frailty/.style={
    circle, draw=DarkEmerald, 
    minimum size=10mm,
    inner sep=0pt},
    unobs/.style={
    circle, draw, 
    minimum size=10mm,
    inner sep=0pt},
    node distance=0.5cm,
    box/.style={draw, thick, rectangle, inner sep=0.2cm},
    allbox/.style={draw, thick, rectangle, inner sep=0.4cm},
    dashedbox/.style={draw=RoyalBlue, dashed, thick, rectangle, inner sep=0.1cm},
    reddashedbox/.style={draw=DarkEmerald, dashed, thick, rectangle, inner sep=0.1cm},
    >=Stealth
]

% terminal event submodel
\node[unobs] (pik) at (-4, 6) {$\pi_{k, i}^{(\gamma)}$}; 
\node[obs, above left=1cm of pik] (XTk) {$\mathbf{X}^{\scriptscriptstyle \text{(T)}}_{i}$};

\node[obs, right=1.8cm of pik] (Nk) {$N_{k, i}$};
% \node[unobs, below right=1cm of pik] (ek) {$e_{k, i}$};

\node[unobs] (pik1) at (3, 6) {$\pi_{k+1, i}^{(\gamma)}$}; 
\node[obs, above left=1cm of pik1] (XTk1) {$\mathbf{X}^{\scriptscriptstyle \text{(T)}}_{i}$};

\node[obs, right=1.8cm of pik1] (Nk1) {$N_{k+1, i}$};
% \node[unobs, below right=1cm of pik1] (ek1) {$e_{k+1, i}$};

% Boxes and labels
\node[box, fit=(XTk)(XTk1)(pik)(pik1)(Nk)(Nk1)] (longBox) {};

% Arrows in terminal event 
\draw[->] (pik) -- (Nk);
\draw[->] (XTk) -- (pik);
% \draw[->] (ek) -- (pik);

\draw[->] (pik1) -- (Nk1);
\draw[->] (XTk1) -- (pik1);
% \draw[->] (ek1) -- (pik1);

% Nodes for longitudinal submodel
\node[unobs] (muk) at (-4, 0) {$\mu_{k, i}^{(\gamma)}$};

\node[obs, above left=1cm of muk] (Xk) {$\mathbf{X}^{\scriptscriptstyle \text{(L)}}_{i}$};
\node[obs, below =1cm of muk] (Zk) {$\mathbf{Z}_{i}$};

\node[obs, right=1.8cm of muk] (Yk) {$Y_{k, i}$};
\node[unobs, below=1cm of Yk] (epsk) {$\varepsilon_{k, i}$};

\node[unobs, above right=1cm of muk] (b) {$b_i$};
\node[frailty, above=2.05cm of muk] (gamma) {$\gamma_i$};

\node[unobs] (muk1) at (3, 0) {$\mu_{k+1, i}^{(\gamma)}$};

\node[obs, above left=1cm of muk1] (Xk1) {$\mathbf{X}^{\scriptscriptstyle \text{(L)}}_{i}$};
\node[obs, below=1cm of muk1] (Zk1) {$\mathbf{Z}_{i}$};

\node[obs, right=1.8cm of muk1] (Yk1) {$Y_{k+1, i}$};
\node[unobs, below=1cm of Yk1] (epsk1) {$\varepsilon_{k+1, i}$};

\node[unobs, above right=1cm of muk1] (b1) {$b_i$};
\node[frailty, above=2.05cm of muk1] (gamma1) {$\gamma_i$};

% Boxes and labels
\node[box, fit=(Xk)(Xk1)(muk)(muk1)(Yk)(Yk1)(epsk)(epsk1)(b)(b1)] (longBox) {};

% Arrows in longitudinal
\draw[->] (Xk) -- (muk);
\draw[->] (Zk) -- (muk);
\draw[->] (Xk1) -- (muk1);
\draw[->] (Zk1) -- (muk1);
\draw[->] (muk) -- (Yk);
\draw[->] (muk1) -- (Yk1);

\draw[->] (epsk) -- (Yk);
\draw[->] (epsk1) -- (Yk1);
\draw[->] (b) -- (muk);
\draw[->] (b1) -- (muk1);

% Associations 
\draw[->, DarkEmerald] (gamma) -- (pik) node[midway, above, rotate = 90] {{\tiny scale multiplier $\alpha$}};
\draw[->, DarkEmerald] (gamma1) -- (pik1) node[midway, above, rotate = 90] {{\tiny scale multiplier $\alpha$}};
\draw[->, DarkEmerald] (gamma) -- (muk);
\draw[->, DarkEmerald] (gamma1) -- (muk1);

\draw[->, very thick, BrickRed] (Nk) -- (pik1) node[midway, above] {$N_{k, i} = 0$};
\draw[->, very thick, BrickRed] (Nk) -- (Yk) node[midway, above, rotate = 90] {$N_{k, i} = 0$};

\draw[->, very thick, BrickRed, transform canvas={yshift=3pt}]
    (Yk) -- (muk1)
    node[midway, above] {$N_{k, i} = 0$};

\draw[->, Goldenrod, transform canvas={yshift=-3pt}]
    (Yk) -- (muk1)
    node[midway, below] {$\eta$};

\draw[->, orange] (Yk) -- (pik1) node[midway, above, rotate = 58] {$\vartheta, \boldsymbol \varphi_k$};
% \draw[->, thick, orange] (muk) -- (pik1) node[midway, above, rotate = 58] {$\theta, \varphi_k$};

% % Shared covariates dashed box
\node[dashedbox, fit=(Xk)(XTk)] {};
\node[dashedbox, fit=(Xk1)(XTk1)] {};

% Section labels
\node[rotate=90] at (-6.8, 6) {\textbf{Terminal event}};
\node[rotate=90] at (-6.8, 0) {\textbf{Longitudinal}};

\node[allbox, fit=(Xk)(Xk1)(muk)(muk1)(Yk)(Yk1)(epsk)(epsk1)(b)(b1)(XTk)(XTk1)(pik)(pik1)(Nk)(Nk1), label= above:{$k = 1,\dots,k_i^r-1$}] (longBox) {};

\end{tikzpicture}
\caption{Graphical representation of the bivariate discrete-time framework; $\pi_{k + 1, i}$ and $\mu_{k + 1, i}$ are defined conditionally on not having experienced the terminal event previously, that is $N_{k, i} = 0$ (in red); $Y_{k, i}$ is only observed if $N_{k, i} =0$ (in red); the dependence between the terminal event submodel and the longitudinal submodel are modeled by (1) a global time-invariant ($\vartheta$) or local time-varying ($\boldsymbol \varphi_k$) dependence (in orange), (2) partially shared fixed effect covariates (in blue) and (3) a shared frailty $\gamma_i$ with scale multiplier $\alpha$ (in green).}
\label{fig:regression_structure}
\end{figure}
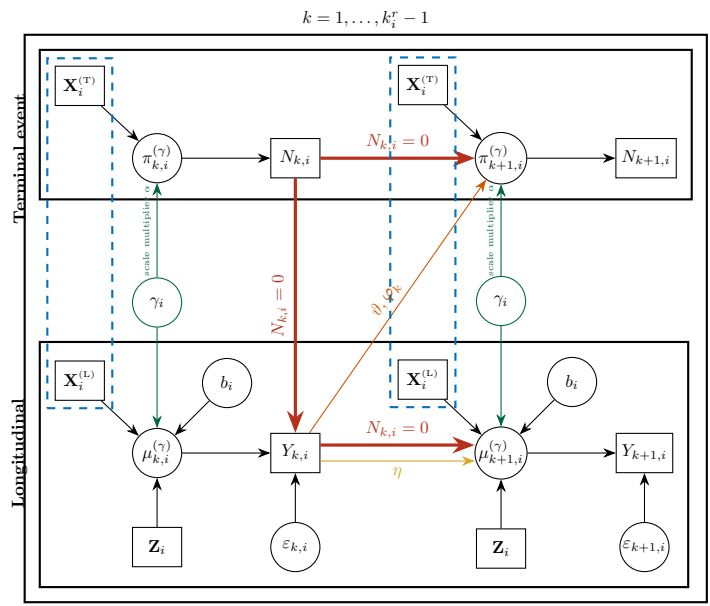

To capture complex time-varying dependence such as that observed between mortality and the MLHFQ score in Section \ref{sec: data}, we propose a framework that enables constructing flexible regression-based submodels for both $\pi_{k, i}$ and $\mu_{k, i}$.
To that end, we consider a variety of model components which capture distinct, complex and time-varying patterns in the bivariate process $(N_{k, i}, Y_{k, i})$; Figure \ref{fig:regression_structure} provides a corresponding visual overview.

\subsubsection{Terminal event submodel}
\label{subsec: terminal event submodel}
We adopt a regression model for $\pi_{k, i}$,
\begin{align*}
    g^{\scriptscriptstyle \text{(T)}}\left(\pi_{k, i} \right) & = \lambda(k) + h^{\scriptscriptstyle \text{(T)}}(\mathcal{H}_{k, i}; \boldsymbol\theta^{\scriptscriptstyle \text{(T)}}(k)),
\end{align*}
Common choices for the link function $g^{\scriptscriptstyle \text{(T)}}(.)$ include the logit, probit or complementary log-log link.
Many options are available to specify the baseline discrete-time hazard function $\lambda(k)$, with one being a piece-wise constant specification and another to use splines, such as restricted cubic splines (RCS), to characterize a smooth function over the discrete-time intervals \citep{HastieT.J2009Teos, harrell2015multivariable}. 
The user-defined function $h^{\scriptscriptstyle \text{(T)}}(.)$, parameterized by $\boldsymbol\theta^{\scriptscriptstyle \text{(T)}}(k)$, represents how the partition-specific terminal event hazard is informed by the history of the bivariate process. 
It may include terms characterizing the dependence on the longitudinal trajectory, and a regression specification for a set of baseline covariates $\mathbf{X}_{i}^{\scriptscriptstyle \text{(T)}} \subseteq \mathbf{X}_{i}$, in terms of (possibly time-varying) coefficients $\boldsymbol\xi(k) \in \boldsymbol\theta^{\scriptscriptstyle \text{(T)}}(k)$, via the linear predictor $\left( \mathbf{X}_{i}^{\scriptscriptstyle \text{(T)}} \right)^\top \boldsymbol\xi(k)$.

\subsubsection{Longitudinal trajectory submodel}
\label{subsec: longitudinal submodel}
We adopt a regression model for the partly-conditional mean $\mu_{k, i}$,
\begin{align*}
    g^{\scriptscriptstyle \text{(L)}}(\mu_{k, i}) & = \zeta(k) + h^{\scriptscriptstyle \text{(L)}}(\mathcal{H}_{k, i} \, | \, N_{k, i} = 0; \boldsymbol\theta^{\scriptscriptstyle \text{(L)}}(k)),
\end{align*}
where the link function $g^{\scriptscriptstyle \text{(T)}}(.)$ depends on the parametric distribution of $Y_{k, i}$; a common observation may be a Gaussian distribution with the identity link. 
Multiple modeling options for the baseline temporal trend of the longitudinal marker, $\zeta(k)$, are available, including a piece-wise definition, or as a smooth function over discrete-time intervals via splines. 
The user-defined function $h^{\scriptscriptstyle \text{(L)}}$ collects the regression and dependence terms and is parameterized $\boldsymbol\theta^{\scriptscriptstyle \text{(L)}}(k)$; it may include baseline longitudinal marker observation, $Y_{0, i}$, and a set of covariates $\mathbf{X}_{i}^{\scriptscriptstyle \text{(L)}} \subseteq \mathbf{X}_{i}$, which can be parameterized in terms of respective (possibly time-varying) coefficients $\beta_0(k)$ and $\boldsymbol{\beta}(k)$. 
Random effects $\boldsymbol b_i$ can also be included for a select covariate set $\mathbf{Z}_{i} \subseteq \mathbf{X}_{i}$, to capture patient-specific deviations from the population average longitudinal trajectory. 
The random effects follow some parametric distribution with density $f_b(. | \boldsymbol \theta^{(b)})$ \citep{diggle2007analysis}; a common choice may be multivariate Gaussian distribution, $\boldsymbol b_i \sim \text{MVN}(\boldsymbol 0, \boldsymbol \Sigma^{(b)})$. 
Together, regression effects and a random effect structure result in the linear predictor terms $Y_{0, i} \cdot \beta_0(k) + \left( \mathbf{X}_{i}^{\scriptscriptstyle \text{(L)}} \right)^\top \boldsymbol\beta(k) + \mathbf{Z}_{i}^\top \boldsymbol b_i$ within the regression function $h^{\scriptscriptstyle \text{(L)}}(\mathcal{H}_{k, i} \, | \, N_{k, i} = 0; \boldsymbol\theta^{\scriptscriptstyle \text{(L)}}(k))$.
Despite the inclusion of random effects, the longitudinal mean $\mu_{k, i}$ retains its interpretation as partly-conditional mean due to the discretization of time and sequential conditioning on survival.

Beyond standard mixed effect specification, one key feature of the proposed framework in the context of joint prediction is the capacity to include an autoregressive component in the longitudinal submodel (in yellow in Figure \ref{fig:regression_structure}). 
An autoregressive component uses past historic longitudinal marker values up to an order $d$ to inform the future trajectory, and can characterize complex outcome trajectories beyond the standard mixed model structure \citep{diggle2007analysis}. 
Denoting autoregression coefficients $(\eta_1,\dots,\eta_d)$, an autoregressive component in 
$h^{\scriptscriptstyle \text{(L)}}(\mathcal{H}_{k, i} \, | \, N_{k, i} = 0; \boldsymbol\theta^{\scriptscriptstyle \text{(L)}}(k))$ might be defined by $\sum_{j=1}^d \eta_j \cdot \left[ Y_{k-j,i} - \boldsymbol{Z}_{k, i}^\top \boldsymbol b_i \right] $. 
Alternative parametrizations of autoregressive components in combination with random effects can be adopted, as discussed in detail in \citep{diggle2007analysis, funatogawa2018longitudinal}.

\subsubsection{Dependence structure between terminal event hazard function and longitudinal marker trajectory} 
\label{subsec: dependence structure}
While $\pi_{k, i}$ and $\mu_{k, i}$ can be indirectly linked through shared baseline covariates, the proposed framework also admits complex temporal dependence between the terminal event hazard function and longitudinal trajectory based on (1) direct global and local dependence of the terminal event hazard on the observed marker trajectory history (in orange in Figure \ref{fig:regression_structure}), and (2) a latent shared frailty capturing unobserved residual dependence (in green in Figure \ref{fig:regression_structure}). 

A general association between the longitudinal marker and terminal event hazard function is captured by global, time-invariant dependence parameters; within $h^{\scriptscriptstyle \text{(T)}}(\mathcal{H}_{k, i}; \boldsymbol\theta^{\scriptscriptstyle \text{(T)}}(k))$, such global dependence up to order $p_g$ is defined as $\sum_{p = 1}^{p_g} \vartheta_p \cdot Y_{k-p, i}$.
Further time-variation of such associations can be captured by additional time-varying `local' dependence parameters up to order $p_l$ defined within $h^{\scriptscriptstyle \text{(T)}}(\mathcal{H}_{k, i}; \boldsymbol\theta^{\scriptscriptstyle \text{(T)}}(k))$ via the additional term $\sum_{q = 1}^{p_l} \varphi_{k, q} \cdot Y_{k-q, i}$. 
For each lag order $p_l$, the local dependence can be defined via separate interval-specific coefficients $\boldsymbol\varphi = \{ \boldsymbol \varphi_k \}_{k = 1}^K = \{ (\varphi_{k, 1}, \dots, \varphi_{k,p_l}) \}_{k = 1}^K$, or modeled as a smooth function over time, e.g. time-linear or time-quadratic, or non-parametrically via splines. 
More generally, the global and local dependence might depend on some transformation of the trajectory. 
Aligning with recent extensions to the joint longitudinal-survival model \citep{courcoul2025location, chen2025bayesian, courcoul2026joint}, the terminal event hazard might be driven by patient-specific biomarker variability. 
As such, the terminal event submodel might include a patient-specific summary of the trajectory curvature as informative covariates. 

Finally, a shared frailty $\gamma_i$ can capture additional association between the terminal event risk and the longitudinal marker that is not acknowledged by shared covariates or the global and local dependence.
Clinically, the shared frailty $\gamma_i$ might be interpreted as the latent vulnerability of a patient to experience the terminal event faster and a worse longitudinal trajectory. 
The shared frailty follows some continuous distribution $\gamma_i \sim f_{\gamma}(. | \boldsymbol\theta^{(\gamma)})$, where $\E(\gamma_i) = 0$ is required for identifiability. 
Common choices of frailty distribution are the Gamma and Gaussian distribution (see Appendix C). 
Specification of a shared frailty proceeds as
\begin{align*}
    g^{\scriptscriptstyle \text{(T)}}\left(\pi_{k, i}^{(\gamma)}\right) & := g^{\scriptscriptstyle \text{(T)}} \left(\Prob(N_{k, i} = 1 \; | \; N_{k-1, i} = 0, \mathcal{H}_{k, i}, \gamma_i) \right) = \alpha \cdot \gamma_i + g^{\scriptscriptstyle \text{(T)}}(\pi_{k, i}),  \\
    g^{\scriptscriptstyle \text{(L)}}\left(\mu_{k, i}^{(\gamma)} \right) & :=  \E(Y_{k, i} \; | \; N_{k, i} = 0, \mathcal{H}_{k, i}, \gamma_i) = \gamma_i + g^{\scriptscriptstyle \text{(L)}}(\mu_{k, i}),
\end{align*}
where a scale multiplier $\alpha$ adjusts the shared frailty to the different scales of the terminal event and longitudinal submodel. 
The shared frailty in the bivariate discrete-time framework resembles the shared random effect in joint longitudinal-survival models, where the scale multiplier is framed as a regression coefficient \citep{rizopoulos2012joint}. 
In the bivariate discrete-time framework, the shared frailty reduces to a patient-specific random intercept in the longitudinal submodel when $\alpha=0$, and, as such, care must be taken to avoid issues with identifiability when both a shared frailty and random effect are considered. 
% This consideration is further discussed in Section \ref{subsec: model parameterization}. 

\subsubsection{Summary}
Summarizing the variety of potential model components that have been described, a candidate model consists of the following terminal event and longitudinal submodel, 
\begin{align*}
     g^{\scriptscriptstyle \text{(T)}}\left(\pi_{k, i}^{(\gamma)}\right) ={} & \lambda(k)  + \left( \mathbf{X}_{i}^{\scriptscriptstyle \text{(T)}} \right)^\top \boldsymbol \xi + \sum_{p = 1}^{p_g} \vartheta_p \cdot \left[ Y_{k-p, i} - \mathbf{Z}_{i}^\top \, \boldsymbol b_i - \gamma_i \right]\\ 
     &  + \sum_{q = 1}^{p_l} \varphi_{k, q} \cdot \left[ Y_{k-q, i} -\mathbf{Z}_{i}^\top \, \boldsymbol b_i - \gamma_i \right] +
     \alpha \cdot \gamma_i \\
     g^{\scriptscriptstyle \text{(L)}}\left(\mu_{k, i}^{(\gamma)}\right)  ={} & \zeta(k) +  Y_{0, i} \cdot \beta_0 + \left( \mathbf{X}_{i}^{\scriptscriptstyle \text{(T)}} \right)^\top \boldsymbol \beta + \mathbf{Z}_{i}^\top \boldsymbol b_i   + \sum_{j = 1}^{d} \eta_j \cdot \left[ Y_{k-j, i} - \mathbf{Z}_{i}^\top \boldsymbol b_i - \gamma_i \right] + \gamma_i
\end{align*}
To simplify notation, we suppress possible time dependence in regression coefficients $\boldsymbol\xi$, $\beta_0$, and $\boldsymbol\beta$.
Notably, by subtracting the latent factors in the dependence terms in the terminal event submodel and in the autoregressive terms in the longitudinal submodel, the latent factors retain a marginal interpretation; that is, they represent patient-specific heterogeneity in the longitudinal trajectory and terminal event risk over time. 
As aforementioned, alternative parameterizations are possible, where latent factors describe heterogeneity in the instantaneous change \citep{diggle2007analysis, funatogawa2018longitudinal}. 

Caution is warranted when constructing submodels to avoid weak identifiability due to flat or multimodal likelihood surfaces. 
Importantly, the random effect term must be subtracted from the lagged longitudinal marker values in the global and local dependence terms of the terminal event submodel to ensure identifiability. 
If $\alpha = 0$ truly, a random intercept and shared frailty can be used to capture the latent dependence in the data; including both terms can lead to convergence issues. 
% Hence, an analyst might have to decide to include either a random patient-specific intercept or shared frailty in a data application. 
Similarly, including unnecessary latent factors that result in variance parameters near the boundary (e.g., $\sigma_b = 0$ or $\sigma_{\gamma} = 0$) can result in model instability and convergence issues. 
Such issues warrant exploring candidate models that exclude latent factors. 

\subsection{The regression-based data likelihood}
Let $D = \left\{\mathbf{X}_i, (N_{k, i}, Y_{k, i})_{k = 1}^{k_i^{r}}\right\}_{i = 1}^n$ denote the observed data, and let $\boldsymbol\Psi$ summarize the unknown regression parameters for the terminal event submodel and longitudinal submodel. 
Under non-informative right censoring, the observed likelihood is the product of discrete time intervals to which patient $i$ contributed patient-time.  
\begin{align}
    \mathcal{L}_i(\boldsymbol\Psi) = \left( \prod_{k = 1}^{k_i^r-1} (1 - \pi_{k, i}) \cdot f_Y\left(y_{k, i}\right)\right) \cdot \pi_{k_i^r, i}^{\Delta_i}
\end{align}
If latent factors are included, we consider the conditional observed likelihood, conditional on the unknown patient-specific random effect $\mathbf{b}_i$ and shared frailty $\gamma_i$. 
\begin{align*}
    \mathcal{L}_i(\boldsymbol\Psi, \boldsymbol b_i, \gamma_i) = \left( \prod_{k = 1}^{k_i^r-1} \left(1 - \pi_{k, i}^{(\gamma)}\right) \cdot f_Y\left(y_{k, i} | \boldsymbol b_i, \gamma_i \right)\right) \cdot \left({\pi_{k_i^r, i}^{(\gamma)}}\right)^{\Delta_i}
\end{align*}
Appendix D provides further detail on the likelihood derivation under right censoring.

\section{Bayesian estimation: prior specification, computation and posterior processing}
\label{sec: bayes}
We proceed with estimation via the Bayesian paradigm. 
Modern Markov chain Monte Carlo (MCMC) methods are well-suited to characterize posterior distributions of complex hierarchical models, allow for a comprehensive quantification of joint estimation uncertainty across submodels and facilitate generating joint posterior predictions for time to terminal event and the longitudinal marker trajectory \citep{gelman1995bayesian}.

\subsection{Prior specification and Bayesian computation}
\label{sec: estimation}
The choice of prior distribution $p_{prior}(\boldsymbol\Psi)$ is both substantively as well as in relation to the specific model components chosen to be context-specific; it can incorporate prior knowledge about parameter values or be non-informative to emphasize data-driven parameter estimation \citep{gelman2017prior}.
To help avoid ridge-like posterior geometries, we apply QR decomposition to the regression design matrices \citep{gelman2020bayesian}, and specify weakly informative independent standard Gaussian prior distributions on the QR-transformed regression parameters \citep{gelman1995bayesian, gelman2017prior}.
We adopt independent non-informative uniform priors for the baseline intercepts, $\lambda_0$ and $\zeta_0$, as well as for variance parameters, for the random intercept $\sigma_b^2$, the shared frailty $\sigma^2_{\gamma}$, and for random noise $\sigma_{\varepsilon}^2$ respectively. 
We set a weakly informative prior on the scale multiplier $\alpha \sim N(0, \sigma^2_{\alpha})$, where the prior standard deviation should be informed by the scale difference between the linear predictor in the terminal event and longitudinal submodel. 

For a given model with latent factor and prior specification, the posterior distribution is, 
\begin{align*}
    p_{post}( \boldsymbol\Psi, \boldsymbol b, \gamma | D) \propto \prod_i \mathcal{L}_{i} (\boldsymbol \Psi, \boldsymbol b_i, \gamma_i) \cdot f_{b}(\boldsymbol b_i | \boldsymbol \theta^{(b)}) \cdot f_{\gamma}(\gamma_i | \boldsymbol \theta^{(\gamma)}) \cdot p_{prior}(\boldsymbol \Psi)
\end{align*}
We employ dynamic Hamiltonian Markov chain Monte Carlo (HMC) to efficiently explore this high-dimensional posterior surface. 
Unlike standard random-walk Metropolis samplers, HMC exploits the gradient of the log-likelihood to make more informed proposals \citep{neal2011mcmc, betancourt2017conceptual}.
We use the dynamic HMC algorithm that is implemented in Stan \citep{carpenter2017STAN, betancourt2017conceptual}, accessed within the statistical software R via the package \texttt{rstan}. 
We rely on the rstan default setting for optional tuning parameters \citep{stan2020}. 
Convergence of the Markov chains was assessed by checking trace plots and Gelman's convergence diagnostic \citep{gelman2020bayesian}. 

\subsection{Posterior processing: terminal event risk profiling, longitudinal mean trajectory characterization and joint posterior prediction}
\label{sec:risk profiling}
Towards our primary goal of prediction, we use the posterior parameter samples to produce a wide array of summary quantities describing the joint posterior predictive distribution of the terminal event time and longitudinal marker, as well as to directly generate joint posterior predictions for a new patient $j$ with baseline covariates $\mathbf{x}_j$. 

One posterior summary of interest for patient $j$ is the estimated patient-specific cumulative incidence function (CIF) based on the estimates discrete-time baseline hazard $\hat{\pi}_{k, j}$. 
\begin{align*}
    \hat{\Prob}(T_j \leq \tau_k | D, \Psi, \mathbf{X}_j, \mathbf{b}_j, \gamma_j) & = \hat{\pi}^{(\gamma)}_{1, j}  + \sum_{m = 2}^k \hat{\pi}^{(\gamma)}_{m, j}\cdot \prod_{l = 1}^{m-1} (1 - \hat{\pi}^{(\gamma)}_{l, i}).
\end{align*}
The corresponding quantity summarizing the longitudinal outcome trajectory for patient $j$ is the estimated partly-conditional mean $\hat{\mu}_{k, j}^{(\gamma)}$. 
It characterizes the expected longitudinal marker trajectory for patient $j$ conditional on surviving until that time. 

Furthermore, we can directly generate patient-specific predictions from the joint posterior predictive distribution \citep{vehtari2012survey}, either drawn conditionally on hypothetical fixed latent factor values for $\mathbf{b}_j$ and $\gamma_j$, or marginalized over the posterior latent factor distribution (see Appendix E).

Using the example of a model with a shared frailty term, while an individual's patient-specific shared frailty value $\gamma_j$ is unknown by definition, we can explore how that individual's predicted outlook changes by conditioning on a range of hypothesized values for $\gamma_j$; that is average frailty ($\gamma_j = 0$), or more frail ($\gamma_j > 0$) or less frail ($\gamma_j < 0$) than the average patient. 
Marginal joint posterior predictions represent a patient's average outlook across frailty levels.

Together, this array of posterior summary and predictive quantities moves beyond existing frameworks for predicting a patient's future experience that only consider the terminal event or the longitudinal marker univariately. 
Most importantly, by considering terminal event risk and marker trajectory simultaneously, it can inform a patient about their outlook across multiple dimensions of health.

\section{Model selection in the Bayesian paradigm}
\label{sec: model selection} 
By virtue of the proposed framework's substantial flexibility in possible specification of regression and dependence structures, analysts must make several non-trivial decisions when developing a bivariate discrete-time model, such as which components to include, which link functions to choose, and which covariates to select. 
These decisions mirror challenges of model selection intrinsic to regression modeling generally \citep{steyerberg2019book}. 
One principled strategy is to construct a set of candidate models based on the presented model components, such that they align with clinical context, and to choose the best model based on model selection criterion.

In the Bayesian paradigm, several model selection criteria have been proposed to assess candidate models in term of model fit and model parsimony \citep{vehtari2017practical}. 
We consider three widely-used model selection criteria which are likelihood-based, and directly or indirectly penalize overfitting. 
The Bayesian Information Criterion (BIC) estimates the maximal observed log likelihood and adds a penalty for the number of model parameters, to assess how well a candidate model fits the observed data on which the models were trained \citep{gelman2014understanding}. 
% The BIC is well-studied, simple, and computationally efficient to compute based on posterior summary statistics.

Another model criterion of interest is the expected log posterior predictive density (ELPD), which describes the expected likelihood of observing new patients, conditional on having learned about model parameters based on the training data $D$. 
Selecting based on the ELPD implies selecting the model that minimizes the Kullback-Leibler divergence between the fitted model and the true data generating mechanism from the set of candidate models. 
The Widely Applicable Information Criterion (WAIC) and the Pareto-smoothed-importance-sampling (PSIS) leave-one-out approximation of the ELPD with moment matching ($\text{ELPD}_{\scriptscriptstyle \text{PSIS-MM}}$) are two popular choices to approximate the ELPD.

The WAIC marginalizes the observed data log likelihood over the posterior distribution for the model parameters \citep{watanabe2010asymptotic}. 
The $\text{ELPD}_{\scriptscriptstyle \text{PSIS-MM}}$ uses approximations of the leave-one-out predictive distribution for each observed patient based on importance resampling to quantify the average predictive likelihood of a new patient \citep{vehtari2002bayesian, vehtari2017practical, sivula2025uncertainty}. 
The R package $\texttt{loo}$ provides functions to conveniently estimate the WAIC and $\text{ELPD}_{\scriptscriptstyle \text{PSIS-MM}}$ for Stan models \citep{loo}.

If any of the candidate models include latent factors, model selection incorporates additional nuance \citep{merkle2019bayesian}. 
The model selection criteria can be estimated based on the conditional log likelihood, conditional on patient-specific latent factors, or based on the marginal log likelihood, marginalizing over patient-specific latent factors. 
The conditional model selection criteria describe goodness-of-fit for known patients or for new patients who strongly resemble the observed patients while marginal model selection criteria capture goodness-of-fit for completely unknown patients \citep{piironen2017comparison, merkle2019bayesian}. 
As we are interested in the latter, we focus on marginal model selection criteria; the interested reader might refer to Appendix F for their derivation.

\section{Practical considerations: discrete-time partitioning and attribution}
\label{sec: practical considerations}
Applying the bivariate discrete-time framework warrants two key practical decisions: (1) the choice of discrete-time partition and (2) the attribution of continuous-time observations to discrete-time partition intervals.

\subsection{Choice of discrete-time partition}
\label{subsec: discrete-time partition}
As motivated in Subsection \ref{subsec: joint outcome}, the bivariate discrete-time framework requires a discrete-time partition of the follow-up period $[0, \tau_K]$. 
In some settings, the discretization may follow naturally from the study design underpinning data collection; in SCD-Heft, the 6-month increments. 
In other settings, including in electronic health records, claims or registry-based studies, data is not collected according to a formal follow-up schedule. 
Hence, the analyst must choose the time partition while considering its implications.

Clinically, the choice determines the timescale of outcome predictions; in SCD-HeFT, mortality and quality-of-life predictions are made in 6-month increments. 
Ideally, these align with clinical time horizons; e.g. in terms of disease progression or health care decisions. 
Statistically, the time partition determines the distance between lagged and current values.
In a more granular discrete-time partition, lagged values are closer together, which may increase the salience of global and local dependence structures as well as autoregressive patterns.
Yet, while the key estimands $\pi_{k, i}$ and $\mu_{k,i}$ are well-defined for any discrete-time partition, they can only be meaningfully estimated when enough data is observed in each discrete-time partition interval. 
Granularity is, therefore, functionally limited by data availability. 

In practice, the analyst should consider multiple discrete-time partitions which should be rooted in the clinical context, and compare the corresponding joint predictions for new patients in a sensitivity analysis.

\subsection{Attribution of observations to discrete-time partition intervals, and intermittent missingness}
\begin{figure}[h!]
\centering
\tiny
\begin{tikzpicture}
\node at (-0.5,7.2) {$i = 1$};
% gray full timeline
\draw[very thick, color = gray] (0,7.2) -- (13,7.2);
\foreach \x/\label in {0/$0$, 2/$\tau_1$, 4/$\tau_2$, 6/$\tau_3$, 8/$\tau_4$, 10/$\tau_5$, 12/$\tau_6$} {
    \draw[thick] (\x,7.05) -- (\x,7.35);
    \node at (\x,6.8) {\tiny\label};
}
\node at (-0.5,4.2) {$i = 2$};
% gray full timeline
\draw[very thick, color = gray] (0,4.2) -- (13,4.2);
\foreach \x/\label in {0/$0$, 2/$\tau_1$, 4/$\tau_2$, 6/$\tau_3$, 8/$\tau_4$, 10/$\tau_5$, 12/$\tau_6$} {
    \draw[thick] (\x,4.05) -- (\x,4.35);
    \node at (\x,3.8) {\tiny\label};
}
% --- \{Y_{k, i}\}_k rows for i = 1
\node[anchor=west] at (-1.5, 6.3) {\scriptsize NN};
\node[anchor=west] at (-1.5, 5.7) {\scriptsize Ceiling};
\node[anchor=west] at (-1.5, 5.1) {\scriptsize Floor};
% --- \{Y_{k, i}\}_k rows for i = 2
\node[anchor=west] at (-1.5, 3.3) {\scriptsize NN};
\node[anchor=west] at (-1.5, 2.7) {\scriptsize Ceiling};
\node[anchor=west] at (-1.5, 2.1) {\scriptsize Floor};
% event labels for i = 1
\foreach \x [count=\k] in {1.6, 2.3, 4.5, 5.9, 7.8, 9.1, 9.7, 10.8} {
    \fill[RoyalBlue] (\x, 7.2) circle (3pt);
    % Color D_{1,i} in gray
    \node[RoyalBlue] at (\x, 7.5) {\tiny $Y_{\k, i}$};
}
\draw[thick] (11.2,6.95) -- (11.2,7.45);
\node[DarkEmerald] at (11.2,6.65) {\scriptsize $C_i$};
% event labels for i = 2; remove position 5 at 7.2
\foreach \x [count=\k] in {1.3, 2.5, 4, 5.7, 8.5, 9.2, 10.2} {
    \fill[RoyalBlue] (\x, 4.2) circle (3pt);
    % Color D_{1,i}
    \node[RoyalBlue] at (\x, 4.5) {\tiny $Y_{\k, i}$};
}
\fill[BrickRed] (10.7, 4.2) circle (3pt);
\node[BrickRed] at (10.7, 4.5) {\scriptsize $T_i$};
\draw[thick, BrickRed] (10.7,3.95) -- (10.7,4.45);
% ---------------- i = 1: Nearest / Ceiling / Floor (add M_{j,i}=1 second rows)
\foreach \x/\a/\b/\c in {
0/{\tiny{\textcolor{gray}{$Y_{0,i}$}}}
   /{\tiny{\textcolor{gray}{$Y_{0,i}$}}}
   /{\tiny{\textcolor{gray}{$Y_{0,i}, Y_{1,i}$}}},
 2/{\tiny{{$Y_{1,i}, Y_{2,i}$}}}
   /{\tiny{{$Y_{1,i}$}}}
   /{\tiny{{$Y_{2,i}$}}}, 
 4/{\tiny{{$Y_{3,i}$}}}
   /{\tiny{{$Y_{2,i}$}}}
   /{\tiny{{$Y_{3,i}, Y_{4,i}$}}}, 
 6/{\tiny{{$Y_{4,i}$}}}
   /{\tiny{{$Y_{4,i}$}}}
   /{\tiny{{$Y_{5,i}$}}}, 
 8/{\tiny{{$Y_{5,i}$}}}
   /{\tiny{{$Y_{5,i}$}}}
   /{\tiny{{$Y_{6,i}, Y_{7,i}$}}}, 
 10/{\tiny{{$Y_{6,i}, Y_{7,i}, Y_{8,i}$}}}
    /{\tiny{{$Y_{6,i}, Y_{7,i}$}}}
    /{\tiny{{$Y_{8,i}$}}}
} {
    \node[text=RoyalBlue] at (\x,6.3) {\a};
    \node[text=RoyalBlue] at (\x,5.7) {\b};
    \node[text=RoyalBlue] at (\x,5.1) {\c};
}
\node[text=DarkEmerald] at (12,6.3) {{\tiny$C_i$}}; 
\node[text=DarkEmerald] at (12,5.7) {{\tiny$C_i$}}; 
\node[text=DarkEmerald] at (12,5.1) {{\tiny$C_i$}}; 
% ---------------- i = 2: Nearest / Ceiling / Floor (add M_{j,i}=1 second rows)
\foreach \x/\a/\b/\c in {
0/{\tiny{\textcolor{gray}{$Y_{0,i}$}}}
   /{\tiny{\textcolor{gray}{$Y_{0,i}$}}}
   /{\tiny{\textcolor{gray}{$Y_{0,i}, Y_{1,i}$}}},
 2/{\tiny{{$Y_{1,i}, Y_{2,i}$}}}
   /{\tiny{{$Y_{1,i}$}}}
   /{\tiny{{$Y_{2,i}$}}}, 
 4/{\tiny{{$Y_{3,i}$}}}
   /{\tiny{{$Y_{2,i}, Y_{3,i}$}}}
   /{\tiny{{$Y_{3,i}, Y_{4,i}$}}}, 
 6/{\tiny{{$Y_{4,i}$}}}
   /{\tiny{{$Y_{4,i}$}}}
   /{\tiny{{}}}, 
 8/{\tiny{{$Y_{5,i}$}}}
   /{\tiny{{}}}
   /{\tiny{{$Y_{5,i}, Y_{6,i}$}}}, 
 10/{\tiny{{$Y_{6,i}, Y_{7,i}$}}}
    /{\tiny{{$Y_{5,i}, Y_{6,i}$}}}
    /{\tiny{{$Y_{7,i}$}}}, 
} {
    \node[text=RoyalBlue] at (\x,3.3) {\a};
    \node[text=RoyalBlue] at (\x,2.7) {\b};
    \node[text=RoyalBlue] at (\x,2.1) {\c};
}
\node[text=BrickRed] at (12,3.3) {\tiny$N_{6, i} = 1$};
\node[text=BrickRed] at (12,2.7) {\tiny$N_{6, i} = 1$};
\node[text=BrickRed] at (12,2.1) {\tiny$N_{6, i} = 1$};
\end{tikzpicture}
\caption{Illustration of the different approaches (Nearest Neighbor (NN), Ceiling, and Floor) to attribute observations $Y_i(t)$ (in blue) beyond baseline measurements at time origin (in gray) from continuous time to its discrete-time partition, until a patient was either censored ($C_i$, in green) or truncated by the terminal event ($T_i$, in red).}
\label{fig:discrete_time_strategies}
\end{figure}
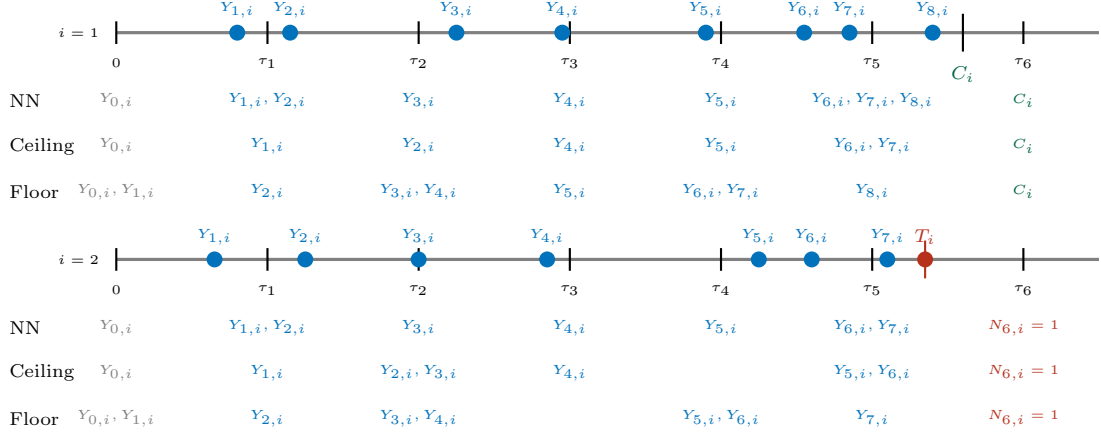

From the development of the notation in Section \ref{sec: method}, recall that we conceptualize an observation to have been recorded at the end of the discrete time interval $k$.
Consider Figure \ref{fig:discrete_time_strategies}; patient $i = 1$ is right censored in interval $k = 6$ ($N_{6, 1} = 0$) while patient $i = 2$ died in interval $k = 6$ ($N_{6, 2} = 1$).  
The longitudinal marker was measured at irregular times; in such a setting, it is unclear how to attribute continuous-time data to a chosen discrete-time partition. 
To resolve this, we consider three approaches. 
We attribute the joint observation to (1) its closest discrete time boundary (nearest neighbor, NN), (2) its ceiling interval or (3) its floor interval \citep{nevo2022modeling, ross2026discretizing}. 
For patient $i = 1$, the first measurement $Y_{1, i}$ is attributed to the first interval $(0, \tau_1]$ for NN and ceiling approach, while it is attributed to the time origin for the floor approach. 
For patient $i = 2$, for the ceiling approach, $Y_{4, i}$ is attributed to $(\tau_2, \tau_3]$ and $Y_{5, i}$ is attributed to $(\tau_4, \tau_5]$. 
For the floor approach, $Y_{4, i}$ is attributed to $(\tau_1, \tau_2]$ and $Y_{5, i}$ is attributed to $(\tau_3, \tau_4]$.

For all three approaches within the example, the analyst has to contend how to summarize multiple continuous-time measurements that are attributed to a single interval; e.g. for the $5^{th}$ interval in the NN and ceiling approach for patient $i = 1$. 
Potential options include estimating the partition-specific mean or using the latest within-partition observation. 
If the measurements are summarized by a partition-specific mean, the partly-conditional mean $\mu_{k, i}$ describes the expected longitudinal marker average, conditional on the patient not yet having experienced the terminal event, over the period of continuous time which is attributed to $\tau_k$. 
If the last observation within the partition interval is chosen, the partly-conditional mean $\mu_{k, i}$ describes the survival-conditional expected longitudinal marker average at the end of the partition interval; that is at time point $\tau_k$. 
Recall that the discrete-time hazard function $\pi_{k, i}$ describes the cumulative probability that the terminal event happens between $\tau_{k-1}$ and $\tau_k$; as such, the coarser the discrete-time partition is constructed, the larger the cumulative probability is.

If an individual does not have any measurements within a partition interval, the result is instead intermittent missingness with respect to the longitudinal marker; e.g. patient $i = 2$ is missing a longitudinal marker record attributed to interval $(\tau_3, \tau_4]$ in the ceiling approach; following the floor approach, patient $i = 2$ is missing a record attributed to interval $(\tau_2, \tau_3]$. 
Practically, there are a number of strategies to handle non-informative intermittent missingness; that is missingness that is independent of both the time to death and the true longitudinal marker value, conditional on the baseline covariates. 
We consider four strategies: (1) adopting a monotone missingness structure where observations are right censored once missingness occurs; (2) imputing missing marker values with the last observed value (last-one-carried-forward; LOCF); (3) interpolating the missing values linearly based on the observed values \citep{little2019statistical}; and (4) a `partial information' approach, which leverages the discrete structure of the bivariate discrete-time framework by sequentially conditioning on the first observation recorded after the intermittent missingness event, and computing likelihood contributions omitting time intervals with missing measurements.
The ability to rely on partial information in the bivariate discrete-time framework resembles how standard linear mixed effects models enable valid estimation under missingness at random \citep{laird1988missing, cnaan1997using}.
Yet, contrary to linear mixed models, the present framework does not implicitly impute values beyond truncation by the terminal event \citep{laird1988missing, rouanet2019interpretation}. 

\section{Simulation study}
\label{sec: simulation}
We performed a simulation study to assess two aspects of performance: (1) operating characteristics of regression parameter estimates for the terminal event submodel and the longitudinal submodel and (2) usage of model selection criteria to select correctly specified models from a set of candidate models. 
We note that we do not compare the proposed framework to any existing method, since the bivariate discrete-time framework targets joint prediction, and, as argued before, serves a different scientific purpose than established methods. 

We provide an R package \texttt{JointBivDiscTime} which generates, compiles and fits Stan models for bivariate discrete-time models that include components discussed in Section \ref{sec: method}. 
More specifically, the package adopts the logistic link function for the discrete-time hazard function $\pi_{k, i}$, Gaussian longitudinal marker trajectories and the identity link function for the partly-conditional marker mean $\mu_{k, i}$, as well as Gaussian latent factors. 
The package, as well as the simulation code are available online in the Github repository of the first author.

\subsection{Simulation scenarios}
We investigate parameter estimation under three different data scenarios (Table \ref{tab:scenario_settings}).  
We simulate a bivariate process $(N_{k, i}, Y_{k, i})$ for $n = 300$ patients over a discrete time period $\{0, 0.5, \dots, 30 \}$. 
The discrete-time terminal event indicator $N_{k,i}$ was sampled from a Bernoulli distribution with logit link, and the longitudinal trajectory was sampled from a discrete-time Gaussian distribution with identity link. 
Latent factors $\mathbf{b}_i$ and $\gamma_i$ were sampled from a Gaussian distribution.
The marker value informed the terminal event risk and once the terminal event occurred, the longitudinal observation was truncated. 
The data generating mechanism is summarized in Appendix G. 
All time basis functions were centered at the mean time point over the discrete-time partition.

\subsection{Bivariate discrete-time candidate models}
For each scenario, we fit a set of bivariate discrete-time candidate models, which include the correctly specified model, misspecified models by omitting key regression model components, and flexible (over-) parametrized models by using RCS and including additional regression model components; see details in Appendix H1. 
We set weakly informative prior on $\boldsymbol \Psi$ (see Subsection \ref{sec: estimation}) and initially ran 3 chains with $5,000$ MCMC iterations each and discarded the first $1,000$ as warm-up. 
To estimate empirical bias and coverage, we ran $500$ repetitions for each candidate model across simulation scenarios. 

\begin{table}[h!]
    \caption{Summary of settings across simulation scenarios; AR = autoregressive component.}
    \centering
    \tiny
    \renewcommand{\arraystretch}{2}
    \begin{tabular}{l l || c c c c c}
    \toprule
         & \textbf{Component} & Scenario 1 & Scenario 2 & Scenario 3 \\
        \midrule
        \multirow{4}{*}{\rotatebox{90}{\textit{Terminal event}}} 
            & Baseline hazard $\lambda(k)$ & time-constant & time-constant & time-linear \\
            & Covariates & \textcolor{green!70!black}{\checkmark} & \textcolor{green!70!black}{\checkmark} & \textcolor{green!70!black}{\checkmark} \\
            & Global dependence & linear ($p_g = 1$) & linear ($p_g = 1$) & linear ($p_g = 1$) \\
            & Local dependence & \textcolor{red!70!black}{\ding{55}} & \textcolor{red!70!black}{\ding{55}} & time-quadratic ($p_l = 1$) &  \\
        \midrule
        \multirow{3}{*}{\rotatebox{90}{\textit{Longitudinal}}} 
            & Baseline trend $\zeta(k)$ & time-linear & time-quadratic & time-linear \\
            & Covariates & \textcolor{green!70!black}{\checkmark} & \textcolor{green!70!black}{\checkmark} & \textcolor{green!70!black}{\checkmark}\\
            & AR & linear ($p = 1$) & \textcolor{red!70!black}{\ding{55}} & \textcolor{red!70!black}{\ding{55}} \\
        \midrule
        \multirow{2}{*}{\rotatebox{90}{\textit{Latent}}} 
            & Latent factor & \textcolor{green!70!black}{\checkmark} & \textcolor{green!70!black}{\checkmark} & \textcolor{green!70!black}{\checkmark}  \\
            & Scale multiplier $\alpha$ & \textcolor{green!70!black}{\checkmark} & $\alpha = 0$ & \textcolor{green!70!black}{\checkmark} \\
        \bottomrule
    \end{tabular}
    \label{tab:scenario_settings}
\end{table}

\subsection{Parameter estimation}
We assessed MCMC convergence by investigating trace plots, and by checking that the Gelman's convergence diagnostic, $\hat{R}$, is smaller than the common threshold $1.01$ for all model parameters \citep{gelman1995bayesian}. 
Any convergence issues were resolved by increasing the number of MCMC iterations and re-initializing the starting values at random (see Appendix H.2 for details on required reruns). 

Correctly specified and flexibly parameterized bivariate discrete-time models showed small bias for posterior Monte Carlo mean $\hat{\boldsymbol\Psi} \approx \E_{post}(\boldsymbol\Psi)$ and close to nominal coverage for the 95\% highest posterior density credible intervals (HPDCI) for model parameters.
Model misspecification resulted in biased parameter estimation and undercoverage. 
In particular, misspecification of the longitudinal submodel resulted in bias in the parameter estimates in the longitudinal submodel, while misspecification of the terminal event submodel and of the dependence structure led to bias in parameter estimates in both submodels. 
% Of note, misspecification of the dependence structure caused larger bias in dependence parameters and the parameters in the longitudinal submodel than in the terminal event submodel. 
When the correct model is nested in an overparameterized model, the posterior distribution for the additional model parameters were correctly centered around zero. 
A detailed discussion is presented in Appendix H.3.1.

\subsection{Model selection criteria}
As premised in Section \ref{sec: model selection}, we compared the candidate models based on the marginal BIC, WAIC and $\text{ELPD}_{\scriptscriptstyle \text{PSIS-MM}}$. 
The $\text{ELPD}_{\scriptscriptstyle \text{PSIS-MM}}$ was reported on the deviance scale, to ensure comparability across criteria. 
For all criteria, the smaller the value, the "better" the model fit.

Across simulation scenarios, the conditional BIC, WAIC and $\text{ELPD}_{\scriptscriptstyle \text{PSIS-MM}}$ failed to pass diagnostic checks, and as such, rendered crude approximations of predictive performance. 
Such behavior for models with patient-specific latent factors has been discussed previously \citep{vehtari2017practical, merkle2019bayesian}. 
Consequently, we refrained from interpreting the values, and focused on the marginal model selection criteria as a more accurate assessment of model fit. 

Across scenarios, either the correctly specified or flexibly specified model achieved the lowest averaged marginal BIC, marginal WAIC and marginal $\text{ELPD}_{\scriptscriptstyle \text{PSIS-MM}}$. 
Overly simplistic misspecified models exhibited poorer performance. 
All model selection criteria selected the correct model specification the majority of times, followed by flexibly specified models with few additional parameters.
The degree to which overparameterization is penalized depended on the number of additional parameters.
Consequently, all three model selection criteria can be used to select a parsimonious model from a set of candidate models that aligned the closest with the correct model specification across scenarios.
A detailed discussion is included in Appendix H.3.2.

\section{Development of a joint prediction model for quality of life and mortality in SCD-HeFT}
\label{sec: application}
Here, we revisit the clinical context of SCD-HeFT, described in Section \ref{sec: data}. 
Specifically, we consider a hypothetical clinical scenario where a patient has received an ICD implantation, and the physician wishes to use their anticipated prognosis to inform their continued management of progressive heart failure and ICD therapy. 
We use the bivariate discrete-time framework to develop a prediction tool that jointly predicts quality of life and mortality, based on data from the SCD-HeFT trial.

\subsection{Discrete-time partition, covariates and sample size}
As aforementioned, we chose 6-month increments as discrete-time partition. 
We attributed continuous time to death to the discrete-time partition following the ceiling approach (see Section \ref{subsec: discrete-time partition}). 
This key decision followed from the unique combination of discrete-time and continuous-time observations in SCD-HeFT. 
It ensures no data was discarded; rather than excluding some marker observations recorded in the last discrete-time interval before death, all measurements were retained by carrying time to death forward to the next discrete-time interval. 
In the context of this paper, we restrict our data sample to $N = 612$ patients for whom the relevant baseline covariates were fully observed.
In practice, analysts should carefully consider alternative options to complete-case analysis which make fuller use of the available data and require fewer assumptions.

\subsection{Intermittent missingness in MLHFQ score trajectory}
Patients might have missing MLHFQ scores for some scheduled visits; $N = 124$ ($ 15.2 \%$) patients missed at least one follow-up visit, and $N = 509$ ($62.3\%$) patients did not answer at least one MLHFQ question over time (Appendix \ref{app: intermittent missingness}). 
We consider intermittent missingness in the SCD-HeFT data to be non-informative; that is the intermittent missingness is independent of time to death and the MLHFQ score conditional on the baseline covariates (Appendix I.1). 
Resulting sample sizes for the aforementioned strategies to handle intermittent missingness are indicated in Appendix I.2.

\subsection{Bivariate discrete-time candidate models}
To highlight the modeling options the bivariate discrete-time framework offers, we constructed four candidate models that had varying flexibility in baseline hazard function, longitudinal trend and dependence structure. 

All candidate models, summarized in Table \ref{tab:data model specifications}, include a global dependence parameter; that is the MLHFQ score measured at $\tau_k$ informs the discrete-time mortality hazard function $\pi_{k+1, i}$. 
The mortality and MLHFQ score submodel share two covariates - patient age and biological sex - and a time-interaction for each covariate in all candidate models. 
Across candidate models, the mortality submodel additionally includes clinical comorbidities and diagnostic test results as informative covariates, as well as a time-interaction for diabetes and atrial fibrillation.  
The base model adopts a time-linear baseline hazard function, a time-linear baseline marker trend and a shared frailty term.
The exploratory data analysis in Section \ref{sec: data} suggested the dependence between the MLHFQ score and mortality risk changes over time. 
Model-LocDep adds a flexible local dependence to the base model to capture the time-varying nature of the direct dependence between the MLHFQ score and mortality risk. 
Model-Flex and Model-RE introduce flexibility to the base model by 4-knot RCS specification of the discrete-time hazard function in the terminal event submodel and baseline trend in the longitudinal submodel, for the time-interaction with age and for the local dependence. 
Model-RE restricts Model-Flex by substituting the shared frailty term with a patient-specific random intercept. 
Finally, Model-AR revises the base model by removing the shared frailty and introducing a lag-1 autoregressive term to characterize the longitudinal marker trajectory. 

The candidate models were fit in Stan by running 4 chains with $7,500$ iterations each and discarding the first $1,000$ as warm-up. 
All candidate models converged; convergence was assessed based on trace plots and Gelman's convergence diagnostic $\hat{R} < 1.01$. 

\begin{table}[h!]
    \caption{Specification of the bivariate discrete-time framework candidate models and marginal model selection criteria for candidate models for SCD-HeFT data, using the partial information strategy to handle intermittent missingness; all terminal event submodels include demographic, clinical and diagnostic covariates and a time-linear interaction for biological sex, diabetes and atrial fibrillation diagnosis; all longitudinal submodels include demographic covariates and a time-linear interaction with biological sex; the pointwise estimates for the marginal model selection criteria are reported on the deviance scale; the Monte Carlo standard error reported for {$\text{ELPD}_{\scriptscriptstyle \text{PSIS-MM}}^*$}, and is not available for BIC; Note: the lower the criteria, the better the fit; fct = function, diab. = Diabetes,  at. fib. = atrial fibrillation, RCS = restricted cubic splines.
    }
    \centering
    \tiny
    \renewcommand{\arraystretch}{2.2}
    \begin{tabular}{l l || c || c c c c}
    \toprule
         & \textbf{Component} & \textbf{Base model} & Model-LocDep & Model-Flex & Model-RE & Model-AR  \\
        \midrule
        \multirow{4}{*}{\rotatebox{90}{\textit{Terminal event}}} 
            & Baseline hazard fct$\lambda(k)$ & time-linear & time-linear & RCS $(k = 3)$ & RCS $(k = 3)$ & RCS $(k = 3)$ \\
            & Interaction Time:Age  & time-linear & time-linear & RCS $(k = 3)$ & RCS $(k = 3)$ &  time-linear \\
            & Global dependence $\vartheta$ & linear ($p_g = 1$) & linear ($p_g = 1$) & linear ($p_g = 1$) & linear ($p_g = 1$)& linear ($p_g = 1$) \\
            & Local dependence $\varphi_k$ & \textcolor{red!70!black}{\ding{55}}  & \makecell[c]{linear ($p_l = 1$) \\ RCS $(k=3)$} & \makecell[c]{linear ($p_l = 1$) \\ RCS $(k=3)$} & \makecell[c]{linear ($p_l = 1$) \\ RCS $(k=3)$} &  \textcolor{red!70!black}{\ding{55}}  \\
        \midrule
        \multirow{3}{*}{\rotatebox{90}{\textit{Longitudinal}}} 
            & Baseline trend $\zeta(k)$ & time-linear & time-linear & RCS $(k = 3)$ & RCS $(k = 3)$ & time-linear \\
            & Interaction Time:Age  & time-linear & time-linear & RCS $(k = 3)$ & RCS $(k = 3)$ &  time-linear \\
            & AR $\eta$ & \textcolor{red!70!black}{\ding{55}}  & \textcolor{red!70!black}{\ding{55}}  & \textcolor{red!70!black}{\ding{55}}  & \textcolor{red!70!black}{\ding{55}} & linear ($p = 1$) \\
        \midrule
        \multirow{2}{*}{\rotatebox{90}{\textit{Latent}}} 
            & Latent factor $\mathbf{b}_i, \; \gamma_i$ & \textcolor{green!70!black}{\checkmark} & \textcolor{green!70!black}{\checkmark} & \textcolor{green!70!black}{\checkmark} & \textcolor{green!70!black}{\checkmark} &  \textcolor{red!70!black}{\ding{55}}\\
            & Scale multiplier $\alpha$ & \textcolor{green!70!black}{\checkmark} & \textcolor{green!70!black}{\checkmark} & \textcolor{green!70!black}{\checkmark} & $\alpha = 0$ & \textcolor{red!70!black}{\ding{55}} \\
            \midrule 
            \midrule 
& \multicolumn{3}{l}{\textbf{Marginal model selection criterion}} \\ 
  \midrule 
& \textit{BIC} & \textbf{23915.23} & 23937.27 & 24006.19 & 24002.86 & 24276.09 \\ 
& \textit{$\text{ELPD}_{\scriptscriptstyle \text{PSIS-MM}}^*$} 
  & \makecell[c]{23796.11 \\ (527.55)} 
  & \makecell[c]{23798.95 \\ (527.48)} 
  & \makecell[c]{23811.07 \\ (527.48)} 
  & \makecell[c]{23811.75 \\ (527.49)} 
  & \makecell[c]{\textbf{23784.46} \\ (530.25)} \\
   \bottomrule
    \end{tabular}
    \label{tab:data model specifications}
\end{table}

% \FloatBarrier

\subsection{Results: parameter estimation}
Posterior means and 95\% credible intervals across candidate models and strategies for handling intermittent missingness are summarized in Figures A.4-A12 in Appendix I.3. 
The estimates for the covariate coefficients $\hat{\boldsymbol \xi}$ in the terminal event submodel, for the global dependence $\hat{\vartheta}$ and for the variance $\hat{\sigma}_{\gamma}^2$ of the shared frailty aligned across candidate models and strategies to handle intermittent missingness.
Considering the estimates for the covariate coefficients $\hat{\boldsymbol \beta}$ in the longitudinal submodel, we expected a difference in covariate coefficients between Model-AR and all other models, given the autoregressive model captures the covariate effects differently; see Figure A6. 
Of note, LOCF and linear interpolation showed larger estimates for the autoregression parameter $\hat{\eta}$ than monotone missingness and partial information.
By construction, LOCF and linear interpolation impute missing values as functions of neighboring observed marker values. 
This might induce stronger autocorrelation in the longitudinal trajectory, which might results in overestimating the autoregressive parameter $\hat{\eta}$. 

A patient's age, heart failure class and diabetes diagnosis were important risk factors for mortality. 
The effect of diabetes remained constant over time, while as time progressed, atrial fibrillation was associated with a higher mortality risk.  
A patient's age and being male were positively associated with the MLHFQ score trajectory.
The time interaction showed, that, as a patient aged, their MLHFQ score increased. 
Corroborating what we expected based on the exploratory data analysis, the MLHFQ score was positively associated with mortality risk; that is $\hat{\vartheta} > 0$ for all models. 
The local dependence showed a change in association over time, where the MLHFQ score had its largest positive effect on the mortality risk at month 24; see Figure A6. 
However, the pointwise 95\% credible intervals for the local dependence indicate that, amid random noise, a local dependence does not add additional insight into the association between the MLHFQ score trajectory and mortality risk. 
As we will see in Subsection \ref{subsec: model criteria results}, this conclusion is corroborated by the model selection criteria. 
The flexibly parametrized discrete-time longitudinal trend in Model-Flex and Model-RE indicated approximate time-linearity beyond month 30; see Figure A6.
In Model-AR, the estimate $\hat{\eta} = 0.79$ indicated a strong autoregressive pattern in the MLHFQ score trajectory; however, the model selection criteria will illustrate that such pattern seemed to be captured almost equally well by the base model.
The 95\% credible interval for the scale parameter $\alpha$ covered zero for the base model and were close to zero for Model-LocDep and Model-Flex, suggesting the shared frailty had little effect on the mortality risk, and primarily functioned as a patient-specific random intercept on the longitudinal trajectory; see Figure \ref{fig: postdens alpha}. 
The parameter estimate $\hat{\sigma}_{\gamma}$ in Model-Flex and $\hat{\sigma}_{b}$ in Model-RE aligned closely, indicated that the shared frailty and random intercept functionally captured the same patient-specific heterogeneity in the longitudinal MLFHQ trajectory; see Figures A7 and A8. 

Moving forward, we focus on partial information as strategy to handle intermittent missingness. 
Compared to monotone missingness, the partial information approach excluded less data. 
Compared to LOCF and linear interpolation, it does not overexaggerate autoregressive patterns. 

\subsection{Results: marginal model selection criteria}
\label{subsec: model criteria results}
Table \ref{tab:data model specifications} also provides the marginal model selection criteria for each candidate model. 
In practice, the analyst would choose the model that performs best in terms of model selection criteria. 
The base model outperformed the other candidate models, even if only slightly, in terms of BIC. 
Model-AR achieved the lowest value for $\text{ELPD}_{\scriptscriptstyle \text{PSIS-MM}}^*$, closely followed by the base model. 
The WAIC is not reported as it did not pass diagnostic checks; the Pareto k diagnostic indicated crude approximation \citep{vehtari2017practical}. 
Considering the parameter estimates and the marginal model selection criteria, we conclude that the SCD-HeFT data does not warrant including time-varying local dependence structures, a flexible baseline discrete-time hazard function or a flexible trend. 
Incorporating an autoregressive component does not decisively improve the model fit. 
Consequently, and aiming for model parsimony, we continue investigating the base model.

\subsection{Results: joint prediction of time to death and MLHFQ score trajectory}
Now, consider example patients who are at high risk for SCD and who, hypothetically, are scheduled to undergo surgery to implant an ICD device (Table \ref{tab: example patients}). 
\begin{table}[h!]
\tiny
\centering
\caption{Covariate profiles for example patients; NYHA = New York Heart Association Functional Class, At. fib. = atrial fibrillation, insuff. = insufficiency, Isch. cardio. = ischemic cardiomyopathy.}
\begin{tabular}{lcccc}
\toprule
Characteristic & Patient 1 & Patient 2 & Patient 3 & Patient 4 \\
\midrule
\multicolumn{2}{l}{\textit{Demographic}} \\
Biological sex & Male & Female & Male & Female \\
Age & 50 & 42 & 80 & 91 \\[0.25cm]
\multicolumn{2}{l}{\textit{Clinical}} \\
NYHA III & Yes & No & No & Yes \\
At. fib. & Yes & No & Yes & Yes \\
Diabetes & Yes & No & No & Yes \\
Renal insuff. & No & No & Yes & Yes \\
Isch. cardio. & Yes & No & Yes & No \\[0.25cm]
\multicolumn{2}{l}{\textit{Diagnostic}} \\
LVEF (\%) & 25 & 15 & 19 & 19 \\
Creatinine [10mg/dL] & 1.5 & 0.5 & 0.4 & 0.7 \\
Sodium [mEq/L] & 167 & 110 & 140 & 145 \\
QRS [10ms] & 120-150 & $<$ 120 & $<$ 120 & $>$150 \\[0.25cm]
\textit{Baseline MLHFQ score} & 10 & 75 & 55 & 88 \\
\bottomrule
\end{tabular}
\label{tab: example patients}
\end{table}

Following Subsection \ref{sec:risk profiling}, we estimate the conditional CIF for mortality and partly-conditional MLHFQ score mean, based on the patients' covariate profile and adopting that the patients have average frailty; that is $\gamma_i = 0$ (see Figure A9 in Appendix I.4). 
For patients 1, 3 and 4, quality of life is predicted to decrease over time; that is, on average, the MLHFQ score increases over time, while patient 2 is predicted to remain stable. 
Patient 1 is predicted to have the best quality of life, that is the lowest MLHFQ score, and lowest mortality risk, while patient 4 is predicted to have the worst quality of life and highest mortality risk. 
As time progresses, the mortality risk increases at different rates across the four patients.
Specifically, it steadily increases for patient 2 as their quality of life remains stable, but rapidly increases for patient 3 and patient 4 as their quality of life deteriorates. 
For patient 4, the mortality rate plateaus off at the end of follow-up. 
The 95\% HPDCI are indicated pointwise and their width depend on the number of deaths observed.

\begin{figure}[h!]
    \centering
    \includegraphics[width=0.8\linewidth]{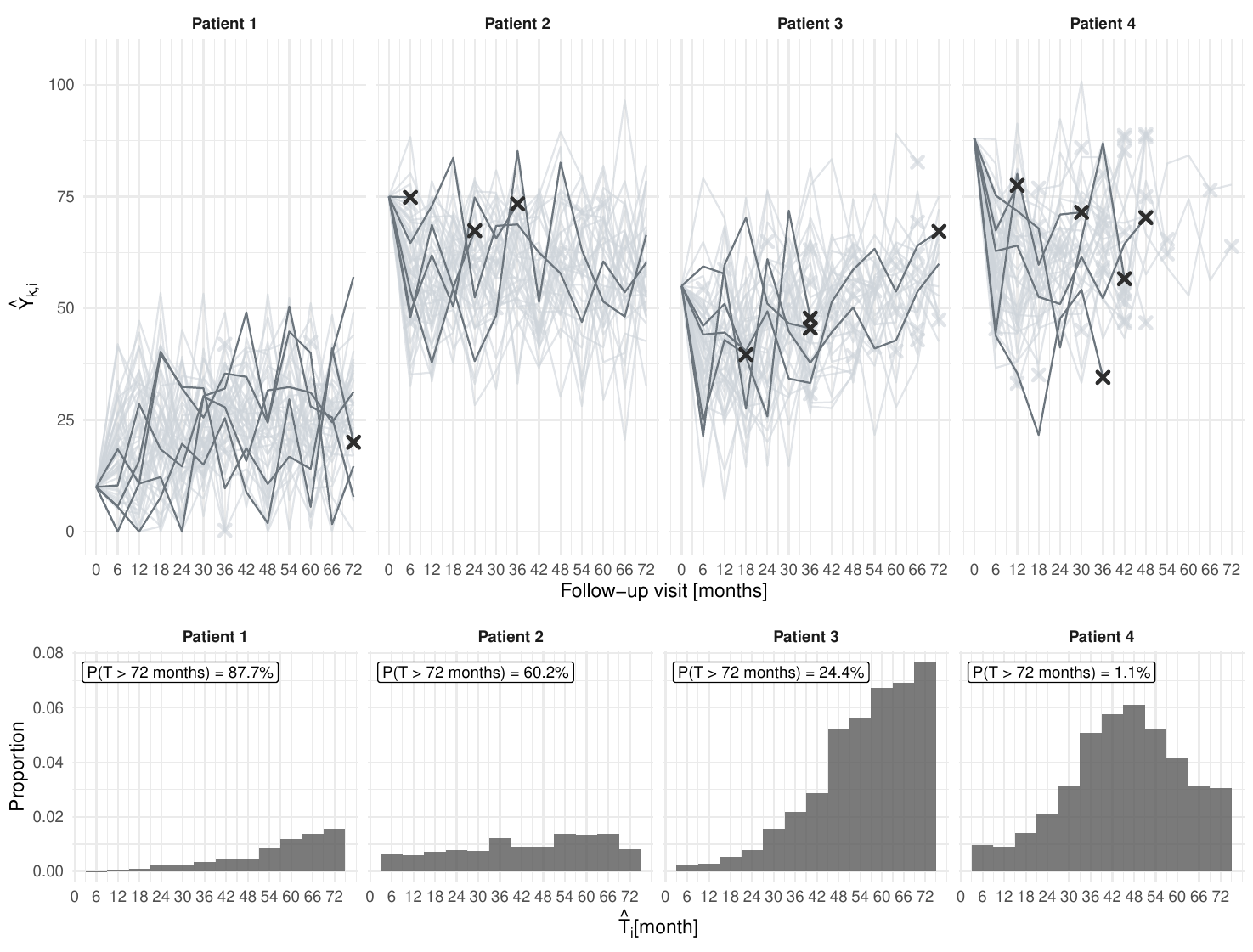}
    \caption{Patient-specific conditional joint posterior predictions for the base model under partial information strategy for intermittent missingness for shared frailty $\gamma_j = 0$; the spaghetti plots show a set of predicted future MLHFQ score trajectories of which a random subset are in bold, and the histograms visualize distribution of predicted time to death  $\hat{T}_i$; predicted longitudinal trajectories are truncated at simultaneously predicted time to death .}
    \label{fig: joint prediction}
\end{figure}

In addition, we sample joint predictions from the conditional joint posterior predictive distribution for a patient's future time-to-event and longitudinal trajectory from the base model. 
Recall from Subsection \ref{sec:risk profiling} that conditioning on a range of shared frailty values $\gamma_j$ query a patient's prediction if we assumed different physical frailty. 
First, we consider patients to be average frail; that is we condition on $\gamma_j = 0$. 
We generated 50 draws from the joint posterior predictive distribution to visualize how different joint predictions of the MLHFQ score trajectory and time to death arise, based on some baseline measurement. 
The top row in Figure \ref{fig: joint prediction} shows these 50 trajectories in gray; its heterogeneity visualizes predictive uncertainty. 
Each predicted trajectory is either truncated by death (bold cross mark) or right censored at the end of the follow-up time at 72 months.  

The lower row in Figure \ref{fig: joint prediction} illustrates the distribution of 2000 patient-specific time to death predictions over discrete time, and gives the average patient-specific predicted probability to survive beyond 72 months post randomization. 
The variance in predicted time to death mirrors the predictive uncertainty; due to the joint nature of the predictions, it is partly driven by uncertainty in the longitudinal marker prediction that is propagated to the mortality prediction.

Considering Figures A9 and \ref{fig: joint prediction} in more detail, we can distinguish four distinct patient journeys. 
Patient 1 is predicted to be in good health with high quality of life and low mortality; the quality of life is predicted to slightly decline over time while, simultaneously, the mortality risk is predicted to increase only little. 
In context with longitudinal decision-making for ICD patients, this profile would support clinical recommendations to continue ICD therapy, including potential ICD replacement when the current device generator reaches the end of its battery life.  

By contrast, patient 2 is predicted to remain in poor quality of life as mortality risk is predicted to steadily increase over time. 
Similarly, the MLHFQ score trajectory for patient 3 is predicted to deteriorate quickly over time, and mortality risk is predicted to increase quickly, particularly towards the end of the follow-up. 
Patient 4 has bad quality of life at baseline, and is predicted to deteriorate at similar speed as patient 3 over time, while their mortality risk is predicted to be elevated from baseline onward, and to increase quickly before plateauing off towards the end of follow-up.
Patient profiles such as 2, 3, and 4 might indicate a need for more structured management of progressive heart failure, but also - crucially, as no current models provide guidance for these discussions – prompt consideration of ICD deactivation or non-replacement in light of patients’ deteriorating quality of life, which ongoing ICD therapy will not improve and may prolong contrary to patients’ evolving goals of care. 

\begin{figure}[h!]
    \centering
    \includegraphics[width=0.85\linewidth]{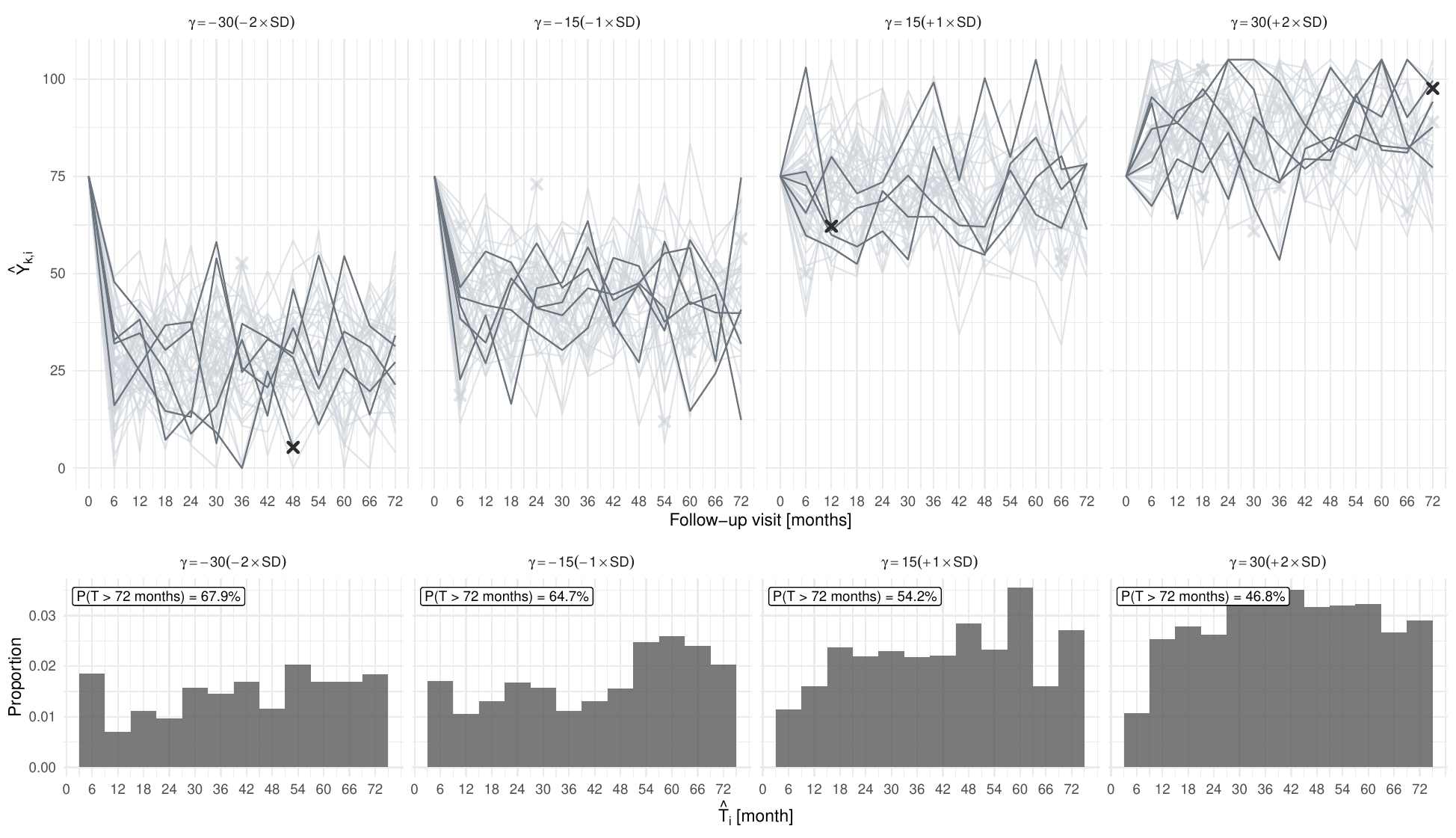}
    \caption{Conditional joint prediction for patient 2 under partial information strategy for intermittent missingness for a range of shared frailty values $\gamma_j$; spaghetti plots show a set of predicted future MLHFQ score trajectories and histograms visualize distribution of predicted time to death  $\hat{T}_i$; predicted longitudinal trajectories are truncated at jointly predicted time to death .}
    \label{fig: joint prediction shared frailty}
\end{figure}

Focusing on patient 2, we jointly predict future MLHFQ scores and mortality risk across a range of hypothetical shared frailty values, given a patient's true frailty is unknown. 
We set this range at one or two times the estimated standard deviation $\hat{\sigma}_{\gamma}$ from the average frailty $\gamma_i = 0$, where negative values represent a patient less frail than average and positive values one who is more frail. 
Figure \ref{fig: joint prediction shared frailty} shows these joint predictions, from left to right, for patient 2 as less frail through more frail than average. 
Their quality of life is predicted either to improve strongly, if physically robust, or to remain at its poor baseline level. 
Across all frailty values, patient 2 is not predicted to deteriorate substantially and is very likely to survive beyond 72 months, though at varying predicted quality-of-life levels.

\section{Discussion} 
\label{sec: discussion}
In chronic heart failure care, quality of life and mortality risk co-evolve over years of follow-up.
The ensemble of jointly predicted trajectory and mortality profile in Figure \ref{fig: joint prediction} may be of direct use in clinical settings, as it conveys this co-evolution and visualizes the predictive uncertainty. 
For individuals like patient 2, 3 and 4, joint prediction may clarify whether an individual's prognosis warrants intensified patient monitoring or adjustment to support medication; when goals-of-care and advance care planning conversations should be initiated given the predicted mortality risk; and at what point symptom-directed and palliative management deserve equal weight to continuing or ceasing ICD therapy \citep{kramer2017d}. 
In this way, joint prediction may support shared, patient-centered decision making across the longer time horizon of heart failure care.

Key to the proposed framework is that the discretization of time eliminates extrapolation of the longitudinal trajectory beyond the terminal event, and consequently, provides predictions with a clear clinical relevance. 
A suitable discrete-time partition either follows naturally from the study design, or may be chosen by the analyst who must balance health care cadence, temporal resolution of dependence and estimability of estimands.  
Regardless of the choice of discretization, the second decision is how to attribute observations recorded in continuous time to discrete time. 
Approaches discussed in this paper include the NN, ceiling and floor strategy.
In settings where there is uncertainty around these choices, an analyst could engage in a sensitivity analysis over a set of discrete-time partitions to allow for additional insight into the temporal development of covariate effects, and dependence structures.

the bivariate discrete-time framework allows for flexible model construction, including modeling autoregressive patterns in the longitudinal trajectory, and distinguishing global and local dependence, to capture time-varying association  between the terminal event and longitudinal marker. 
Latent frailties, shared across the longitudinal and terminal event submodels, can then be used to capture some of the remaining dependence.
In settings where it is unclear which regression components to include, or multiple model specifications are plausible, established Bayesian model selection criteria may be useful.
In this paper, we considered the BIC, the WAIC and the $\text{ELPD}_{\scriptscriptstyle \text{PSIS-MM}}$. 
From a clinical perspective, it is desirable to assess the predictive accuracy of a prediction model based on a metric that has some clinically intuitive interpretation.
As a likelihood-based metric, the ELPD is challenging to interpret clinically. 
Widely-used predictive accuracy metrics, such as the Brier score and the mean squared error, focus univariately on a single outcome \citep{haider2020effective, hyndman2006another}.  
To the best of our knowledge, there is no predictive accuracy metric that readily assesses how well a joint prediction tool predicts a longitudinal marker trajectory and time to terminal event. 
Future work will develop a novel metric that quantifies joint predictive accuracy in a clinically interpretable manner. 

The modeling scope of the bivariate discrete-time framework can be extended in several directions. 
Existing methods often face computational challenges when fitting models with such complex dependence structures, as they rely on computationally intensive marginalization over multidimensional latent factors \citep{rizopoulos2012fast}. 
By contrast, Stan operates on the joint log likelihood, which avoids such marginalization and, hence, facilitates considering higher dimensional covariates to perform data-driven variable selection. 
Future work could explore adopting shrinkage priors to select informative covariates within the bivariate discrete-time framework.  
The framework could also be adopted to model multiple longitudinal marker trajectories, and to accommodate competing and semi-competing risks with time-varying dependence structures, while accounting for correlation between markers and event risks through shared latent factor vectors.

% Well-established Bayesian model selection criteria are used to distinguish correctly specified models, and functionally equivalent models, from misspecified models \citep{vehtari2012survey}.
% Because we rely on ELPD approximations, comparing candidate models necessarily involves grappling with the uncertainty of such estimate and the implications this uncertainty has for identifying the best-performing model. 
% In practice, the variance of ELPD approximations, like WAIC and $\text{ELPD}_{\scriptscriptstyle \text{PSIS-MM}}$, is typically quantified using a Gaussian approximation, but recent work has shown that this approach systematically underestimates the variability in ELPD differences between models \citep{sivula2025uncertainty}. 
% % Simulation studies by \cite{sivula2025uncertainty} demonstrate such underestimation occurs in settings with small sample sizes, with model outliers, or when competing models exhibit similar predictive performance.
% This, in turn, leads to unstable or arbitrary model choices, especially when the candidate models are nested or structurally similar, thereby introducing selection-induced bias \citep{piironen2017comparison, mclatchie2024efficient}.

%%%%%%%%%%%%%%%%%%%%%%%%%%%%%%%%%%%%%%%%%%%%%%
%% Support information, if any,             %%
%% should be provided in the                %%
%% Acknowledgements section.                %%
%%%%%%%%%%%%%%%%%%%%%%%%%%%%%%%%%%%%%%%%%%%%%%
\section*{Acknowledgments}

\section*{Funding}
This framework was developed in the context of and with financial support of an ongoing collaboration under the Prospective Investigation of Palliative and End-of-Life ExpeRiences with ICDs (PIPER-ICD) Study (R01 5R01HL161697-03; PI Dr. Kramer). 
The first author was supported by Grant 1F31HL182267-01.

%%%%%%%%%%%%%%%%%%%%%%%%%%%%%%%%%%%%%%%%%%%%%%
%% Supplementary Material, including data   %%
%% sets and code, should be provided in     %%
%% {supplement} environment with title      %%
%% and short description. It cannot be      %%
%% available exclusively as external link.  %%
%% All Supplementary Material must be       %%
%% available to the reader on Project       %%
%% Euclid with the published article.       %%
%%%%%%%%%%%%%%%%%%%%%%%%%%%%%%%%%%%%%%%%%%%%%%
% \begin{supplement}
% \stitle{???}
% \sdescription{???.}
% \end{supplement}
\clearpage

%%%%%%%%%%%%%%%%%%%%%%%%%%%%%%%%%%%%%%%%%%%%%%
%% Appendix---Please move all appendices to %%
%% a Supplementary file.                    %%
%%%%%%%%%%%%%%%%%%%%%%%%%%%%%%%%%%%%%%%%%%%%%%
\appendix

\setcounter{table}{0}
\renewcommand{\thetable}{A\arabic{table}}

\setcounter{figure}{0}
\renewcommand{\thefigure}{A\arabic{figure}}

\section{SCD-HeFT data}
\label{app: table 1}
We consider the subpopulation of patients in SCD-HeFT who received an ICD device. 

\begin{table}[h!]
\centering
\scriptsize
\caption{Characteristics of SCD-HeFT Study subpopulation of patients who received an ICD; complete-case analysis; sd = standard deviation; NYHA = New York Heart Association, LVEF = left ventricular ejection fraction. }
\begin{tabular}{l | l}
  \toprule
  &  \textbf{Total} N =  612 \\ 
  \midrule
  \multicolumn{2}{l}{\textit{Demographic information}} \\
  Men; n(\%) &  475 (77.6) \\ 
  Age [years]; mean (sd) &  58.61 (11.62) \\ [0.25cm]
  \multicolumn{2}{l}{\textit{Clinical comorbidities}} \\
  NYHA Class III; n(\%) & 183 (29.9) \\ 
  Attrial fibrillation; n(\%) & 43 (7.0) \\ 
  Diabetes mellitus; n(\%) & 181 (29.6) \\ 
  Renal insufficiency; n(\%) & 145 (23.7) \\ 
  Ischemic cardiomyopathy; n(\%) &  321 (52.5) \\ [0.25cm]
  \multicolumn{2}{l}{\textit{Diagnostic test results}} \\
  LVEF [\%]; mean (sd) & 23.43 (6.80) \\ 
  Serum creatinine [mg/dL]; mean (sd) & 1.20 (0.44) \\ 
  Serum sodium [mEq/L]; mean (sd) & 139.20 (3.23)\\ 
  Duration of QRS complex; n(\%) &  \\ 
  \hspace{0.5cm} $< 120$ [in 10ms] & 363 (59.3) \\
  \hspace{0.5cm} $120 - 150 $  [in 10ms] & 136 (22.2)\\
  \hspace{0.5cm} $< 150$  [in 10ms] & 113 (18.5)\\
   \bottomrule
\end{tabular}
\label{tab: table 1}
\end{table}

\clearpage

\section{Possible distribution for longitudinal marker trajectory}
\label{app: long. distributions}
The longitudinal marker has a parametric distribution for which the mean is well defined.

\begin{table}[h!]
\caption{Possible parametric distributions of a longitudinal marker and a potential choice of link function.}
\scriptsize
\centering
\renewcommand{\arraystretch}{1.4}
\begin{tabular}{C{7.5cm} C{5cm}}
\toprule
\makecell{\textbf{Longitudinal marker distribution} \\
$Y_{k,i}\,|\,N_{k,i}=0,H_{k,i} \sim f_Y\left( .\right)$}
&
\makecell{\textbf{Link function} $g^{\scriptscriptstyle \text{(L)}}(\mu)$} \\
\midrule
\multicolumn{2}{l}{\textit{Continuous distributions}} \\
\midrule
% Gaussian
\makecell{Gaussian distribution \\
$Y_{k,i}\sim N\left( \mu_{k, i}, \sigma^2 \right)$}
&
\makecell{Identity \\
$g^{\scriptscriptstyle \text{(L)}}(\mu)=\mu$}
\\
\addlinespace[0.3em]
\multicolumn{2}{C{12.5cm}}{\textbf{Clinical examples:} Blood pressure, 6-minute walking test, physical frailty score}
\\
\midrule
% Gamma
\makecell{Gamma distribution \\
$Y_{k,i}\sim \Gamma\left(\alpha_{k, i}, \beta_{k, i} \right)$}
&
\makecell{Inverse \\
$g^{\scriptscriptstyle \text{(L)}}(\mu)=\dfrac{1}{\mu}$}
\\
\addlinespace[0.3em]
\multicolumn{2}{C{12.5cm}}{\textbf{Clinical examples:} Hormone levels, C-reactive protein levels, Troponin levels}
\\
\midrule
% Beta
\makecell{Beta distribution \\
$Y_{k,i}\sim \mathrm{Beta}\left(\alpha_{k,i},\beta_{k,i}\right)$}
&
\makecell{Logit \\
$g^{\scriptscriptstyle \text{(L)}}(\mu)=\log\left(\dfrac{\mu}{1-\mu}\right)$}
\\
\addlinespace[0.3em]
\multicolumn{2}{C{12.5cm}}{\textbf{Clinical examples:} Ejection fraction, fractional flow reserve}
\\
\midrule
\multicolumn{2}{l}{\textit{Discrete distributions}} \\
\midrule
% Bernoulli
\makecell{Bernoulli distribution \\
$Y_{k,i}\sim \mathrm{Ber}(\mu_{k,i})$}
&
\makecell{Logit \\
$g^{\scriptscriptstyle \text{(L)}}(\mu)=\log\left(\dfrac{\mu}{1-\mu}\right)$}
\\
\addlinespace[0.3em]
\multicolumn{2}{C{12.5cm}}{\textbf{Clinical example:} Heart failure hospitalization}
\\
\midrule
% Poisson
\makecell{Poisson distribution \\
$Y_{k,i}\sim \mathrm{Poisson}\left(\mu_{k,i}\right)$}
&
\makecell{Log \\
$g^{\scriptscriptstyle \text{(L)}}(\mu)=\log(\mu)$}
\\
\addlinespace[0.3em]
\multicolumn{2}{C{12.5cm}}{\textbf{Clinical examples:} Atrial fibrillation episodes, plaque counts}
\\
\bottomrule
\end{tabular}
\label{tab:marker_distribution}
\end{table}

\clearpage

\section{Possible distribution for shared frailty and random effects}
\label{app: latent distributions}
The shared frailty and random effects are independently and identically distributed across patients. 
The shared frailty - and similarly, if univariate, the random effect - can follow a range of distributions, which inherently carry certain clinical interpretations. 

\begin{table}[h!]
\scriptsize
    \centering
    \caption{Shared frailty distribution}
    \renewcommand{\arraystretch}{1.5} 
    \begin{tabular}{L{4.5cm} | L{4.5cm} | L{4.5cm}}
    \toprule
        \makecell{\textbf{Distribution} \\ $\gamma_i \sim F_{\gamma}(\boldsymbol \theta^{(\gamma)})$} & \makecell{\textbf{Interpretation}} & \makecell{\textbf{Clinical examples}} \\
        \midrule 
        \makecell{Gaussian distribution \\ $\gamma_i \sim N(0, {\sigma}^2_{\gamma})$} & \makecell{$\gamma_i \in \mathbb{R}$ allows for equally \\ many positive and negative \\ patient-specific effects} & Resting heart rate variability, polygenic risk, vascular resilience \\
        \midrule
        \makecell{Gamma distribution \\ $\gamma_i - 1 \sim \Gamma\left( \theta_1^{(\gamma)},  \theta_2^{(\gamma)}\right)$ } & \makecell{$\gamma_i \geq -1$ is skewed to the left. \\ All patients have positive effects; \\ few patients have large positive \\ shared frailties.} & Inflammatory damage, structural myocardial damage, physical frailty, comorbidity severity \\
        \bottomrule
    \end{tabular}
    \label{tab:frailty_distribution}
\end{table}

\section{Assumption of non-informative right censoring}
\label{app: right censoring}
We assume non-informative right censoring \citep{klein2003survival}. 
\begin{align*}
    \forall k \in \{0, \dots, k_i^r \}: (N_{k, i}, Y_{k, i}) \indep C_i \; | \; \mathbf{X}_{i}
\end{align*}
Furthermore, any random effects $\boldsymbol b_i$ or shared frailty $\gamma_i$ are conditionally independent of right censoring. 
\begin{align*}
    \forall k \in \{0, \dots, k_i^r \}: \boldsymbol b_i, \gamma_i \indep C_i \; | \; \mathbf{X}_{i}
\end{align*}

The conditional factorization of the likelihood under the assumption of non-informative right censoring renders right censoring ignorable. 
\begin{align*}
     & p\left( N_{k, i} = n_{k, i}, Y_{k, i} = y_{k, i}, C_i > \tau_k \; | \; N_{k-1, i} = 0, \mathcal{H}_{k,i}\right) \\
     & \hspace{0.7cm} = p\left( N_{k, i} = n_{k, i}, Y_{k, i} = y_{k, i} \; | \; N_{k-1, i} = 0, \mathcal{H}_{k,i} \right) \cdot \Prob(C_i > \tau_k \; | \; N_{k-1, i} = 0, \mathcal{H}_{k,i})  \\
\end{align*}

\section{Joint prediction}
\label{app: joint prediction}

Joint posterior predictions conditional on fixed values of can be generated $\mathbf{b}_j$ and $\gamma_j$, $(\hat{N}_{k, j}, \hat{Y}_{k, j})_{k = 1}^K \mid Y_{0, j}, \mathbf{X}_j,   \boldsymbol\ b_j, \gamma_j \sim p_{post}(. | D, Y_{0, j}, \mathbf{X}_j,  \boldsymbol\ b_j, \gamma_j)$ for patient $j$ from the posterior
\begin{align*}
    p_{post}\left( (N_{k, j}, Y_{k, j})_k \mid D, \mathbf{X}_j, \boldsymbol b_j, \gamma_j \right) & \propto \int \mathcal{L}_j\left(\boldsymbol \Psi, \boldsymbol b_j, \gamma_j \right) \cdot p_{post}(\boldsymbol \Psi \mid D) \, \partial \boldsymbol \Psi 
\end{align*}
If we wished to predict a patient's mortality risk and longitudinal trajectory, across any patient-specific deviation and frailty, we generate marginal joint predictions for patient $j$ directly from the posterior predictive distribution defined by
\begin{align*}
    & p_{post}\left( (N_{k, j}, Y_{k, j})_{k = 1}^K \mid D, \mathbf{X}_j \right) \\
    & \hspace{2.5cm} \propto \int \mathcal{L}_j\left(\boldsymbol \Psi, \gamma_j \right) \cdot p_{post}(\boldsymbol\Psi, \boldsymbol b_j, \gamma_j \mid D) \, \partial \boldsymbol \Psi \, \partial \boldsymbol b_j \, \partial \gamma_j \\
    & \hspace{2.5cm} \propto \int \mathcal{L}_j\left(\boldsymbol \Psi,\boldsymbol b_j, \gamma_j \right) \cdot f_b(\boldsymbol b_j \mid \boldsymbol \theta^{(b)}) \cdot f_{\gamma}(\gamma_j | \boldsymbol \theta^{(\gamma)}) \cdot p_{post}(\boldsymbol \Psi \mid D) \, \partial \boldsymbol \Psi \, \partial \boldsymbol b_j \, \partial \gamma_j
\end{align*}

For notational simplicity, we suppress time-varying fixed and random effects; the joint prediction algorithm follows identically to the time-invariant case. 

\begin{algorithm}[h!]
\caption{Bayesian joint posterior prediction based on bivariate discrete-time framework.}
\label{alg:BivariateDiscreteTimePred}
\begin{algorithmic}[1]
\REQUIRE A discrete-time partition $\{\tau_0, \dots, \tau_K \}$ and a collection of posterior samples for the model parameters $\{\boldsymbol \Psi_s\}_{s = 1}^S$. 
\ENSURE Patient-specific prediction of bivariate process $(N_{k, i}, Y_{k, i})$ for patient $i$ over a discrete-time partition, drawn from posterior predictive distribution. 
\FOR{$s = 1, \dots, S$}
\STATE Let $\boldsymbol \Psi^s$ be a posterior sample. For simplicity, we drop the superscript $s$ to denote the posterior sample for each model parameter. 
\IF{\textit{Conditional joint prediction}}
  \STATE Set $\boldsymbol b_i$ and $\gamma_i$ to some values. 
  \FOR{$k = 1, \dots, K$}
    \STATE Estimate conditional discrete-time hazard. 
    \[
      \begin{aligned}
        \pi_{k, i}^{(\gamma)}
          &= {g^{\scriptscriptstyle \text{(T)}}}^{(-1)}\Big( \lambda(k)
             + h^{\scriptscriptstyle \text{(T)}}\big(\mathcal{H}_{k, i}, \hat{Y}_{k-1, i};
               {\boldsymbol \theta^{\scriptscriptstyle \text{(T)}}}, \boldsymbol b_i, \gamma_i \big) \Big)
      \end{aligned}
    \]
    \STATE Sample $\hat{N}_{k, i}$ from conditional distribution, $\hat{N}_{k, i} \sim \text{Ber}\left( {\pi_{k, i}^{(\gamma)}} \right)$.
    \STATE \textbf{Break} if terminal event occurs ($\hat{N}_{k, i} = 1$).
    \STATE Estimate conditional longitudinal marker expectation.
    \[
      \begin{aligned}
        {\mu_{k, i}^{(\gamma)}}
          &= {g^{\scriptscriptstyle \text{(L)}}}^{(-1)}\Big( \zeta(k)
             + h^{\scriptscriptstyle \text{(L)}}\big(\mathcal{H}_{k, i}, \hat{Y}_{k-1, i};
               {\boldsymbol \theta^{\scriptscriptstyle \text{(L)}}}, \boldsymbol b_i, \gamma_i \big) \Big) 
      \end{aligned}
    \]
    \STATE Sample $\hat{Y}_{k, i}$ from conditional distribution, $\hat{Y}_{k,i} \sim F_Y\left(\mu_{k, i}^{(\gamma)}\right)$.
  \ENDFOR
\ENDIF
\IF{\textit{Marginal joint prediction}}
\FOR{$k = 1, \dots, K$}
    \STATE Draw $\boldsymbol b_i \sim f_b(. | \boldsymbol \theta^{(b)})$ and $\gamma_i \sim f_{\gamma}(. | \boldsymbol \theta^{(\gamma)})$
    \STATE Estimate discrete-time hazard. 
    \[
      \begin{aligned}
        \pi_{k, i}
          &= {g^{\scriptscriptstyle \text{(T)}}}^{(-1)}\Big( \lambda(k)
             + h^{\scriptscriptstyle \text{(T)}}\big(\mathcal{H}_{k, i}, \hat{Y}_{k-1, i};
               {\boldsymbol \theta_{i}^{\scriptscriptstyle \text{(T)}}}, \boldsymbol b_i, \gamma_i \big) \Big)
      \end{aligned}
    \]
    \STATE Sample $\hat{N}_{k, i}$ from distribution, $\hat{N}_{k, i} \sim \text{Ber}\left( {\pi_{k, i}} \right)$. 
    \STATE \textbf{Break} if terminal event occurs ($\hat{N}_{k, i} = 1$).
    \STATE Estimate longitudinal marker expectation.
    \[
      \begin{aligned}
        \mu_{k, i}
          &= {g^{\scriptscriptstyle \text{(L)}}}^{(-1)}\Big( \zeta(k)
             + h^{\scriptscriptstyle \text{(L)}}\big(\mathcal{H}_{k, i}, \hat{Y}_{k-1, i};
               {\boldsymbol \theta^{\scriptscriptstyle \text{(L)}}}, \boldsymbol b_i, \gamma_i \big) \Big) 
      \end{aligned}
    \]
    \STATE Sample $\hat{Y}_{k, i}$ from marginal distribution $f_Y(Y_{k, i} | \mu_{k, i})$.
  \ENDFOR
\ENDIF
\ENDFOR
\RETURN Set of conditional or marginal joint predictions, ${\Bigg \{} \left(\hat{N}_{k, i}, \hat{Y}_{k, i} \right)_{k} {\Bigg \}}_{s = 1}^S$. 
\end{algorithmic}
\end{algorithm}

\clearpage

\section{Model selection criteria in Bayesian paradigm}
\label{app: marginal selection criteria}

\subsection{Common choices of Bayesian model selection criteria}
The Bayesian Information Criterion (BIC) approximates the observed data likelihood, marginalized over the model parameters, by computing the observed data likelihood evaluated at the maximum likelihood estimate (MLE) $\hat{\boldsymbol \Psi}$ and adding an explicit penalty for the number of model parameters to avoid rewarding overfitting \citep{schwarz1978estimating}.
In our setting, we approximate $\hat{\boldsymbol \Psi}$ by performing a grid search over the posterior samples and selecting $\boldsymbol \Psi^s$ that maximizes the observed data likelihood.
\begin{align*}
    \hat{\boldsymbol \Psi} & = \argmax\limits_{\boldsymbol \Psi \in \{\boldsymbol \Psi\}_s} \; \left[\prod_i \mathcal{L}_i(\boldsymbol \Psi, \boldsymbol b_i, \gamma_i) \right]
\end{align*}

The BIC assesses how well candidate models fit the observed data \citep{gelman2014understanding}. 
It relies on a point estimate, and, consequently, is not fully Bayesian and does not account for uncertainty in parameter estimation.

Alternatively, the logarithmic score is a common prediction-focused choice of utility function which assesses goodness-of-fit based on the expected log posterior predictive density (ELPD) for future out-of-sample data  \citep{gelman2014understanding, piironen2017comparison}. 
The larger the ELPD, the better the model \citep{piironen2017comparison}. 
The logarithmic score is a proper scoring rule. 
In short, a proper scoring rule guarantees that the true model that aligns perfectly with the data generating mechanism (DGM) achieves the best score \citep{gneiting2007strictly, gelman2014understanding}. 
\begin{align*}
    \text{ELPD} & = \sum_i \int p(\boldsymbol N^*_i, \boldsymbol Y^*_i) \cdot \log\left[p((\boldsymbol N^*_i, \boldsymbol Y^*_i) | D) \right] \, \partial \boldsymbol N^*_i \partial \boldsymbol Y^*_i
\end{align*}
The key assumption is conditional independence between the future out-of-sample data, $(N^*_i, Y^*_i)$, and the observed data, conditional on model parameters \citep{merkle2019bayesian}. 
Since the data generating distribution, $p(N^*_i, Y^*_i)$, for future out-of-sample data is unknown, multiple approaches have been devised to approximate the ELPD. 
The Widely Applicable Information Criterion (WAIC) approximates the ELPD with the log pointwise predictive density (LPD) for observed data; that is the log of the marginal observed data likelihood marginalized over the posterior distribution of $(\boldsymbol \Psi, \boldsymbol b_i, \gamma_i)$.
\begin{align*}
     \text{lpd}  = \sum_i \log\left( \int \mathcal{L}_i(\boldsymbol \Psi, \boldsymbol b_i, \gamma_i) \cdot p_{post}(\boldsymbol \Psi, \boldsymbol b_i, \gamma_i | D) \, \partial \boldsymbol b_i \partial \gamma_i \partial \boldsymbol \Psi \right)
\end{align*}
The WAIC adjusts for overestimating the predictive performance due to double dipping by subtracting a penalty term informed by effective number of model parameters \citep{watanabe2010asymptotic}. 

The WAIC might suffer from numerical instability. 
Consequently, \cite{vehtari2002bayesian} developed a leave-one-out (LOO) approximation of the ELPD.
The log leave-one-out predictive likelihood, also known as the log pseudo-marginal likelihood, describes the marginal likelihood of the leave-one-out patient $i$, integrated over the posterior distribution of $(\boldsymbol \Psi, \boldsymbol b_i, \gamma_i)$ conditional on the data $D_{-i}$ which excludes observations $\{(N_{k,  i}, Y_{k, i})\}_k$ for patient $i$. 
\begin{align*}
    \text{ELPD}_{\tiny \text{LOO}} & = \sum_i \log \left( f(D_i | D_{-i}) \right) \\
    & = \sum_i \log \left( \int p(D_i | \boldsymbol \Psi, \boldsymbol b_i, \gamma_i) \cdot p_{post}(\boldsymbol \Psi, \boldsymbol b_i, \gamma_i | D_{-i}) \, \partial \boldsymbol b_i \, \partial \gamma_i \, \partial \boldsymbol \Psi \right) 
\end{align*}
Estimating the leave-one-out predictive likelihood is computationally intensive, even for a small sample size. 
Hence, \cite{vehtari2002bayesian} proposed an importance sampling approach to use the posterior sample $\boldsymbol \Psi^s \sim p_{post}(.| D)$ to Monte Carlo approximate the expectation in terms of the LOO posterior density $p_{post}(\boldsymbol \Psi | D_{-i})$. 
To stabilize importance sampling weights, \cite{vehtari2017practical} proposed a Pareto smoothed importance sampling (PSIS) scheme. 
The PSIS approach is sensitive to heterogeneity between groups where leaving out particularly influential subjects changes the posterior distribution $p_{post}(\boldsymbol \Psi, \boldsymbol b_i, \gamma_i | D_{-i})$ strongly such that even a weighted sample of $\boldsymbol \Psi^s$ is a bad pseudo-sample for full LOO posterior distribution $p_{post}(\boldsymbol \Psi | D_{-i})$.  
To reduce bias in the ELPD approximation, \cite{sivula2025uncertainty} proposed moment matching based on mean and variance to improve representativeness of the modified draws $\tilde{\boldsymbol \Psi}^s$ for a sample from $p(\boldsymbol \Psi | D_{-i})$ ($\text{ELPD}_{\scriptscriptstyle \text{PSIS-MM}}$). 
Under strong heterogeneity, bias in ELPD approximation remains despite moment matching. 
In such cases, k-fold cross validation (CV) can approximate the full posterior LOO distribution by leaving out data subsets \citep{vehtari2002bayesian}. 
Computational constraints dictate if k-fold CV can be applied and what maximum k can be chosen.
Due to long runtime to fit Stan models with larger sample sizes, we refrained from implementing k-fold CV in our simulations.

\subsection{Marginalization of common Bayesian model selection criteria}
The ELPD approximation methods only provide goodness-of-fit assessment for known patients in the trainign data set or under limited variability between patients \citep{piironen2017comparison, merkle2019bayesian}.  
If, however, we are interested in assessing goodness of fit for new patients, we should investigate model selection criteria based on marginalized likelihoods, $\mathcal{L}_i(\boldsymbol \Psi)$, marginalized over latent factors. 
Such marginal model selection criteria describes how well the models fit data from new patients. 

The marginal BIC requires the marginal ML estimate which we approximate as the parameter value that maximizes the marginal log likelihood.
\begin{align*}
    \hat{\boldsymbol \Psi} & = \argmax\limits_{\boldsymbol \Psi \in \{\boldsymbol \Psi\}_s} \; \left[ \prod_i \mathcal{L}_i(\boldsymbol \Psi) \right]\\ 
    & = \argmax\limits_{\boldsymbol \Psi \in \{\boldsymbol \Psi\}_s} \left[\prod_i \int \mathcal{L}_i(\boldsymbol \Psi, \boldsymbol b_i, \gamma_i) \cdot f_b(\boldsymbol b_i | \boldsymbol \theta^{(b)}) \cdot f_{\gamma}(\gamma_i | \boldsymbol \theta^{(\gamma)}) \, \partial \boldsymbol b_i \, \partial \gamma_i  \right] 
\end{align*}
For the marginal WAIC, we marginalize the log posterior distribution over the posterior distribution of $\boldsymbol b_i$ and $\gamma_i$. 
\begin{align*}
    \mathcal{L}_{post, i}(D_i | D) =  \int {\bigg [} \int \mathcal{L}_i(\boldsymbol \Psi, \boldsymbol b_i,  \gamma_i)\cdot p_{post}(\gamma_i, \boldsymbol b_i \mid \boldsymbol \Psi, D) \, \partial \boldsymbol b_i \, \partial \gamma_i {\bigg ]} \cdot p_{post}(\boldsymbol \Psi \mid D)  \, \partial \boldsymbol \Psi 
\end{align*}
According to Bayes Theorem, the posterior distribution for $\boldsymbol b_i$ and $\gamma_i$ can be rewritten as, 
\begin{align*}
p_{post}(\gamma_i, \boldsymbol b_i \mid \boldsymbol \Psi, D)
&= \frac{ \mathcal{L}_i(\boldsymbol \Psi, \boldsymbol b_i, \gamma_i)\, p(\gamma_i, \boldsymbol b_i \mid \boldsymbol \Psi)}{\mathcal{L}_i(\boldsymbol \Psi)}
\end{align*}
Consequently, we can express the data likelihood, marginalized over the posterior distribution for $\boldsymbol b_i$ and $\gamma_i$ in terms of the marginal likelihood $\mathcal{L}_i(\boldsymbol \Psi)$. 
\begin{align*}
& \int \mathcal{L}_i(\boldsymbol \Psi, \boldsymbol b_i,  \gamma_i)\cdot p_{post}(\gamma_i, \boldsymbol b_i \mid \boldsymbol \Psi, D) \, \partial \boldsymbol b_i \, \partial \gamma_i \\
& \hspace{1.5cm} = \frac{1}{\mathcal{L}_i(\boldsymbol \Psi)} \int \mathcal{L}_i(\boldsymbol \Psi, \boldsymbol b_i, \gamma_i)^2 \cdot f_b(\boldsymbol b_i | \boldsymbol \theta^{(b)}) \cdot f_{\gamma}(\gamma_i | \boldsymbol \theta^{(\gamma)}) \, \partial \boldsymbol b_i \, \partial \gamma_i
\end{align*}
We numerically integrate the observed data log likelihood over the latent factors using 20-node Gauss–Hermite quadrature \citep{abramowitz1948handbook}.

For the marginal $\text{ELPD}_{\scriptscriptstyle \text{PSIS-MM}}$, the LOO posterior density factorizes in terms of $\boldsymbol b_i$ and $\gamma_i$, since $\boldsymbol b_i$ and $\gamma_i$ are conditionally independent of the LOO sample $D_{-i}$, given the model parameters $\boldsymbol \Psi$.
\begin{align*}
    p_{post}(\boldsymbol \Psi, \boldsymbol b_i, \gamma_i | D_{-i}) & = p_{post}(\boldsymbol \Psi | D_{-i}) \cdot f_b(\boldsymbol b_i | \boldsymbol \theta^{(b)}) \cdot f_{\gamma}(\gamma_i | \boldsymbol \theta^{(\gamma)})
\end{align*}

\clearpage

\section{Data Generating Mechanism for simulations}
\label{app: data generating algorithm}
The longitudinal marker follows a Gaussian process. 
If included, the random effect and shared frailty have a Gaussian distribution.
Within each discrete-time partition interval, the probability of experiencing the terminal event, conditional on not having experienced it up to the prior partition interval, follows a Bernoulli distribution.\footnote{For the sake of brevity, the notation in the algorithm insinuates a shared frailty. For model construction without shared frailty, substitute the estimands $\left(\pi_{k, i}^{(\gamma)}, \mu_{k, i}^{(\gamma)}\right)$ with $(\pi_{k, i}, \mu_{k, i})$. The data generating mechanisms applies accordingly.}. 
The discrete time terminal event hazard function depends on covariates, the historic longitudinal marker trajectory via global or local dependence and a shared frailty.  
The longitudinal marker trajectory depends on time-invariant fixed-effect and random-effect random variables, an autoregressive component and a shared frailty.

\begin{algorithm}[h!]
\caption{Data generating mechanism for simulations in the bivariate discrete-time framework.}
\label{alg:BivariateDiscreteTimeSim}
\begin{algorithmic}[1]
\REQUIRE A discrete-time partition $\{\tau_0, \dots, \tau_K \}$. 
\ENSURE Simulation of a bivariate process $\{(N_{k,i}, Y_{k,i})\}_{k = 1}^K$ for each subject $i$. 
\FOR {$i = 1, \dots, n$}
\STATE \hspace{1em} Sample baseline covariates $\mathbf{X}_i^{\scriptscriptstyle \text{(T)}}, \mathbf{X}_i^{\scriptscriptstyle \text{(L)}} \in \mathbb{R}^p$.
\STATE \hspace{1em} \textit{If included:} sample random effect $ b_i \sim N(0,\sigma_b^2)$; otherwise set $b_i = 0$.
\STATE \hspace{1em} \textit{If included:} sample shared frailty $\gamma_i \sim N(0,\sigma_\gamma^2)$; otherwise set $\gamma_i = 0$.
\STATE \hspace{1em} All patients are at risk and observed from time origin onward; set $N_{0,i} = 0$ and patient entry time $L_i = 0 = 0$. 
\STATE \hspace{1em} Generate baseline longitudinal marker. 
\[
Y_{0,i} \sim N\!\left(\zeta_0 + (\mathbf{X}_i^{\scriptscriptstyle \text{(L)}})^\top \beta + b_i + \gamma_i,\; \sigma_{\varepsilon}^2\right)
\]

\FOR{$k = 1$ to $K$}
    \IF{$N_{k-1,i} = 0$}
        \STATE Sample terminal event indicator $N_{k,i} \sim \mathrm{Ber}(\pi_{k,i}^{(\gamma)})$.
            \[
            \pi_{k,i}^{(\gamma)} =
            \mathrm{expit}\!\left(
            \lambda(k) + 
            (\mathbf{X}_i^{\scriptscriptstyle \text{(T)}})^\top \xi +
            \sum_j^{p_g} \vartheta_j \cdot (Y_{k-j,i} - b_i - \gamma_i) +
            \sum_j^{p_l} \varphi_{k, j} \cdot (Y_{k-j,i} - b_i - \gamma_i) +
            \alpha \gamma_i
            \right)
            \]

            \IF{$N_{k,i} = 0$}
                \STATE Sample longitudinal marker.
                \[
                Y_{k,i} \sim N\!\left(
                \zeta(k) +
                (\mathbf{X}_i^{\scriptscriptstyle \text{(L)}})^\top \beta +
                \sum_j^d \eta_j \cdot (Y_{k-j,i} - b_i - \gamma_i) +
                b_i + \gamma_i,\;
                \sigma_{\varepsilon}^2
                \right)
                \]
            \ELSE
                \STATE Set $Y_{k,i} = \mathrm{NA}$ 
                \STATE Set $k_i^r = k$
            \ENDIF
    \ELSE
        \STATE \textit{break}
    \ENDIF
\ENDFOR
\ENDFOR
\RETURN $\forall i: \left(N_{k,i}, Y_{k,i}\right)_{k = 0}^{k_i^r}$
\end{algorithmic}
\end{algorithm}

\clearpage 

\clearpage

\section{Simulation studies}
\subsection{Simulation settings}
\label{app: simulation settings}
Each simulation is run for $R = 500$ repetitions, with a data set of size $n = 300$. 
Covariates are simulated to be constant over time. 

In the simulation scenarios, the discrete-time hazard function is either constant or linear over (centered) discrete time. 
\begin{align*}
    \lambda(k) = \tilde{\lambda}_1 + \tilde{\lambda}_2 \cdot (\tau_k - 15)
\end{align*}
The simulated longitudinal trend might either be linear or quadratic over discrete time. 
\begin{align*}
    \zeta(k) = \tilde{\zeta}_1 + \tilde{\zeta}_2 \cdot (\tau_k - 15) + \tilde{\zeta}_3 \cdot (\tau_k - 15)^2
\end{align*}
Similarly, the local dependence might be a function of discrete time. 
\begin{align*}
    \varphi_k & = \tilde{\varphi}_1 \cdot (\tau_k - 15) + \tilde{\varphi}_2 \cdot (\tau_k - 15)^2 
\end{align*}

\FloatBarrier

\begin{table}[ht] {
\caption{Overview of simulation settings for parameter estimation.} 
\scriptsize 
\renewcommand{\arraystretch}{1.8} 
\begin{tabular}{l|c|c|c} 
\toprule 
\textbf{Component} & \textbf{Scenario 1} & \textbf{Scenario 2} & \textbf{Scenario 3} \\ 
\midrule 
\multicolumn{3}{l}{discrete-time partition $\{0, 0.5, \dots, 30 \}$} \\
\midrule 
\multicolumn{3}{l}{\textit{Terminal event submodel}} \\ 
\midrule Baseline & $\tilde{\lambda}_1 = -7$ & $\tilde{\lambda}_1 = -7$ & $\tilde{\lambda}_1 = -8$; $\tilde{\lambda}_2 = 0.25$ \\ 
Covariates & \makecell{$X_1^{\scriptscriptstyle \text{(T)}} \sim \text{N}(0,1)$ \\ $X_2^{\scriptscriptstyle \text{(T)}} \sim \text{N}(0,4)$} & $X_1^{\scriptscriptstyle \text{(T)}}, X_2^{\scriptscriptstyle \text{(T)}} \sim \text{N}(0,1)$ & $X_1^{\scriptscriptstyle \text{(T)}}, X_2^{\scriptscriptstyle \text{(T)}} \sim \text{N}(0,1)$ \\ 
Fixed effects & $\xi = (0.3, 0.2)$ & $\xi = (-0.5, -0.2)$ & $\xi = (0.5, 0.2)$ \\ 
Global dependence & $\vartheta = -0.5$ & $\vartheta = 0.2$ & $\vartheta = 0.05$ \\ 
Local dependence & & $\tilde{\varphi}_1 = 0.009; \tilde{\varphi}_2 = 1 / 160^2$ & \\ 
\midrule 
\multicolumn{3}{l}{\textit{Longitudinal submodel}} \\ 
\midrule Baseline & $\tilde{\zeta}_1 = -1.05$; $\tilde{\zeta}_2 = -0.07$ & \makecell{$\tilde{\zeta}_1 = 4.9$; $\tilde{\zeta}_2 = 0.12$; \\ $\tilde{\zeta}_3 = 0.004$} & $\tilde{\zeta}_1 = 90$; $\tilde{\zeta}_2 = 1.5$ \\ Covariates & $X_1^{\scriptscriptstyle \text{(L)}}, X_2^{\scriptscriptstyle \text{(L)}} \sim \text{N}(0,1)$ & $X_1^{\scriptscriptstyle \text{(L)}}, X_2^{\scriptscriptstyle \text{(L)}} \sim \text{N}(1,0.25)$ & \makecell{$X_1^{\scriptscriptstyle \text{(L)}}, X_2^{\scriptscriptstyle \text{(L)}} \sim \text{N}(1,0.25)$ \\ $X_3^{\scriptscriptstyle \text{(L)}} \sim \text{Ber}(0.6)$} \\ 
Fixed effects & $\beta = (-0.2, 0.6)$ & $\beta = (5, 4)$ & $\beta = (3, 3, 10)$ \\ 
AR parameter & $\eta = 0.75$ & & \\
Error variance & $\sigma_{\varepsilon} = 0.8$ & $\sigma_{\varepsilon} = 1$ & $\sigma_{\varepsilon} = 2$ \\ 
\midrule 
\multicolumn{3}{l}{\textit{Latent factors}} \\ 
\midrule 
RE / Shared frailty & $\sigma_{\gamma} = 1.5$ & $\sigma_b = 2.1$ & $\sigma_{\gamma} = 2.5$ \\ Scale multiplier & $\alpha = -0.25$ & & $\alpha = 0.5$ \\ 
\bottomrule 
\end{tabular} 
\label{tab:simulation_settings}} 
\end{table}

\begin{table}[h!]
\caption{Overview of candidate models considered for Scenario 1, 2 and 3.}
\scriptsize
\centering
\renewcommand{\arraystretch}{1.25}
\begin{tabular}{p{2.5cm} p{11cm}}
\toprule
\textbf{Scenario} & \textbf{Model specifications} \\
\midrule

\textbf{Scenario 1} &
\makecell[l]{
\textbf{Model 1 (Correct):} Constant discrete-time terminal hazard; time-linear longitudinal \\ 
trend; AR(1); global dependence; shared frailty with prior on scale multiplier \\
$\alpha \sim N(0,20^2)$. \\[4pt]
\textbf{Models 2.x (Misspecified):} \\
\quad \textit{2.1} Flexible longitudinal trend via RCS ($k=3$), shared frailty with prior \\
\quad on scale multiplier $\alpha \sim N\left(0,20^2\right)$; no AR component. \\
\quad \textit{2.2} No shared frailty. \\ [4pt]
\textbf{Models 3 (Flexible):} Longitudinal trend via RCS ($k=3$); AR(1); shared frailty \\ 
with prior on scale multiplier $\alpha \sim N\left(0,20^2\right)$. 
} \\
\midrule
\textbf{Scenario 2} &
\makecell[l]{
\textbf{Model 1 (Correct):} Constant discrete-time terminal hazard; time-quadratic longitudinal \\ trend; random patient-specific intercept; global dependence; local time-quadratic \\ dependence. \\ [4pt]
\textbf{Models 2.x (Misspecified):} \\
\quad \textit{2.1} No local dependence, shared frailty with prior on scale 
multiplier $\alpha \sim N\left(0, 20^2\right)$, \\ \quad no random effect. \\
\quad \textit{2.2} No random effect. \\[4pt]
\textbf{Model 3 (Flexible):} Shared frailty with prior on scale multiplier $\alpha \sim N\left(0, 20^2\right)$, no \\
random effect.
} \\
\midrule
\textbf{Scenario 3} &
\makecell[l]{
\textbf{Model 1 (Correct):} Time-linear discrete-time terminal hazard; \\ 
time-linear longitudinal trend; global dependence; shared frailty with prior on scale \\ 
multiplier $\alpha \sim N\left(0, 20^2\right)$. \\[4pt]
\textbf{Models 2 (Misspecified):}  Constant discrete-time hazard. \\
\textbf{Models 3 (Flexible / Overparameterized):} \\
\quad \textit{3.1} Discrete-time terminal hazard via RCS ($k = 3$). \\
\quad \textit{3.2} AR(1) component. \\
\quad \textit{3.3} Time-varying local dependence via RCS ($k = 1$). \\[4pt]
} \\
\bottomrule
\end{tabular}
\label{tab:scenario_model_overview}
\end{table}

\clearpage

\subsection{Reruns for convergence}
\label{app: reruns}
The following table summarizes how often, out of $R = 500$ repetitions, the dynamic HMC sampler had to be rerun with an increased number of iterations and a new random initialization. 
The initial number of iterations was set to $5,000$ iterations with $1,000$ iterations for warm-up. 
We set the simulations such that $+1$ rerun caused $+5,000$ iterations. 

\begin{table}[h!]
    \caption{Overview of iterations run across candidate models for Scenario 1, 2, and 3 over $R = 500$ repetitions; \# = number; it. = iterations; Prop. = proportion.}
    \scriptsize
    \centering
    \begin{tabular}{l rrr}
        \toprule 
        & Prop. rerun [\%] & Average \# it. & Maximum \# it. \\
        \midrule 
        \multicolumn{3}{l}{\textbf{Scenario 1}} \\ 
        \hspace{0.2cm} Model 1 & 10.0 & 5580 & 20000 \\ 
        \hspace{0.2cm} Model 2.1 & 98.2 & 13360 & 25000 \\ 
        \hspace{0.2cm} Model 2.2 & 1.8 & 5090 & 10000 \\ 
        \hspace{0.2cm} Model 3 & 0.2 & 5010 & 10000 \\
        \midrule 
        \multicolumn{3}{l}{\textbf{Scenario 2}} \\ 
        \hspace{0.2cm} Model 1 & 98.4 & 14920 & 30000 \\ 
  \hspace{0.2cm} Model 2.1 &  99.20 & 16600 & 35000 \\ 
  \hspace{0.2cm} Model 2.2 & 0.4 & 5020 & 10000 \\ 
  \hspace{0.2cm} Model 3 & 98.8 & 15430 & 30000 \\ 
  \midrule 
        \multicolumn{3}{l}{\textbf{Scenario 3}} \\ 
        \hspace{0.2cm} Model 1 & 64.8 & 8700 & 15000 \\ 
  \hspace{0.2cm} Model 2 & 81.6 & 9810 & 20000 \\ 
  \hspace{0.2cm} Model 3.1 & 24.0 & 6210 & 15000 \\ 
  \hspace{0.2cm} Model 3.2 & 49.0 & 7560 & 15000 \\ 
  \hspace{0.2cm} Model 3.3 & 59.4 & 8170 & 15000 \\ 
        \bottomrule 
    \end{tabular}
    \label{tab:rerun_overview}
\end{table}

\clearpage

\subsection{Simulation results}
\label{app: simulation results}

\subsubsection{Parameter estimation for candidate models}
\label{app: parameter estimation}
The posterior standard deviation $\overline{\text{SD}}$ is averaged over $R = 500$ repetitions. 
The empirical standard deviation (ESE) is the standard deviation of the posterior mean over $R = 500$ repetitions.

\begin{figure}[h!]
    \centering
    \begin{subfigure}[t]{0.31\linewidth}
        \vspace{0pt}
        \includegraphics[width=\linewidth]{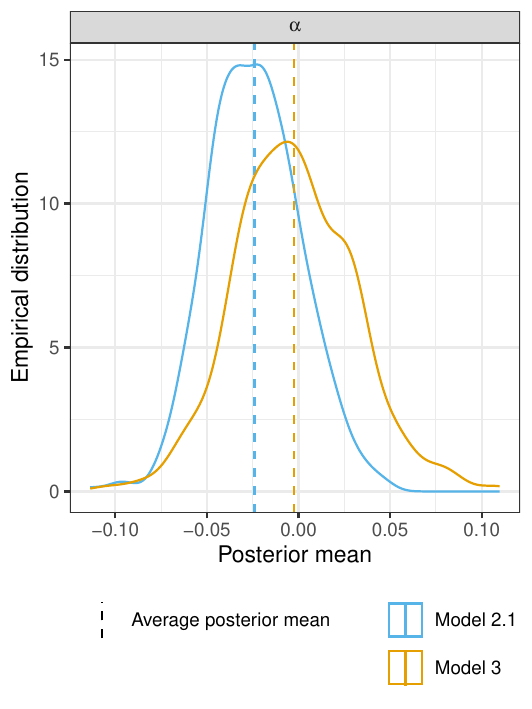}
        \caption{Empirical distribution of posterior mean for scale multiplier $\alpha$ for Scenario 2.}
        \label{fig: postdens alpha}
    \end{subfigure}
    \hfill
    \begin{subfigure}[t]{0.68\linewidth}
        \vspace{0pt}
        \includegraphics[width=\linewidth]{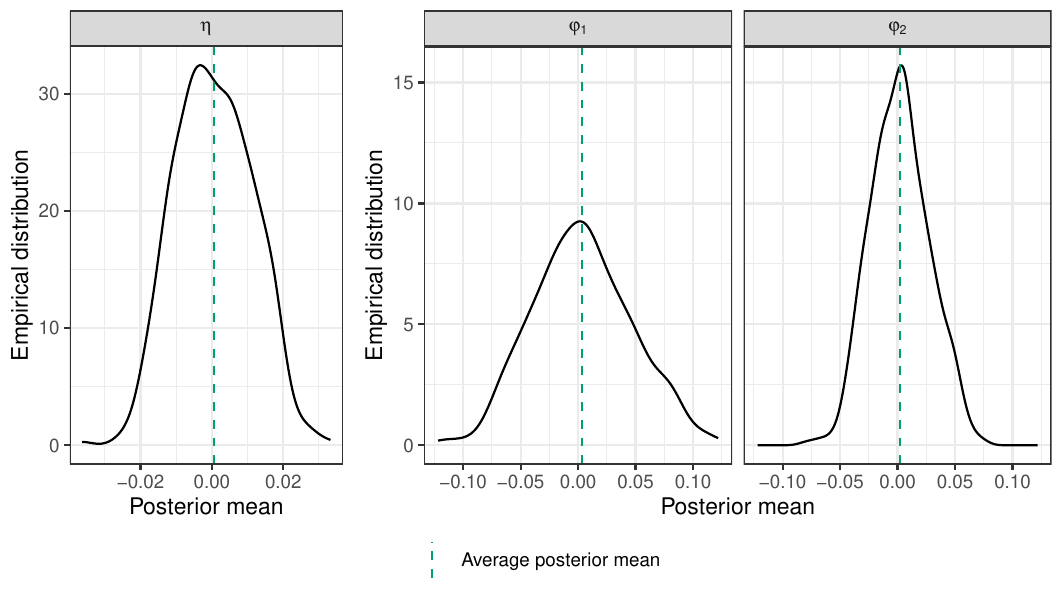}
        \vspace{0.4em}
        \caption{Empirical distribution of posterior mean for autoregressive parameter $\eta$ in Model 3.2 and for local dependence $\varphi$ (RCS with $k = 1$) in Model 3.3 for Scenario 3.}
        \label{fig: post density add components}
    \end{subfigure}
    \caption{Empirical distribution of posterior means for Scenarios 2 and 3 over $R = 500$ repetitions; dashed vertical marks the average posterior mean.}
    \label{fig: simulation scenarios}
\end{figure}

We draw the following conclusions in terms of parameter estimation from the simulation study:
\begin{itemize}
    \item Across scenarios, the posterior standard deviation matched the empirical standard deviation for the correct and flexible model specifications (Tables \ref{tab:correct_scenario1}-\ref{tab:rest_scenario3}). 
    \item Correctly specified and flexibly specified bivariate discrete-time models displayed small bias for posterior Monte Carlo mean $\hat{\boldsymbol\Psi} \approx \E_{post}(\boldsymbol\Psi)$ and close to nominal coverage for the 95\% highest posterior density credible intervals for all model parameters (Tables \ref{tab:correct_scenario1}, \ref{tab:correct_scenario2}, \ref{tab:correct_scenario3}, and Model 3 for Scenario 1 and 2 in Tables \ref{tab:rest_scenario1} and \ref{tab:rest_scenario2}; Model 3.1-3.3 for Scenario 3 in Table \ref{tab:rest_scenario3}). 
    \item Model misspecification resulted in biased parameter estimation, and undercoverage. 
    \begin{itemize}
        \item Misspecification of the longitudinal submodel seemed to bias the parameters associated with the longitudinal submodel (Model 2.1 for Scenario 1 in Table \ref{tab:rest_scenario1}).
        \item Misspecification of the terminal event submodel led to bias in parameters for both submodels (Model 2.1 for Scenario 2 in Table \ref{tab:rest_scenario2}, Model 2 for Scenario 3 in Table \ref{tab:rest_scenario3}). 
        \item Misspecification in latent factor structure led to bias in parameters for both submodels (Model 2.1 and 2.2 for Scenario 2 in Table \ref{tab:rest_scenario2}). 
        \item Misspecification of the dependence structure resulted in bias across parameters for both submodels. It influenced the dependence parameters and the parameters in the longitudinal submodel the most (Model 2.2 for Scenario 1 in Table \ref{tab:rest_scenario1}, Model 2.2 for Scenario 2 in Table \ref{tab:rest_scenario2}). 
    \end{itemize}
    \item Including a shared frailty term in a joint bivariate discrete-time framework model can become functionally equivalent to including a random effect if the scale multiplier $\alpha$ has value zero (Model 3 for Scenario 2). The posterior distribution for the scale multiplier, fit in Model 3 for Scenario 2, centered around zero, as expected (Figure \ref{fig: postdens alpha}).
    \item When the correct model is nested in an overparameterized model (Model 3.1-3.3 for Scenario 3), the posterior distribution for the additional model parameters were centered around zero (Figure \ref{fig: post density add components}). 
\end{itemize}

% correct model specification for Scenario 1
\begin{table}[h!]
\caption{Results for Model 1 (Correct) for Scenario 1, averaged over $R = 500$ repetitions; Sim. = simulation parameter; PM = posterior mean; SD = standard deviation; ESE = empirical standard error; Abs. = absolute; Rel. = relative; Cov. = coverage for highest posterior density credible interval.}
\scriptsize
\centering
\begin{tabular}{l r r rr rr r}
\multicolumn{8}{c}{\textbf{Scenario 1 - correct model specification (Model 1)}} \\
\toprule
& Sim. & PM & \multicolumn{2}{c}{SD} & \multicolumn{2}{c}{Bias} & Cov. [\%] \\
 \cmidrule(lr){4-5} \cmidrule(lr){6-7} 
&  &  & $\overline{\text{\textit{SD}}}$ & \textit{ESE} & \textit{abs.} & \textit{rel.} [\%] &  \\ 
\midrule
\multicolumn{8}{l}{\textit{Terminal event submodel}} \\
$\tilde{\lambda}_1$      & -7 & -6.88 & 0.23 & 0.22 & 0.12 & -1.72 & 92.0 \\ 
$\xi_1$        & 0.30  & 0.30  & 0.08 & 0.08 & -0.004 & -1.42 & 96.0 \\ 
$\xi_2$        & 0.20  & 0.19  & 0.04 & 0.04 & -0.007 & -3.61 & 94.6 \\
$\vartheta$    & -0.50 & -0.49 & 0.03 & 0.03 & 0.01 & -2.96 & 93.0 \\  
\midrule
\multicolumn{8}{l}{\textit{Longitudinal submodel}} \\
$\tilde{\zeta}_1$      & -1.05 & -1.03 & 0.03 & 0.03 & 0.02 & -1.51 & 92.8 \\ 
$\tilde{\zeta}_2$      & -0.07 & -0.07 & 0.002 & 0.002 & 0.001 & -1.28 & 93.4 \\ 
$\beta_1$      & -0.20 & -0.20 & 0.02 & 0.02 & 0.001 & -0.54 & 95.2 \\ 
$\beta_2$      & 0.60 & 0.60 & 0.03 & 0.03 & -0.009 & -1.43 & 94.0 \\
$\eta$         & 0.75 & 0.75 & 0.006 & 0.006 & 0.004 & 0.47 & 91.4 \\ 
$\sigma_{\varepsilon}$ & 0.8 & 0.8 & 0.005 & 0.005 & -0.0000 & -0.003 & 93.0\\ 
\midrule
\multicolumn{8}{l}{\textit{Latent factors}} \\
$\alpha$ & -0.25 & -0.24 & 0.06 & 0.06 & 0.01 & -2.88 & 95.2 \\
$\sigma_{\gamma}$ & 1.50 & 1.50 & 0.07 & 0.07 & -0.002 & -0.15 & 96.4 \\
\bottomrule
\end{tabular}
\label{tab:correct_scenario1}
\end{table}

% incorrect / flexible model specification for Scenario 1
\begin{table}[h!]
\caption{Results for Model 2.1, 2.2 and 3 for Scenario 1, averaged over $R = 500$ repetitions; Cov. = coverage of 95\% highest posterior density credible interval; SD = standard deviation; ESE = empirical standard error.}
\scriptsize
\centering
\begin{tabular}{l rr rr rrrr}
\multicolumn{9}{c}{\textbf{Scenario 1}} \\
\toprule
& \multicolumn{4}{c}{\textit{incorrect model specification}} 
& \multicolumn{4}{c}{\textit{flexible model specification}} \\
\cmidrule(lr){2-5} \cmidrule(lr){6-9}
 & \multicolumn{2}{c}{\textbf{Model 2.1}} 
 & \multicolumn{2}{c}{\textbf{Model 2.2}} 
 & \multicolumn{4}{c}{\textbf{Model 3}} \\
\cmidrule(lr){2-3} \cmidrule(lr){4-5} \cmidrule(lr){6-9}
 & \multicolumn{1}{c}{abs. Bias}
 & \multicolumn{1}{c}{Cov. [\%]}
 & \multicolumn{1}{c}{abs. Bias}
 & \multicolumn{1}{c}{Cov. [\%]}
 & \multicolumn{1}{c}{abs. Bias}
 & \multicolumn{1}{c}{Cov. [\%]} 
 & $\overline{\text{\textit{SD}}}$ 
 & \textit{ESE}
 \\
\midrule
\multicolumn{7}{l}{\textit{Terminal event submodel}} \\
$\xi_1$     
 & 0.01 & 95.4
 & -0.02 & 96.0
 & 0.01 & 95.0 & 0.08 & 0.08 \\

$\xi_2$     
 & 0.001 & 94.6
 & -0.01  & 93.4
 & 0.002 & 94.4 & 0.04 & 0.04 \\

$\vartheta$ 
 & -0.02 & 93.2
 & 0.06  & 41.2
 & -0.01 & 94.0 & 0.03 & 0.03 \\

\midrule
\multicolumn{7}{l}{\textit{Longitudinal submodel}} \\

$\beta_1$   
 & -0.55 & 0.0
 & 0.11  & 0.0
 & 0.001 & 95.0 & 0.02 & 0.02 \\

$\beta_2$   
 & 1.61 & 0.0
 & -0.33 & 0.0
 & -0.01 & 93.4 & 0.03 & 0.03 \\

$\eta$      
 &  & 
 & 0.15 & 0.0
 & 0.003 & 91.8 & 0.006 & 0.006 \\

$\sigma_{\varepsilon}$ 
 & 0.36 & 0.0
 & 0.03 & 0.0
 & -0.0001 & 93.6 & 0.005 & 0.005 \\

\midrule
\multicolumn{7}{l}{\textit{Latent factors}} \\

$\alpha$    
 & 0.01 & 94.0
 &  & 
 & -0.002 & 94.0 & 0.06 & 0.06 \\

$\sigma_{\gamma}$ 
 & 0.05 & 90.6
 &  & 
 & -0.01 & 95.6 & 0.07 & 0.07 \\

\bottomrule
\end{tabular}
\label{tab:rest_scenario1}
\end{table}

% correct model specification for Scenario 2 
\begin{table}[h!]
\caption{Results for Model 1 (Correct) for Scenario 2, averaged over $R = 500$ repetitions; Sim. = simulation parameter; PM = posterior mean; SD = standard deviation; ESE = empirical standard error; Abs. = absolute; Rel. = relative; Cov. = coverage for highest posterior density credible interval.}
\scriptsize
\centering
\begin{tabular}{l r r rr rr r}
\multicolumn{8}{c}{\textbf{Scenario 2 - correct model specification (Model 1)}} \\
\toprule
& Sim. & PM & \multicolumn{2}{c}{SD} & \multicolumn{2}{c}{Bias} & Cov. [\%] \\
 \cmidrule(lr){4-5} \cmidrule(lr){6-7} 
&  &  & $\overline{\text{\textit{SD}}}$ & \textit{ESE} & \textit{abs.} & \textit{rel.} [\%] &  \\ 
\midrule
\multicolumn{8}{l}{\textit{Terminal event submodel}} \\
$\tilde{\lambda}_1$ & -7.00 & -7.07 & 0.51 & 0.51 & -0.07 & 1.04 & 93.0 \\ 
$\xi_1$        & -0.5 & -0.5 & 0.07 & 0.08 & -0.005 & 0.94 & 94.6 \\ 
$\xi_2$        & -0.2 & -0.2 & 0.07 & 0.07 & -0.003 & 1.29 & 94.4 \\
$\vartheta$    & 0.2 & 0.2 & 0.03 & 0.03 & 0.004 & 2.07 & 92.2 \\  
$\varphi_1$    & 0.009 & 0.009 & 0.003 & 0.004 & 0.0002 & 1.94 & 92.6 \\
$\varphi_2$    & $\frac{1}{160^2}$ & 0.0001 & 0.0004 & 0.0004 & 0.0000 & 102 & 93.8 \\
\midrule
\multicolumn{8}{l}{\textit{Longitudinal submodel}} \\
$\tilde{\zeta}_1$ & 4.9 & 5.0 & 0.4 & 0.4 & 0.06 & 1.14 & 94.4 \\
$\tilde{\zeta}_2$ & 0.12 & 0.12 & 0.001 & 0.001 & 0.00 & -0.006 & 94.4 \\ 
$\tilde{\zeta}_3$ & 0.004 & 0.004 & 0.0002 & 0.0002 & 0.0000 & 0.25 & 95.0 \\
$\beta_1$      & 5.00 & 4.97 & 0.24 & 0.22 & -0.03 & -0.7 & 96.4 \\ 
$\beta_2$      & 4.00 & 3.98 & 0.23 & 0.22 & -0.02 & -0.58 & 95.2 \\
$\sigma_{\varepsilon}$ & 1.00 & 1.00 & 0.006 & 0.007 & -0.0004 & -0.04 & 94.2 \\
\midrule
\multicolumn{8}{l}{\textit{Latent factors}} \\
$\sigma_{b}$ & 2.1 & 2.1 & 0.09 & 0.08 & 0.004 & 0.17 & 96.4 \\
\bottomrule
\end{tabular}
\label{tab:correct_scenario2}
\end{table}

% incorrect / flexible model specification for Scenario 2
\begin{table}[h!]
\caption{Results for Model 2.1, 2.2 and 3 for Scenario 2, averaged over $R = 500$ repetitions; Cov. = coverage of 95\% highest posterior density credible interval; SD = standard deviation; ESE = empirical standard error.}
\scriptsize
\centering
\begin{tabular}{l rr rr rrrr}
\multicolumn{9}{c}{\textbf{Scenario 2}} \\
\toprule
& \multicolumn{4}{c}{\textit{incorrect model specification}} 
& \multicolumn{4}{c}{\textit{flexible model specification}} \\
\cmidrule(lr){2-5} \cmidrule(lr){6-9}
 & \multicolumn{2}{c}{\textbf{Model 2.1}} 
 & \multicolumn{2}{c}{\textbf{Model 2.2}} 
 & \multicolumn{4}{c}{\textbf{Model 3}} \\
\cmidrule(lr){2-3} \cmidrule(lr){4-5} \cmidrule(lr){6-9}
 & \multicolumn{1}{c}{abs. Bias}
 & \multicolumn{1}{c}{Cov. [\%]}
 & \multicolumn{1}{c}{abs. Bias}
 & \multicolumn{1}{c}{Cov. [\%]}
 & \multicolumn{1}{c}{abs. Bias}
 & \multicolumn{1}{c}{Cov. [\%]} 
 & $\overline{\text{\textit{SD}}}$ 
 & \textit{ESE}
 \\
\midrule
\multicolumn{7}{l}{\textit{Terminal event submodel}} \\
$\xi_1$ &  0.23 & 3.0 & 0.04 & 87.6 & -0.006 & 94.6 & 0.07 & 0.07 \\ 
$\xi_2$ &  0.09 & 76.6 & 0.02 & 93.6 & -0.003 & 94.4 & 0.07 & 0.07\\
$\vartheta$ &  0.06 & 60.2 & -0.06 & 39.8 & 0.004 & 92.4 & 0.03 & 0.03\\
$\varphi_1$ &  &  & -0.004 & 73.6 & 0.0002 & 93.0 & 0.003 &  0.004 \\
$\varphi_2$ &  &  & -0.0001 & 93.0 & 0.0000 & 93.6 & 0.0004 & 0.0004 \\
\midrule
\multicolumn{7}{l}{\textit{Longitudinal submodel}} \\
$\beta_1$  & -0.52 & 55.0 & -0.02 & 25.6 & -0.03 & 95.8 & 0.24 & 0.22 \\ 
$\beta_2$  & -0.41 & 66.6 & 0.001 & 27.6 & -0.02 & 95.6 & 0.23 & 0.22 \\ 
$\sigma_{\varepsilon}$ & -0.0004 & 94.0 & 1.31 & 0.0 & -0.0004 & 94.2 & 0.006 & 0.007 \\ 
\bottomrule
\end{tabular}
\label{tab:rest_scenario2}
\end{table}

% correct model specification for Scenario 3 
\begin{table}[h!]
\caption{Results for Model 1 (Correct) for Scenario 3, averaged over $R = 500$ repetitions; Sim. = simulation parameter; PM = posterior mean; SD = standard deviation; ESE = empirical standard error; Abs. = absolute; Rel. = relative; Cov. = coverage for highest posterior density credible interval.}
\scriptsize
\centering
\begin{tabular}{l r r rr rr r}
\multicolumn{8}{c}{\textbf{Scenario 3 - correct model specification (Model 1)}} \\
\toprule
& Sim. & PM & \multicolumn{2}{c}{SD} & \multicolumn{2}{c}{Bias} & Cov. [\%] \\
 \cmidrule(lr){4-5} \cmidrule(lr){6-7} 
&  &  & $\overline{\text{\textit{SD}}}$ & \textit{ESE} & \textit{abs.} & \textit{rel.} [\%] &  \\ 
\midrule
\multicolumn{8}{l}{\textit{Terminal event submodel}} \\
$\tilde{\lambda}_1$     -8 & -7.25 & 1.56 & 1.58 & 0.75 & -9.38 & 92.4 \\ 
$\tilde{\lambda}_2$     & 0.25 & 0.26 & 0.03 & 0.03 & 0.01 & 5.53 & 93.6 \\ 
$\xi_1$        & 0.5 & 0.5 & 0.07 & 0.07 & 0.003 & 0.68 & 92.8 \\  
$\xi_2$        & 0.2 & 0.2 & 0.06 & 0.06 & 0.003 & 1.33 & 95.6 \\  
$\vartheta$    & 0.05 & 0.04 & 0.02 & 0.02 & -0.007 & -14.41 & 92.4 \\   
\midrule
\multicolumn{8}{l}{\textit{Longitudinal submodel}} \\
$\tilde{\zeta}_1$       & 90 & 90.113 & 0.41 & 0.42 & 0.13 & 0.15 & 92.4 \\
$\tilde{\zeta}_2$       & 1.5 & 1.5 & 0.004 & 0.004 & 0.0001 & 0.01 & 93.1 \\
$\beta_1$      & 3 & 2.97 & 0.21 & 0.22 & -0.03 & -0.96 & 92.8  \\ 
$\beta_2$      & 3 & 2.97 & 0.21 & 0.22 & -0.03 & -0.98 & 95.4 \\
$\beta_3$      & 10 & 9.92 & 0.27 & 0.27 & -0.08 & -0.76 & 93.4 \\
$\sigma_{\varepsilon}$ & 2 & 2 & 0.015 & 0.015 & -0.0007 & -0.04 & 93.2 \\ 
\midrule
\multicolumn{8}{l}{\textit{Latent factors}} \\
$\alpha$ & 0.5 & 0.51 & 0.04 & 0.04 & 0.005 & 1.08 & 93.0 \\ 
$\sigma_{\gamma}$ & 2.5 & 2.51 & 0.11 & 0.11 & 0.01 & 0.52 & 94.6 \\
\bottomrule
\end{tabular}
\label{tab:correct_scenario3}
\end{table}

\clearpage

\begin{landscape}
    % incorrect / flexible model specification for Scenario 3 
\begin{table}[h!]
\caption{Results for Model 2, 3.1, 3.2 and 3.3 for Scenario 3, averaged over $R = 500$ repetitions; Cov. = coverage of 95\% highest posterior density credible interval; SD = standard deviation; ESE = empirical standard error.}
\scriptsize
\centering
\begin{tabular}{l rr rrrr rrrr rrrr}
\multicolumn{15}{c}{\textbf{Scenario 3}} \\
\toprule
& \multicolumn{2}{c}{\makecell{\textit{incorrect model } \\ \textit{specification}}} 
& \multicolumn{12}{c}{\textit{flexible model specification}} \\
\cmidrule(lr){2-3} \cmidrule(lr){4-15}
 & \multicolumn{2}{c}{\textbf{Model 2}} 
 & \multicolumn{4}{c}{\textbf{Model 3.1}} 
 & \multicolumn{4}{c}{\textbf{Model 3.2}} 
 & \multicolumn{4}{c}{\textbf{Model 3.3}} \\
\cmidrule(lr){2-3} \cmidrule(lr){4-7} \cmidrule(lr){8-11} \cmidrule(lr){12-15}
 & \multicolumn{1}{c}{abs. Bias}
 & \multicolumn{1}{c}{Cov. [\%]}
 & \multicolumn{1}{c}{abs. Bias}
 & \multicolumn{1}{c}{Cov. [\%]}
 & $\overline{\text{\textit{SD}}}$ 
 & \textit{ESE}
 & \multicolumn{1}{c}{abs. Bias}
 & \multicolumn{1}{c}{Cov. [\%]}
 & $\overline{\text{\textit{SD}}}$ 
 & \textit{ESE}
 & \multicolumn{1}{c}{abs. Bias}
 & \multicolumn{1}{c}{Cov. [\%]} 
 & $\overline{\text{\textit{SD}}}$ 
 & \textit{ESE} \\
\midrule
\multicolumn{9}{l}{\textit{Terminal event submodel}} \\
$\xi_1$     
 & -0.11 & 61.2 & 0.01 & 92.4 & 0.07 & 0.07 & 0.003 & 92.8 & 0.07 & 0.07 & 0.01 & 91.8 & 0.07 & 0.08 \\ 
$\xi_2$ 
 & -0.04 & 89.0 & 0.005 & 95.2 & 0.07 & 0.07 & 0.003 & 95.6 & 0.06 & 0.06 & 0.005 & 95.6 & 0.07 & 0.07 \\
$\vartheta$ 
 & 0.12 & 0.00 & 0.00 & 94.4 & 0.02 & 0.02 & -0.007 & 92.6 & 0.02 & 0.02 & -0.003 & 94.8 & 0.03 & 0.03 \\
\midrule
\multicolumn{9}{l}{\textit{Longitudinal submodel}} \\
$\beta_1$ 
 & 0.41 & 63.2 & 0.003 & 93.2 & 0.2 & 0.2 & -0.03 & 92.8 & 0.2 & 0.2 & 0.002 & 93.4 & 0.2 & 0.2 \\ 
$\beta_2$ 
 & 0.42 & 63.6 & 0.003 & 94.2 & 0.2 & 0.2 & -0.03 & 95.0 & 0.2 & 0.2 & 0.002 & 94.8 & 0.2 & 0.2 \\ 
$\beta_3$   
 & 1.32 & 1.2 & 0.006 & 93.6 & 0.3 & 0.3 & -0.08 & 93.8 & 0.3 & 0.3 & 0.002 & 93.2 & 0.3 & 0.3 \\ 
$\sigma_{\varepsilon}$ 
 & -0.0007 & 93.0 & -0.0009 & 92.6 & 0.015 & 0.015 & -0.0007 & 92.8 & 0.015 & 0.015 & -0.0009 & 93.2 & 0.015 & 0.015 \\ 
\midrule
\multicolumn{9}{l}{\textit{Latent factors}} \\
$\alpha$    
 & -0.06 & 59.0 & 0.01 & 92.4 & 0.04 & 0.04 & 0.006 & 93.0 & 0.04 & 0.04 & 0.01 & 91.8 & 0.04 & 0.04 \\ 
$\sigma_{\gamma}$ 
 & 0.11 & 85.6 & -0.0002 & 94.4 & 0.1 & 0.1 & 0.01 & 94.6 & 0.1 & 0.1 & 0.0003 & 94.6 & 0.1 & 0.1 \\ 
\bottomrule
\end{tabular}
\label{tab:rest_scenario3}
\end{table}
\end{landscape}

\clearpage

\subsubsection{Marginal model selection criteria for candidate models}
\label{app: model selection results}
The empirical mean for the marginal model selection criteria is reported, as well as their empirical standard deviation over $R = 500$ repetitions.
Across simulation scenarios, we also fit the conditional model selection criteria. 
However, the conditional BIC, WAIC and $\text{ELPD}_{\scriptscriptstyle \text{PSIS-MM}}$ failed to pass diagnostic checks, and as such, rendered crude approximations of predictive performance. 
Such behavior for models with subject-specific latent factors has been discussed previously \citep{vehtari2017practical, merkle2019bayesian}.

We make the following observations in terms of marginal model selection criteria from the simulation study: 
\begin{itemize}
    \item Across simulation scenarios, either the correctly specified or flexibly specified model achieved the lowest averaged marginal BIC, marginal WAIC and marginal $\text{ELPD}_{\scriptscriptstyle \text{PSIS-MM}}$.
    \item Overly simplistic misspecified models exhibited poorer performance (Model 2.1-2.2 for Scenario 1 and Scenario 2, Model 2 for Scenario 3 in Table \ref{tab: model selection values}). 
    \item Overparameterization is penalized. However, since few additional parameters were included, the penalty was limited in numeric value.  
    \begin{itemize}
        \item For Scenario 1, Model 3 required estimating 4 additional parameters. Model 1 achieved lowest BIC and $\text{ELPD}_{\scriptscriptstyle \text{PSIS-MM}}$ values on average (Table \ref{tab: model selection values}) and in the majority of repetitions (Table \ref{tab: model selection proportions}). 
        \item For Scenario 2, Model 3 required estimating one additional parameter. Model 1 achieved lower BIC but had identical $\text{ELPD}_{\scriptscriptstyle \text{PSIS-MM}}$ values to Model 3 on average (Table \ref{tab: model selection values}) and across repetitions (Table \ref{tab: model selection proportions}) 
        \item For Scenario 3, Model 3.2 added one parameter, Model 3.3 added 2 parameters and Model 3.1 included 3 additional parameters. The model selection criteria show the same ranking, on average (Table \ref{tab: model selection values}) and across repetitions (Table \ref{tab: model selection proportions}). 
    \end{itemize}
    \item All model selection criteria selected the correct model specification the majority of times, followed by flexibly specified models with few additional parameters (Table \ref{tab: model selection proportions}). 
\end{itemize}

\begin{table}[h!]
\caption{Marginal model selection criteria across candidate models for Scenario 1, 2 and 3, averaged over $R = 500$ repetitions; loglik = log likelihood; BF = Bayes Factor; $\text{ELPD}_{\scriptscriptstyle \text{PSIS-MM}}^*$ = $\text{ELPD}_{\scriptscriptstyle \text{PSIS-MM}}$ on divergence scale; empirical standard deviation reported in brackets $()$.}
\scriptsize
\centering
\begin{tabular}{l rrr}
\toprule
 & \multicolumn{3}{c}{Marginal} \\
\cmidrule(lr){2-4}
 & \multicolumn{1}{c}{\textit{BIC}}
 & \multicolumn{1}{c}{\textit{WAIC}}
 & \multicolumn{1}{c}{\textit{$\text{ELPD}_{\scriptscriptstyle \text{PSIS-MM}}^*$}} \\
  \midrule
  \multicolumn{3}{l}{\textbf{Scenario 1}} \\
    \multicolumn{3}{l}{\hspace{0.2cm} \textit{correct model specification}} \\
\hspace{0.5cm} Model 1 
& 36716\,(619) 
& 35898\,(614) 
& 36672\,(619) \\ 
\multicolumn{3}{l}{\hspace{0.2cm} \textit{incorrect model specification}} \\
\hspace{0.5cm} Model 2.1 
& 48331\,(853) 
& 48133\,(872) 
& 49021\,(869) \\ 
\hspace{0.5cm} Model 2.2 
& 37380\,(635) 
& 37216\,(635) 
& 37216\,(635) \\ 
\multicolumn{3}{l}{\hspace{0.2cm} \textit{flexible model specification}} \\
\hspace{0.5cm} Model 3 
& 36731\,(619) 
& 35899\,(614) 
& 36675\,(619) \\ 
\midrule 
\multicolumn{3}{l}{\textbf{Scenario 2}} \\
\multicolumn{3}{l}{\hspace{0.2cm} \textit{correct model specification}} \\
\hspace{0.5cm} Model 1 
& 40260\,(788) & 43247\,(1233) & 43278\,(1043) \\
\multicolumn{3}{l}{\hspace{0.2cm} \textit{incorrect model specification}} \\
\hspace{0.5cm} Model 2.1 
& 40504\,(806) & 43997\,(1468) & 43819\,(1169) \\ 
\hspace{0.5cm} Model 2.2 
& 59065\,(1535) & 58977\,(1539) & 58977\,(1539) \\
\multicolumn{3}{l}{\textit{flexible model specification}} \\
\hspace{0.5cm} Model 3 
& 40264\,(791) & 43245\,(1235) & 43278\,(1045) \\
\midrule
\multicolumn{3}{l}{\textbf{Scenario 3}} \\
\multicolumn{3}{l}{\hspace{0.2cm} \textit{correct model specification}} \\
\hspace{0.5cm} Model 1 
& 44853\,(830) & 44164\,(844) & 45093\,(844) \\ 
\multicolumn{3}{l}{\hspace{0.2cm} \textit{incorrect model specification}} \\
\hspace{0.5cm} Model 2 
& 44981\,(834) & 44369\,(859) & 45286\,(858) \\
\multicolumn{3}{l}{\textit{flexible model specification}} \\
\hspace{0.5cm} Model 3.1 
& 44870\,(830) & 44156\,(844) & 45087\,(844) \\
\hspace{0.5cm} Model 3.2 
& 44857\,(830) & 44164\,(842) & 45093\,(843) \\
\hspace{0.5cm} Model 3.3 
& 44863\,(830) & 44155\,(844) & 45086\,(844) \\
\bottomrule
\end{tabular}
\label{tab: model selection values}
\end{table}

\begin{table}[h!]
\caption{Proportion [\%] of model selected based on marginal model selection criteria across candidate models for Scenario 1, 2 and 3, averaged over $R = 500$ repetitions; marg. loglik = marginal log likelihood; $\text{ELPD}_{\scriptscriptstyle \text{PSIS-MM}}^*$ = $\text{ELPD}_{\scriptscriptstyle \text{PSIS-MM}}$ on divergence scale.}
\scriptsize
\centering
\begin{tabular}{l rrr}
\toprule
 \multirow{2}{3.5cm}{\textbf{Proportion of selection [\%] based on...}} & 
 \multicolumn{3}{c}{Marginal} \\
 \cmidrule(lr){2-4}
 & \multicolumn{1}{c}{\textit{BIC}}
 & \multicolumn{1}{c}{\textit{WAIC}}
 & \multicolumn{1}{c}{\textit{$\text{ELPD}_{\scriptscriptstyle \text{PSIS-MM}}^*$}} \\
  \midrule
  \multicolumn{3}{l}{\textbf{Scenario 1}} \\ 
\multicolumn{3}{l}{\hspace{0.2cm} \textit{correct model specification}} \\
\hspace{0.5cm} Model 1  & 100.00 & 77.60 & 85.60 \\ 
\multicolumn{3}{l}{\hspace{0.2cm} \textit{incorrect model specification}} \\
\hspace{0.5cm} Model 2.1 & 0 & 0 & 0 \\ 
\hspace{0.5cm} Model 2.2 & 0 & 0 & 0 \\ 
\multicolumn{3}{l}{\hspace{0.2cm} \textit{flexible model specification}} \\
\hspace{0.5cm} Model 3 & 0 & 22.40 & 14.40 \\
\midrule 
\multicolumn{3}{l}{\textbf{Scenario 2}} \\ 
\multicolumn{3}{l}{\hspace{0.2cm} \textit{correct model specification}} \\
\hspace{0.5cm} Model 1 & 53.60 & 48.60 & 49.00 \\  
\multicolumn{3}{l}{\hspace{0.2cm} \textit{incorrect model specification}} \\
\hspace{0.5cm} Model 2.1  & 0.20 & 0.20 & 0 \\
\hspace{0.5cm} Model 2.2 & 0 & 0 & 0 \\
\multicolumn{3}{l}{\hspace{0.2cm} \textit{flexible model specification}} \\
\hspace{0.5cm} Model 3  & 46.20 & 51.20 & 51.00 \\
\midrule 
  \multicolumn{3}{l}{\textbf{Scenario 3}} \\ 
\multicolumn{3}{l}{\hspace{0.2cm} \textit{correct model specification}} \\
\hspace{0.5cm} Model 1 & 60.80 & 7.80 & 9.00 \\
\multicolumn{3}{l}{\hspace{0.2cm} \textit{incorrect model specification}} \\
\hspace{0.5cm} Model 2 & 0 & 0 & 0 \\
\multicolumn{3}{l}{\hspace{0.2cm} \textit{flexible model specification}} \\
\hspace{0.5cm} Model 3.1 & 1.40 & 31.80 & 26.40 \\
\hspace{0.5cm} Model 3.2  & 31.00 & 26.40 & 31.40 \\ 
\hspace{0.5cm} Model 3.3 & 6.80 & 34.00 & 33.20 \\ 
\bottomrule
\end{tabular}
\label{tab: model selection proportions}
\end{table}

\clearpage

\section{Development of a joint prediction model for physical frailty and mortality in SCD-HeFT}
\subsection{Intermittent missingness for terminal events and longitudinal trajectories}
\label{app: intermittent missingness}
Intermittent missingness applies to the MLHFQ score trajectory. 
In terms of mortality, the data structure is inherently non-intermittent: either the death time is observed, or it is not. 
In the latter case, the absence of an observed death corresponds to right censoring. 
By contrast, the longitudinal outcome may be intermittently unobserved in certain time periods; that is the intermittent missingness indicator is $M_i(t) = 0$ for some $t \in [0, \tilde{T}_i]$.
The continuous-time indicator process, $M_i(t)$, maps onto the discrete-time intermittent missingness indicator $M_{k, i}$ non-uniquely; it is induced by the choice of discretization and the approach chosen to attribute observations from continuous to discrete time.  

\begin{figure}[h!]
    \centering
    \includegraphics[width=0.7\linewidth]{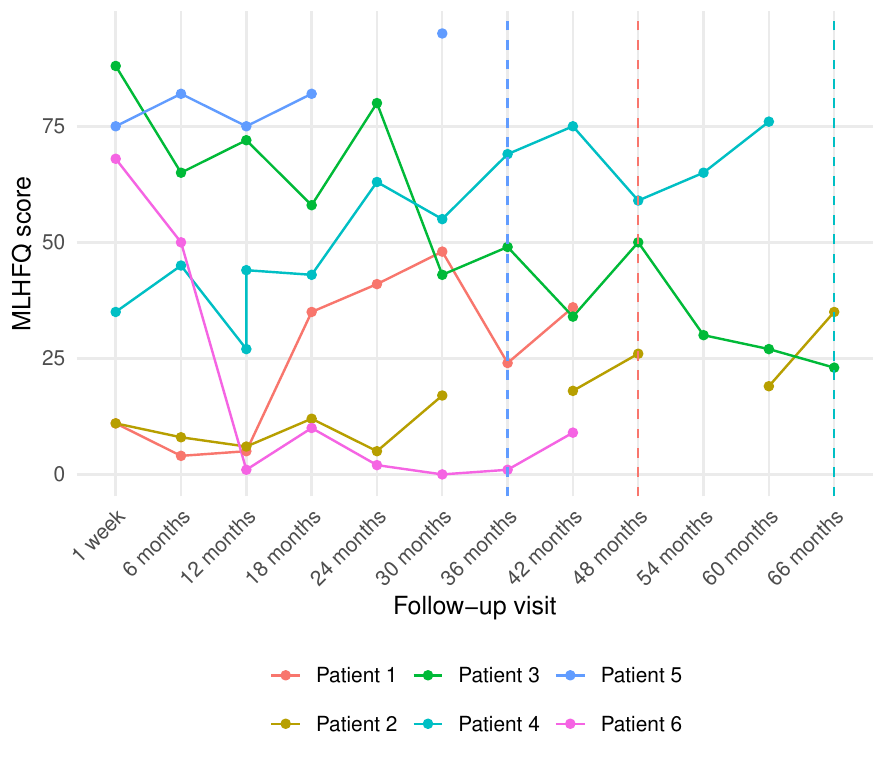}
    \caption{MLHFQ score trajectories for example patients over discrete time; dashed line visualizes patient death for patient 1, 4 and 5.}
    \label{fig:example_traj}
\end{figure}

We assume conditionally independent, or non-informative, intermittent missingness over the full discrete-time partition; that is  
\begin{align*}
    \forall k \in \{0, k_i^r\}: (N_{k, i}, Y_{k, i}) \indep M_{k, i} \; | \; \mathbf{X}_{i}
\end{align*}
Assuming we include a autoregressive component or global or local dependence structure of order 1, the observed data likelihood for patient $i$ follows as a product over observed data 
\begin{align*}
    \prod_{k: M_{k-1, i} = 1} \left[  (1 - \pi_{k, i}) \cdot f_Y\left(y_{k, i} | N_{k, i} = 0, \mathcal{H}_{k, i} \right) \right ] \cdot \pi_{k_i^r, i}^{\Delta_i}
\end{align*}

\clearpage

\subsection{Sample sizes for intermittent missingness strategies}
\label{app: data sample sizes}
Sample sizes depended on the strategy used to account for intermittent missingness. 
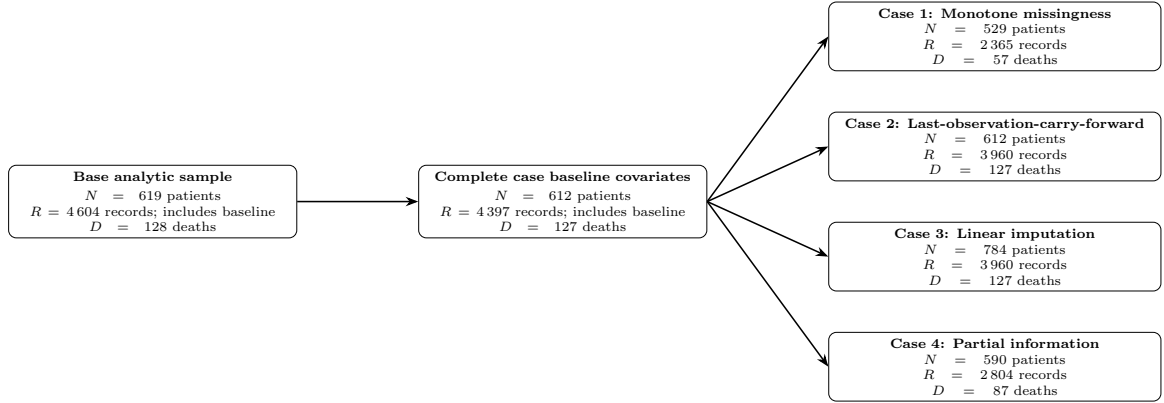
\begin{figure}[h!]
\centering
\resizebox{0.99\textwidth}{!}{%
\begin{tikzpicture}[
  node distance = 18mm and 12mm,
  box/.style = {
    draw,
    rounded corners,
    align=center,
    minimum height=10mm,
    text width=50mm,
    font=\scriptsize
  },
  casebox/.style = {
    draw,
    rounded corners,
    align=center,
    minimum height=10mm,
    text width=58mm,
    font=\scriptsize
  },
  arrow/.style = {-{Stealth[length=2.3mm]}, thick}
]

\node[box] (base) {
  \textbf{Base analytic sample}\\[2pt]
  $N = 619$ patients\\
  $R = 4\,604$ records; includes baseline \\
  $D = 128$ deaths
};

\node[box, right=22mm of base] (cc) {
  \textbf{Complete case baseline covariates}\\[2pt]
  $N = 612$ patients\\
  $R = 4\,397$ records; includes baseline \\
  $D = 127$ deaths
};

% --- Branching cases (2x2 grid to the right)
\node[casebox, right=22mm of cc, yshift=30mm] (c1) {
  \textbf{Case 1: Monotone missingness}\\
  $N = 529$ patients\\
  $R = 2\,365$ records\\ 
  $D = 57$ deaths
};

\node[casebox, right=22mm of cc, yshift=10mm] (c2) {
  \textbf{Case 2: Last-observation-carry-forward}\\
  $N = 612$ patients\\
  $R = 3\,960$ records\\
  $D = 127$ deaths 
};

\node[casebox, right=22mm of cc, yshift=-10mm] (c3) {
  \textbf{Case 3: Linear imputation}\\
  $N = 784$ patients\\
  $R = 3\,960$ records\\
  $D = 127$ deaths 
};

\node[casebox, right=22mm of cc, yshift=-30mm] (c4) {
  \textbf{Case 4: Partial information}\\
  $N = 590$ patients\\
  $R = 2\,804$ records\\
  $D = 87$ deaths
};

% --- Arrows
\draw[arrow] (base.east) -- (cc.west); 
\draw[arrow] (cc.east) -- (c1.west);
\draw[arrow] (cc.east) -- (c2.west);
\draw[arrow] (cc.east) -- (c3.west);
\draw[arrow] (cc.east) -- (c4.west);

\end{tikzpicture}
}
\caption{Flow diagram of analytic sample construction for SCD-HeFT data where patients must have been measured at time origin available, had complete-case for baseline covariates; and were recorded beyond time origin under four strategies to handle intermittent missingness.}
\label{fig:flowchart sample sizes}
\end{figure}

\clearpage

\subsection{Results for strategies to handle intermittent missingness}
\label{app: results IM strategies}
We implemented multiple strategies to handle intermittent missingness. 
When we focus on comparing the results across the four strategies to handle intermittent missingness,  the posterior mean estimates aligned closely for covariate coefficients, the baseline discrete-time hazard and longitudinal baseline marker trend between strategies. 
The imputation artificially caused a stronger autoregressive pattern, leading to potentially biased, higher estimates for the autoregressive parameter $\eta$. 
As expected, the width of the credible intervals was narrower for strategies which excluded less information due to intermittent missingness; that is for LOCF and linear interpolation, followed by partial information. 
The credible intervals were the widest when imposing monotone missingness.

\begin{figure}[h!]
    \centering
    \includegraphics[width=0.95\linewidth]{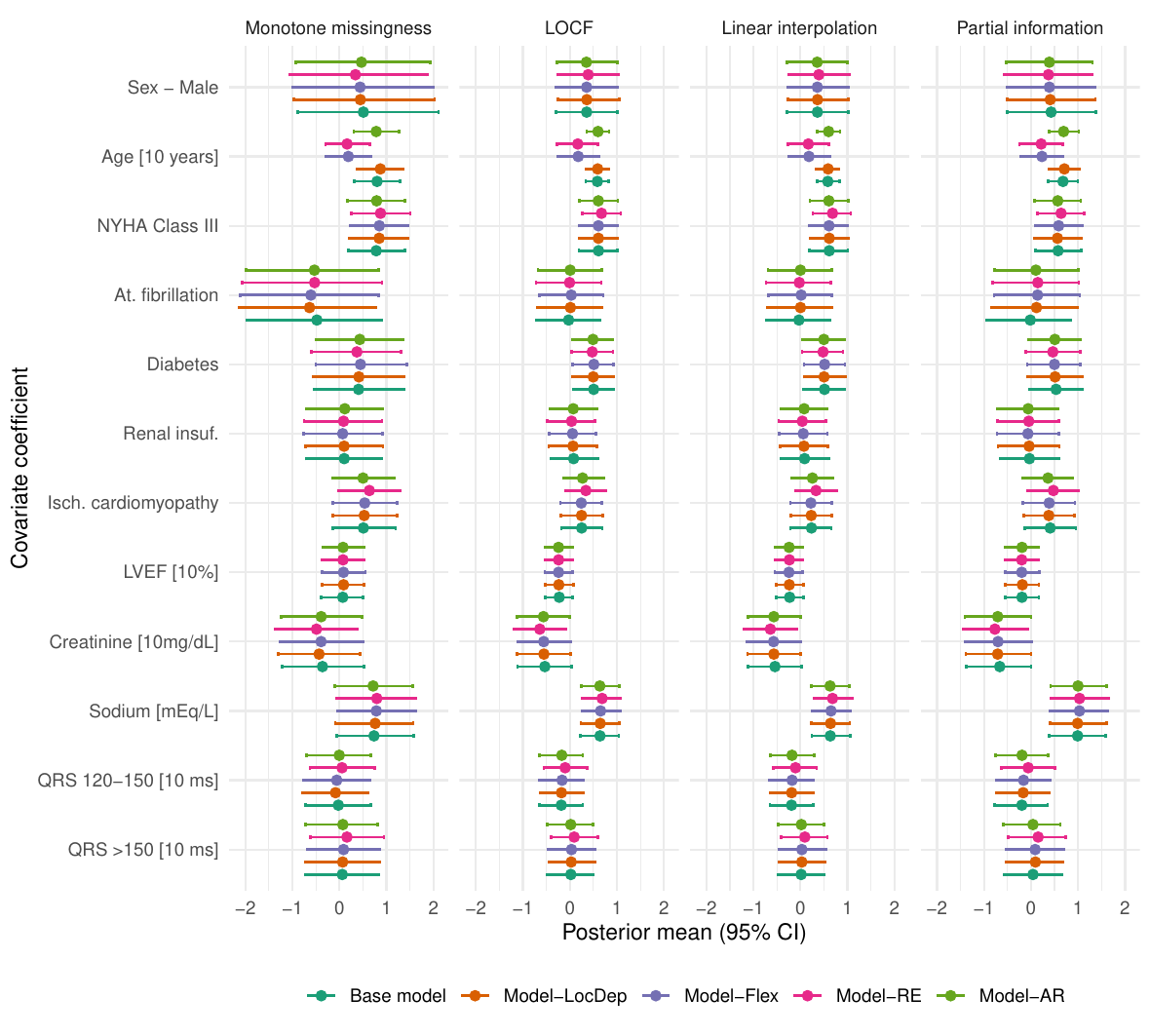}
    \caption{Posterior mean and 95\% highest posterior density credible interval for covariate coefficients in terminal event submodel across intermittent missingness strategies.}
    \label{fig: TE parameter all strategies}
\end{figure}

\begin{figure}
    \centering
    \includegraphics[width=0.95\linewidth]{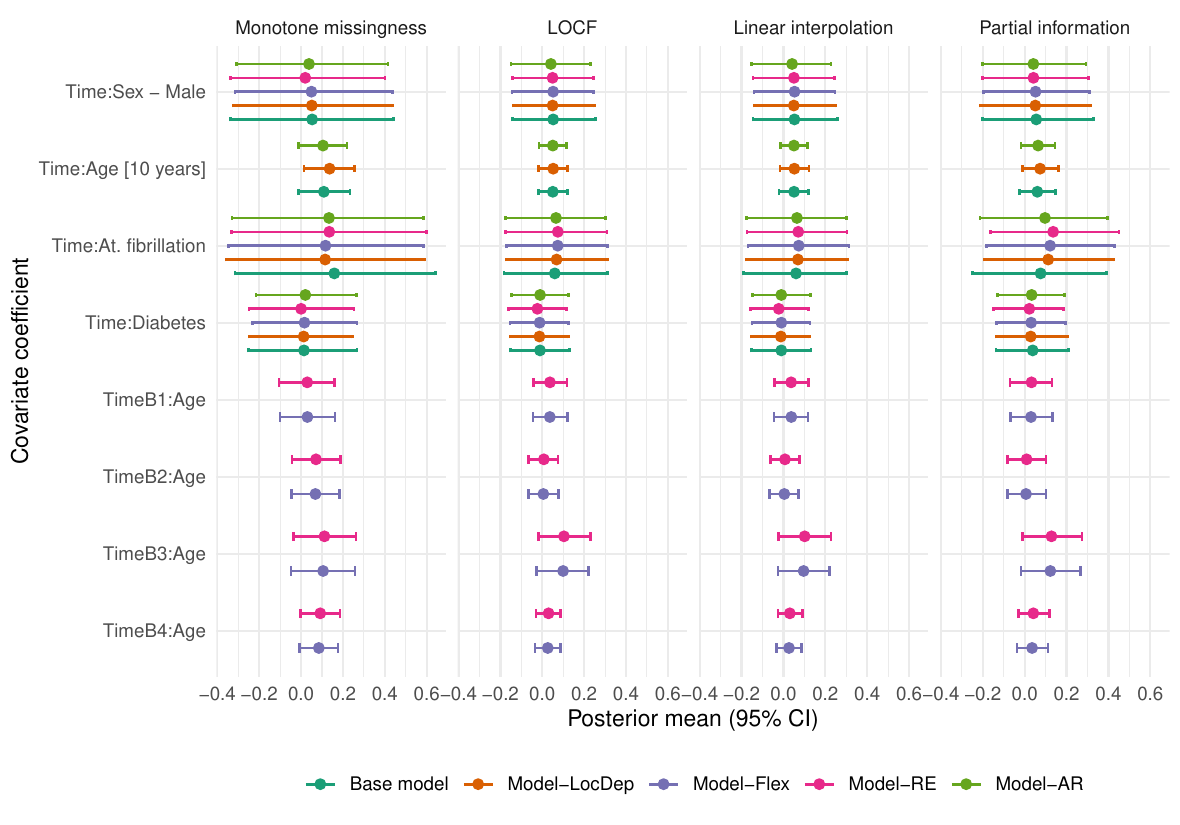}
    \caption{Posterior mean and 95\% highest posterior density credible interval for tiem and covariate interactions in terminal event submodel across intermittent missingness strategies.}
    \label{fig: TE time interaction parameter all strategies}
\end{figure}

\begin{figure}[h!]
    \centering
    \includegraphics[width=0.95\linewidth]{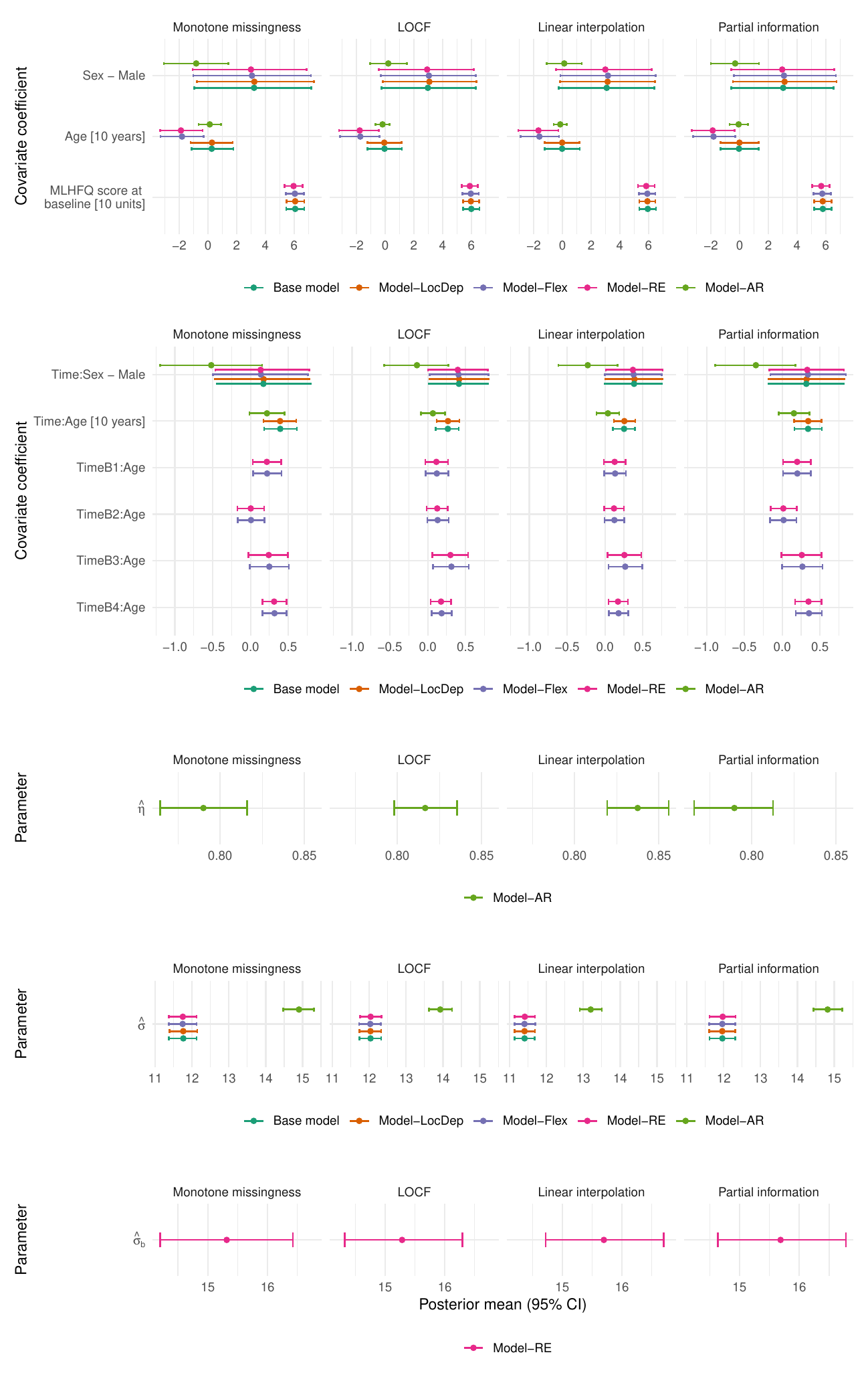}
    \caption{Posterior mean and 95\% highest posterior density credible interval for parameters in longitudinal submodel across intermittent missingness strategies.}
    \label{fig: LM parameter all strategies}
\end{figure}

\begin{figure}[h!]
    \centering
    \includegraphics[width=0.95\linewidth]{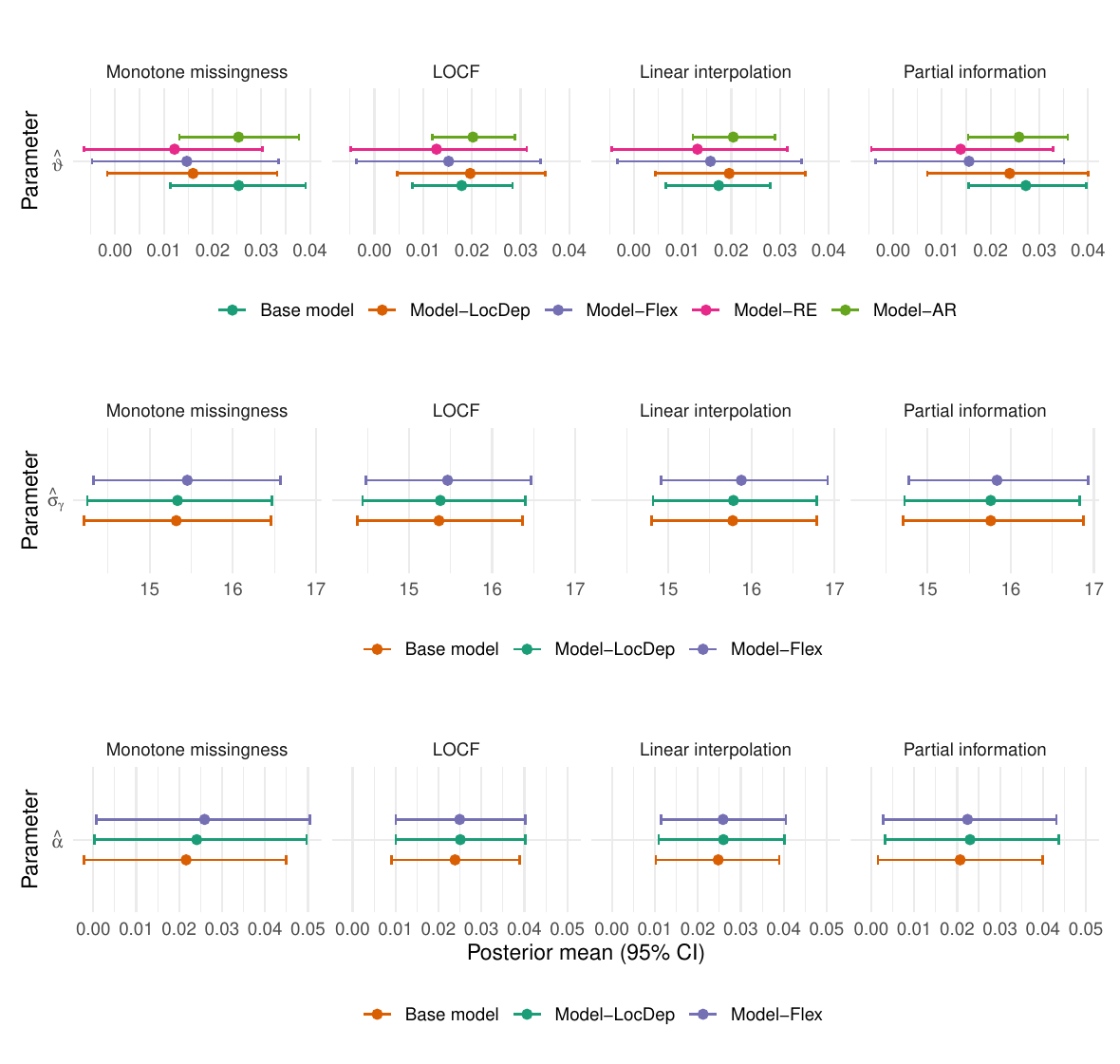}
    \caption{Posterior mean and 95\% highest posterior density credible interval for parameters that represent the dependence structure between the longitudinal trajectory and terminal event risk across intermittent missingness strategies.}
    \label{fig: LF parameter all strategies}
\end{figure}

\begin{figure}[h!]
    \centering
    
    \begin{subfigure}{0.95\linewidth}
        \centering
        \includegraphics[width=\linewidth]{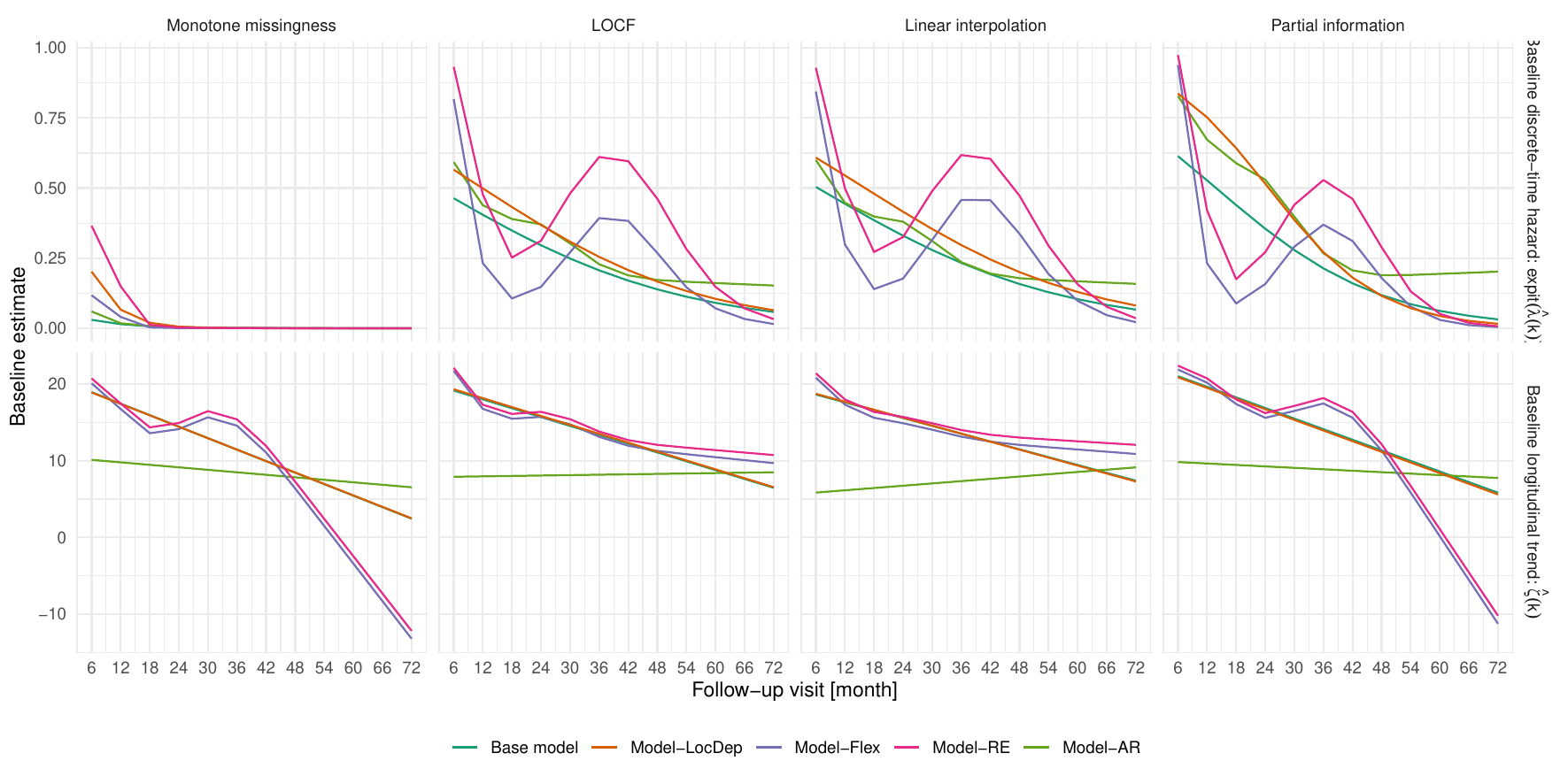}
        \caption{Baseline discrete-time hazard and longitudinal temporal trend.}
    \end{subfigure}
    
    \vspace{0.5em}
    
    \begin{subfigure}{0.95\linewidth}
        \centering
        \includegraphics[width=\linewidth]{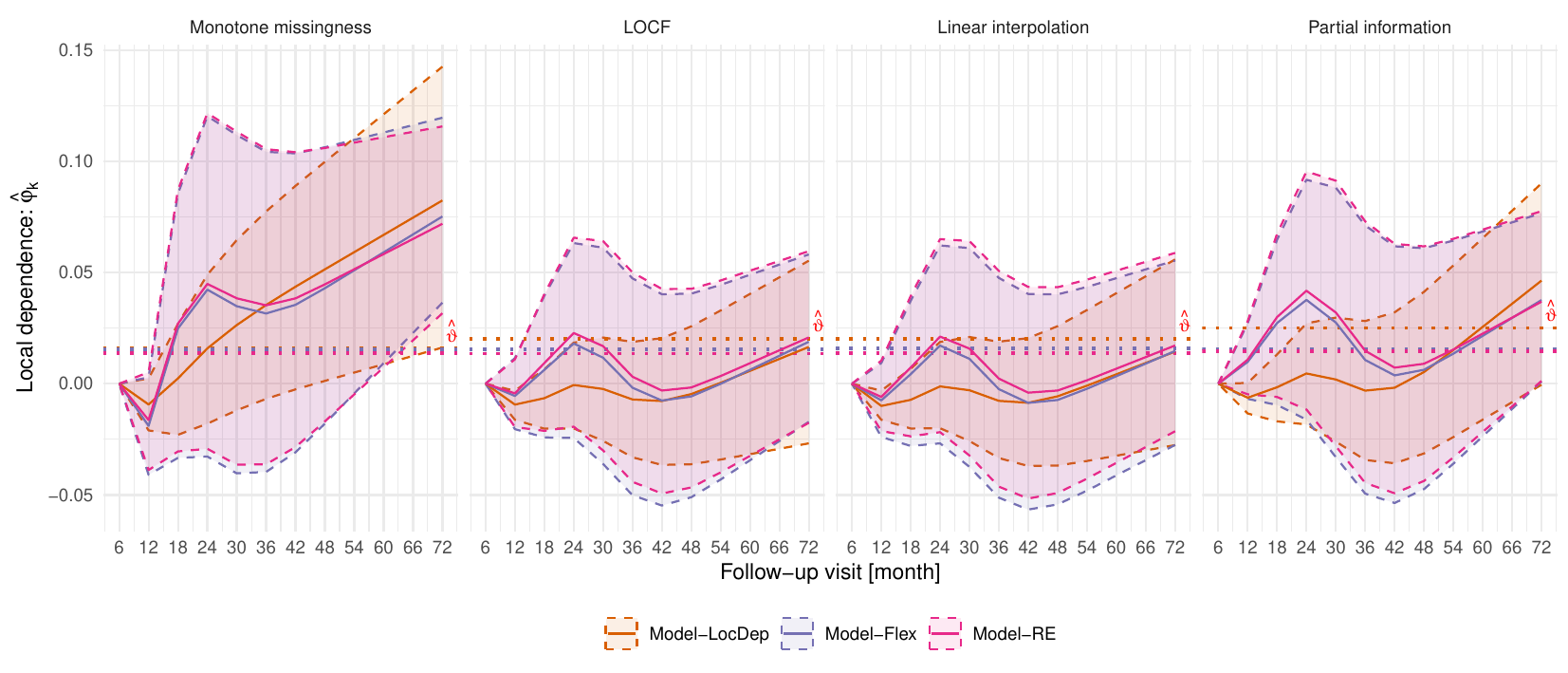}
        \caption{Local dependence $\varphi_k$ over time; red dotted line indicates global dependence $\vartheta$.}
    \end{subfigure}
    
    \caption{Posterior mean and 95\% highest posterior density credible intervals over discrete time across intermittent missingness strategies.}
    \label{fig: BaseHaz LocDep all strategies}
\end{figure}

%%  imsart-nameyear.bst  will be used to                   %%
%%  MR numbers will be added by VTeX.                      %%
%%                                                         %%
%%  Use  \cite{...} to cite references in text.             %%
%%                                                         %%
%%%%%%%%%%%%%%%%%%%%%%%%%%%%%%%%%%%%%%%%%%%%%%%%%%%%%%%%%%%%%

\clearpage
\bibliographystyle{plainnat} % Style BST file
\bibliography{Bib}       % Bibliography file (usually '*.bib')

\end{document}